\pdfoutput=1

\documentclass[12pt,a4paper]{article}

\usepackage[colorinlistoftodos]{todonotes}
\usepackage{pdflscape}
\usepackage{afterpage}

\usepackage{ifthen} 
\newboolean{pdflatex}
\setboolean{pdflatex}{true} 

\newboolean{articletitles}
\setboolean{articletitles}{true} 

\newboolean{uprightparticles}
\setboolean{uprightparticles}{false} 

\newboolean{inbibliography}
\setboolean{inbibliography}{false} 

\def\paperauthors{LHCb collaboration} 
\def\paperasciititle{Dalitz plot analysis of the D->K-K+K+ decay} 
\def\papertitle{Dalitz plot analysis of the \DToKKK decay} 
\def\paperkeywords{{High Energy Physics}, {LHCb}} 
\def\papercopyright{\the\year\ CERN for the benefit of the LHCb collaboration} 
\def\paperlicence{CC-BY-4.0 licence}
\def\paperlicenceurl{https://creativecommons.org/licenses/by/4.0/}

\usepackage[top=1in, bottom=1.25in, left=1in, right=1in]{geometry}

\usepackage{microtype}
\usepackage{lineno}  
\usepackage{xspace} 
\usepackage{caption} 

\usepackage{graphicx}  
\usepackage{color}
\usepackage{colortbl}
\graphicspath{{./figs/}} 

\usepackage{amsmath} 
\usepackage{amssymb}
\usepackage{amsfonts}
\usepackage{upgreek} 

\newcommand*\patchAmsMathEnvironmentForLineno[1]{%
\expandafter\let\csname old#1\expandafter\endcsname\csname #1\endcsname
\expandafter\let\csname oldend#1\expandafter\endcsname\csname
end#1\endcsname
 \renewenvironment{#1}%
   {\linenomath\csname old#1\endcsname}%
   {\csname oldend#1\endcsname\endlinenomath}%
}
\newcommand*\patchBothAmsMathEnvironmentsForLineno[1]{%
  \patchAmsMathEnvironmentForLineno{#1}%
  \patchAmsMathEnvironmentForLineno{#1*}%
}
\AtBeginDocument{%
\patchBothAmsMathEnvironmentsForLineno{equation}%
\patchBothAmsMathEnvironmentsForLineno{align}%
\patchBothAmsMathEnvironmentsForLineno{flalign}%
\patchBothAmsMathEnvironmentsForLineno{alignat}%
\patchBothAmsMathEnvironmentsForLineno{gather}%
\patchBothAmsMathEnvironmentsForLineno{multline}%
\patchBothAmsMathEnvironmentsForLineno{eqnarray}%
}

\usepackage{hyperxmp}

\usepackage[pdftex,
            pdfauthor={\paperauthors},
            pdftitle={\paperasciititle},
            pdfkeywords={\paperkeywords},
            pdfcopyright={Copyright (C) \papercopyright},
            pdflicenseurl={\paperlicenceurl}]{hyperref}

\usepackage[all]{hypcap} 

\def\lhcb {\mbox{LHCb}\xspace}

\def\MagUp {\mbox{\em Mag\kern -0.05em Up}\xspace}

\ifthenelse{\boolean{uprightparticles}}%
{

 \def\PDelta      {\ensuremath{\Delta}\xspace}                 
 \def\PXi      {\ensuremath{\Xi}\xspace}                 
 \def\PLambda      {\ensuremath{\Lambda}\xspace}                 
 \def\PSigma      {\ensuremath{\Sigma}\xspace}                 
 \def\POmega      {\ensuremath{\Omega}\xspace}                 
 \def\PUpsilon      {\ensuremath{\Upsilon}\xspace}                 
 
 \def\PB      {\ensuremath{\mathrm{B}}\xspace}                 
                  
 \def\PD      {\ensuremath{\mathrm{D}}\xspace}

 \def\PK      {\ensuremath{\mathrm{K}}\xspace}

 \def\Pb      {\ensuremath{\mathrm{b}}\xspace}                 
 \def\Pc      {\ensuremath{\mathrm{c}}\xspace}

 \def\Pi      {\ensuremath{\mathrm{i}}\xspace}

}
{

 \mathchardef\PDelta="7101
 \mathchardef\PXi="7104
 \mathchardef\PLambda="7103
 \mathchardef\PSigma="7106
 \mathchardef\POmega="710A
 \mathchardef\PUpsilon="7107
                  
 \def\PB      {\ensuremath{B}\xspace}                 
                  
 \def\PD      {\ensuremath{D}\xspace}

 \def\PK      {\ensuremath{K}\xspace}

 \def\Pb      {\ensuremath{b}\xspace}                 
 \def\Pc      {\ensuremath{c}\xspace}

 \def\Pi      {\ensuremath{i}\xspace}

}

\makeatletter
\ifcase \@ptsize \relax
  \newcommand{\miniscule}{\@setfontsize\miniscule{4}{5}}
\or
  \newcommand{\miniscule}{\@setfontsize\miniscule{5}{6}}
\or
  \newcommand{\miniscule}{\@setfontsize\miniscule{5}{6}}
\fi
\makeatother

\DeclareRobustCommand{\optbar}[1]{\shortstack{{\miniscule (\rule[.5ex]{1.25em}{.18mm})}
  \\ [-.7ex] $#1$}}

\def\cquark    {{\ensuremath{\Pc}}\xspace}

\def\bquark    {{\ensuremath{\Pb}}\xspace}

\def\kaon    {{\ensuremath{\PK}}\xspace}
  \def\Kbar    {{\kern 0.2em\overline{\kern -0.2em \PK}{}}\xspace}

\def\KorKbar    {\kern 0.18em\optbar{\kern -0.18em K}{}\xspace}

\def\Kp      {{\ensuremath{\kaon^+}}\xspace}
\def\Km      {{\ensuremath{\kaon^-}}\xspace}

  \def\Dbar    {{\kern 0.2em\overline{\kern -0.2em \PD}{}}\xspace}
\def\D       {{\ensuremath{\PD}}\xspace}

\def\DorDbar    {\kern 0.18em\optbar{\kern -0.18em D}{}\xspace}

\def\Dp      {{\ensuremath{\D^+}}\xspace}

\def\Bbar    {{\ensuremath{\kern 0.18em\overline{\kern -0.18em \PB}{}}}\xspace}

\def\BorBbar    {\kern 0.18em\optbar{\kern -0.18em B}{}\xspace}

  \def\Y#1S{\ensuremath{\PUpsilon{(#1S)}}\xspace}

\def\Lbar        {{\ensuremath{\kern 0.1em\overline{\kern -0.1em\PLambda}}}\xspace}
\def\LorLbar    {\kern 0.18em\optbar{\kern -0.18em \PLambda}{}\xspace}

\newcommand{\decay}[2]{\ensuremath{#1\!\to #2}\xspace}         
\def\ra                 {\ensuremath{\rightarrow}\xspace}
\def\to                 {\ensuremath{\rightarrow}\xspace}

\def\DToKKK      {\decay{\Dp}{\Km\Kp\Kp}}

\def\AT#1     {\ensuremath{A_{\mathrm{T}}^{#1}}\xspace}           

\def\C#1      {\ensuremath{\mathcal{C}_{#1}}\xspace}                       
\def\Cp#1     {\ensuremath{\mathcal{C}_{#1}^{'}}\xspace}                    
\def\Ceff#1   {\ensuremath{\mathcal{C}_{#1}^{\mathrm{(eff)}}}\xspace}        
\def\Cpeff#1  {\ensuremath{\mathcal{C}_{#1}^{'\mathrm{(eff)}}}\xspace}       
\def\Ope#1    {\ensuremath{\mathcal{O}_{#1}}\xspace}                       
\def\Opep#1   {\ensuremath{\mathcal{O}_{#1}^{'}}\xspace}                    

\newcommand{\tev}{\ifthenelse{\boolean{inbibliography}}{\ensuremath{~T\kern -0.05em eV}\xspace}{\ensuremath{\mathrm{\,Te\kern -0.1em V}}}\xspace}
\newcommand{\gev}{\ensuremath{\mathrm{\,Ge\kern -0.1em V}}\xspace}
\newcommand{\gevgev}{\ensuremath{\mathrm{\,Ge\kern -0.1em V^2}}\xspace}
\newcommand{\mev}{\ensuremath{\mathrm{\,Me\kern -0.1em V}}\xspace}
\newcommand{\kev}{\ensuremath{\mathrm{\,ke\kern -0.1em V}}\xspace}
\newcommand{\ev}{\ensuremath{\mathrm{\,e\kern -0.1em V}}\xspace}
\newcommand{\gevc}{\ensuremath{{\mathrm{\,Ge\kern -0.1em V\!/}c}}\xspace}
\newcommand{\mevc}{\ensuremath{{\mathrm{\,Me\kern -0.1em V\!/}c}}\xspace}
\newcommand{\gevcc}{\ensuremath{{\mathrm{\,Ge\kern -0.1em V\!/}c^2}}\xspace}
\newcommand{\gevgevcccc}{\ensuremath{{\mathrm{\,Ge\kern -0.1em V^2\!/}c^4}}\xspace}
\newcommand{\mevcc}{\ensuremath{{\mathrm{\,Me\kern -0.1em V\!/}c^2}}\xspace}

\def\m    {\ensuremath{\rm \,m}\xspace}

\def\mum  {\ensuremath{{\,\upmu\rm m}}\xspace}

\def\invfb   {\ensuremath{\mbox{\,fb}^{-1}}\xspace}

\newcommand{\chisq}{\ensuremath{\chi^2}\xspace}

\newcommand{\chisqip}{\ensuremath{\chi^2_{\rm IP}}\xspace}

\def\gsim{{~\raise.15em\hbox{$>$}\kern-.85em
          \lower.35em\hbox{$\sim$}~}\xspace}
\def\lsim{{~\raise.15em\hbox{$<$}\kern-.85em
          \lower.35em\hbox{$\sim$}~}\xspace}

\def\ptot       {\mbox{$p$}\xspace}
\def\pt         {\mbox{$p_{\rm T}$}\xspace}

\def\evtgen     {\mbox{\textsc{EvtGen}}\xspace}

\def\geant      {\mbox{\textsc{Geant4}}\xspace}

\def\photos     {\mbox{\textsc{Photos}}\xspace}

\def\pythia     {\mbox{\textsc{Pythia}}\xspace}

\def\tell1  {TELL1\xspace}
\def\ukl1   {UKL1\xspace}

\def\lp {\left( }
\def\rp {\right) }
\def\lb {\left[ }
\def\rb {\right] }
\def\lc {\left\{ }
\def\rc {\right\} }
\def\ra {\rangle }
\def\la {\langle }

\def\rar {\rightarrow}
\def\lrar {\leftrightarrow}

\def\beq{\begin{equation}}
\def\eeq{\end{equation}}
\def\bea{\begin{eqnarray}}
\def\eea{\end{eqnarray}}
\def\nn {\nonumber}

\def\cd {\!\cdot\!}

\def\ct {\tilde{c}}

\def\Ob {\bar{\Omega}}

\def\rtw {\sqrt{2}}
\def\rth {\sqrt{3}}
\def\rts {\sqrt{6}}
\def\sp {\!+\!}
\def\sm {\!-\!}

\def\mr2 {m_\rho^2 }

\def\cK {{\cal{K}}}

\def\Gb {\bar{\Gamma}}
\def\f {\phi}
\def\G {\Gamma}

\def\m{\mu}

\def\p {\pi}

\def\th {\theta}

\usepackage{cite} 
\usepackage{mciteplus}

\usepackage{longtable} 

\begin{document}

\renewcommand{\thefootnote}{\fnsymbol{footnote}}
    \renewcommand{\arraystretch}{1.3}
\setcounter{footnote}{1}


\begin{titlepage}
\pagenumbering{roman}

\vspace*{-1.5cm}
\centerline{\large EUROPEAN ORGANIZATION FOR NUCLEAR RESEARCH (CERN)}
\vspace*{1.5cm}
\noindent
\begin{tabular*}{\linewidth}{lc@{\extracolsep{\fill}}r@{\extracolsep{0pt}}}
\ifthenelse{\boolean{pdflatex}}
{\vspace*{-1.5cm}\mbox{\!\!\!\includegraphics[width=.14\textwidth]{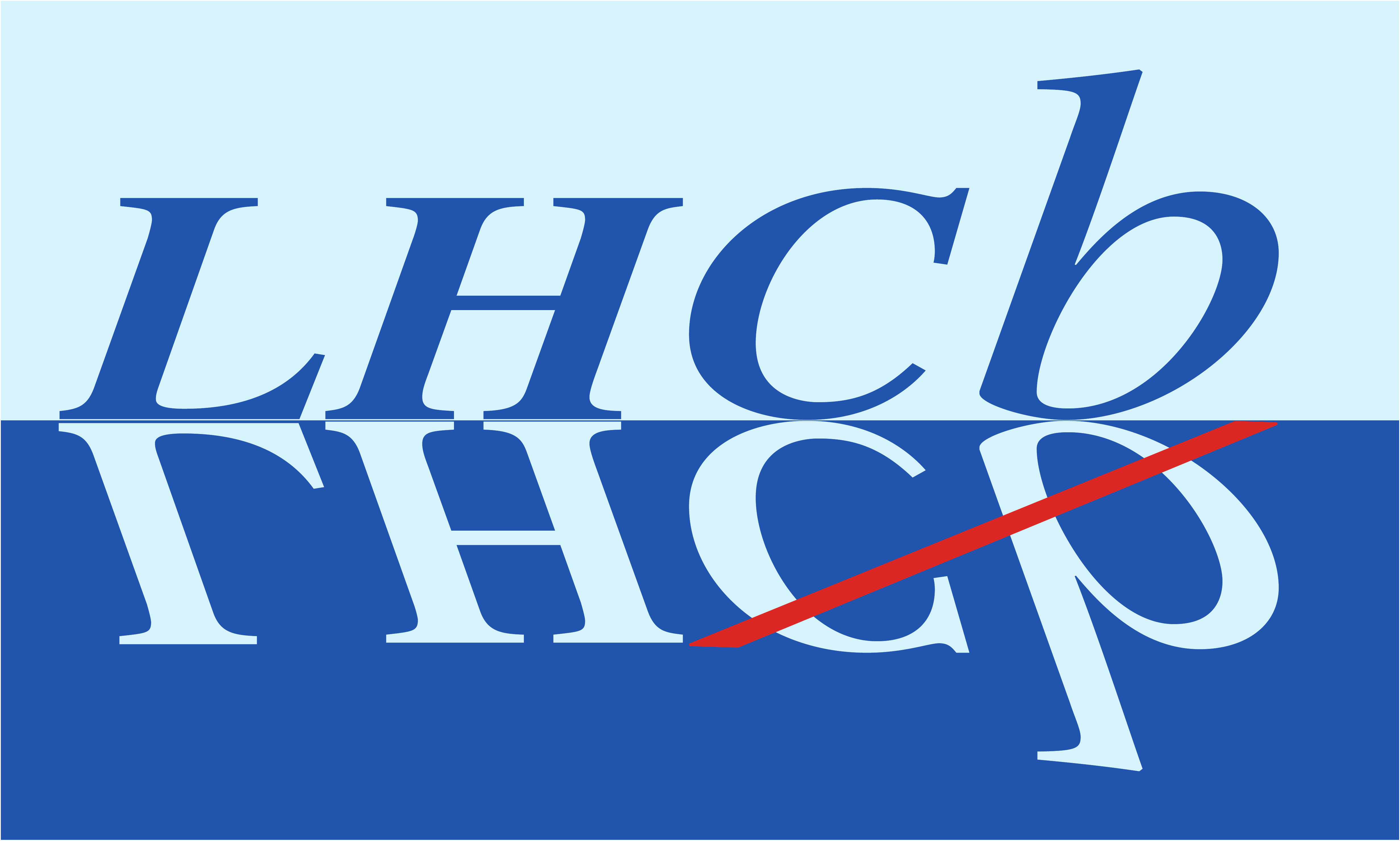}} & &}%
{\vspace*{-1.2cm}\mbox{\!\!\!\includegraphics[width=.12\textwidth]{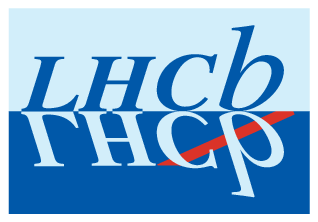}} & &}%
\\
 & & CERN-EP-2018-336 \\  
 & & LHCb-PAPER-2018-039 \\  
 & & \today \\ 
 & & \\
\end{tabular*}

\vspace*{4.0cm}

{\normalfont\bfseries\boldmath\huge
\begin{center}
  \papertitle 
\end{center}
}

\vspace*{2.0cm}

\begin{center}
\paperauthors\footnote{Authors are listed at the end of this paper.}
\end{center}

\vspace{\fill}

\begin{abstract}
  \noindent
The  resonant structure of the doubly Cabibbo-suppressed decay \DToKKK is studied for the first time.
The measurement is based on a sample of $pp$-collision data, collected at a centre-of-mass energy 
of 8 TeV 
with the LHCb detector and corresponding to an integrated luminosity of 2\invfb. The amplitude analysis 
of this decay 
is performed with the isobar model and a phenomenological model based on an effective chiral 
Lagrangian. In both models the  S-wave component in the $K^-K^+$ system is dominant, with a small contribution of the 
$\phi(1020)$ meson and a
negligible contribution from tensor resonances. The $K^+K^-$ scattering amplitudes 
for the considered combinations of spin (0,1) and isospin (0,1) of the two-body system
are obtained from the Dalitz plot fit with the phenomenological decay amplitude.

\end{abstract}

\vspace*{2.0cm}

\begin{center}
  Published in JHEP 04 (2019) 063.
\end{center}

\vspace{\fill}

{\footnotesize 
\centerline{\copyright~\papercopyright. \href{\paperlicenceurl}{\paperlicence}.}}
\vspace*{2mm}

\end{titlepage}


\newpage
\setcounter{page}{2}
\mbox{~}
%
%
%
%

\cleardoublepage


\renewcommand{\thefootnote}{\arabic{footnote}}
\setcounter{footnote}{0}



\pagestyle{plain} 
\setcounter{page}{1}
\pagenumbering{arabic}


%


\section{Introduction}
\label{sec:Introduction}

The theoretical treatment of  weak decays of charm mesons is very challenging.
The charm quark is not light enough for the reliable application of chiral perturbation theory,
which is successfully applied in predictions of kaon decays. The charm quark is also not heavy 
enough for the reliable application of the factorisation approach and heavy-quark expansion tools, 
as used in predictions of properties of {\it b} hadrons. The description of charm meson 
decays relies on approximate symmetries and phenomenological models. For such models, the 
knowledge of 
branching fractions and the resonant structures, in the case of multi-body decays, are key inputs. 
In this paper, the first determination of the resonant structure of the doubly Cabibbo-suppressed 
decay \DToKKK is presented.\footnote{Charge conjugation is implied throughout the paper.} 
The analysis is based on a data sample of $pp$ collisions collected with the LHCb detector,
corresponding to an integrated luminosity of 2 \invfb at a centre-of-mass energy of 8 TeV. The
determination of the resonant structure of this decay is complementary to 
the recent LHCb measurement of its branching fraction \cite{paper:dcs}, based on the same data set.

The amplitude analysis of the \DToKKK decay is performed using two methods. The Dalitz plot is 
fitted with the isobar model, in which the decay amplitude is a coherent sum of resonant and nonresonant
amplitudes~\cite{asner}.
The Dalitz plot is also fitted with a phenomenological model derived from an effective chiral Lagrangian 
with resonances~\cite{mmm}. This phenomenological model, referred to as the multi-meson model, 
or \mbox{\em Triple-M}, includes the effects of coupled channels --- $\pi\pi$, $K^+K^-$, $\pi\eta$, 
$\eta\eta$ and $\rho\pi$ --- in the final state interactions (FSI), in four 
considered combinations of spin {\em J}  and isospin {\em I} ($J\!=\!0,1$; $I\!=\!0,1$). 
Given the small phase space of the \DToKKK decay and the lack of tensor resonances with
significant coupling to $K^+K^-$, the contribution from  D-wave is expected to be
 suppressed.

An additional motivation for the Dalitz plot analysis of the \DToKKK decay is to obtain the $K^+K^-$ 
scattering amplitudes. Most information currently available on $\pi\pi$ and $K\pi$ scattering 
is obtained indirectly 
from meson-nucleon interactions~\cite{grayer,E135,lass}. In the regime where the momentum 
transferred to the nucleon is small enough, the  interaction is assumed to be dominated by the 
one-pion-exchange amplitude. The asymptotically free incoming meson interacts with a virtual pion,
resulting in what is generally referred as $\pi\pi$ and $K\pi$ scattering data. The resulting 
$\pi\pi \to \pi\pi$ and $K\pi \to K\pi$ phase shifts are affected by ambiguities and large 
systematic uncertainties. The $\pi\pi\to K\Kbar$ scattering was studied both in
$\pi p$ and $\pi n$ reactions~\cite{paper:cohen1,paper:cohen2}, and in $p\bar p$ annihilation 
at rest~\cite{paper:obelix}. For the $K\Kbar \to K\Kbar$ scattering, no meson-nucleon
data exists. 

Three-body  decays of $D$ mesons into kaons and pions are an interesting alternative 
for light-meson spectroscopy, as they are complementary to the meson-nucleon 
reactions. Large data sets from the $B$-factories and LHCb exist for 
these decays. However, it is necessary to isolate the physics of two-body systems 
from the rich dynamics of three-body decays, which involve the weak decay of the $c$ 
quark, the formation of the mesons and their FSI. This is achieved
with the Triple-M decay amplitude, in which these three stages are included. 
The FSI are described in terms of the $K^+K^-$ scattering 
amplitudes for the considered spin-isospin combinations, allowing the determination 
of these amplitudes from a fit to the \DToKKK Dalitz plot.

This paper is organised as follows. A brief description of the LHCb detector is presented in 
Sec.~\ref{sec:Detector}. The signal selection is presented in Sec.~\ref{sec:selection}. 
In Sec.~\ref{sec:efficiencies}, the efficiency determination and  background model are discussed. 
The formalism for the Dalitz plot fit is presented in Sec.~\ref{sec:dalitzprocedure}. 
In Sec.~\ref{sec:isobar}, the results of the fit with the isobar model are presented, whilst the results of the
Dalitz plot fit with the Triple-M amplitude are presented in Sec.~\ref{sec:MMM}. Systematic 
uncertainties are discussed in Sec.~\ref{sec:systematics}. A summary and the conclusions are 
presented in Sec.~\ref{sec:conclusions}.

\section{Detector and simulation}
\label{sec:Detector}

The \lhcb detector~\cite{Alves:2008zz,LHCb-DP-2014-002} is a single-arm forward
spectrometer covering the \mbox{pseudorapidity} range $2<\eta <5$,
designed for the study of particles containing \bquark or \cquark
quarks. The detector includes a high-precision tracking system
consisting of a silicon-strip vertex detector surrounding the $pp$
interaction region, a large-area silicon-strip detector located
upstream of a dipole magnet with a bending power of about
$4{\mathrm{\,Tm}}$, and three stations of silicon-strip detectors and straw
drift tubes placed downstream of the magnet.
The tracking system provides a measurement of the momentum, \ptot, of charged particles with
a relative uncertainty that varies from 0.5\% at low momentum to 1.0\% at 
200\gev.\footnote{Natural units with $\hbar=c=1$ are used in this paper.}
The minimum distance of a track to a primary vertex (PV), the impact parameter (IP), 
is measured with a resolution of $(15+29/\pt)\mum$,
where \pt is the component of the momentum transverse to the beam, in GeV.
Different types of charged hadrons are distinguished using information
from two ring-imaging Cherenkov detectors~\cite{LHCb-DP-2012-003}. 
Photons, electrons and hadrons are identified by a calorimeter system consisting of
scintillating-pad and pre-shower detectors, an electromagnetic
and a hadronic calorimeter. Muons are identified by a
system composed of alternating layers of iron and multi-wire
proportional chambers.

The online event selection is performed by a trigger, 
which consists of a hardware stage, based on information from the calorimeter and muon
systems, followed by a software stage, which applies a full event
reconstruction. At the hardware trigger stage, events are required to have a muon with high \pt or a
hadron, photon or electron with high transverse energy in the calorimeters.
The software trigger  is divided into two parts. The first employs a partial reconstruction of the
candidates from the hardware trigger and a cut-based selection. In the second stage,  a
full event reconstruction is applied and various dedicated algorithms are used in the selection of 
specific decays. In this analysis, a dedicated algorithm is used to select 
\DToKKK decay candidates.

 In the simulation, $pp$ collisions are generated using
\pythia~\cite{Sjostrand:2006za,*Sjostrand:2007gs}  with a specific \lhcb
configuration~\cite{LHCb-PROC-2010-056}.  Decays of hadronic particles
are described by \evtgen~\cite{Lange:2001uf}, in which final-state
radiation is generated using \photos~\cite{Golonka:2005pn}. The
interaction of the generated particles with the detector, and its response,
are implemented using the \geant
toolkit~\cite{Allison:2006ve, *Agostinelli:2002hh} as described in
Ref.~\cite{LHCb-PROC-2011-006}.


\section{Candidate selection}
\label{sec:selection}

 The \DToKKK decay candidates are selected offline with requirements that exploit the decay topology 
 by combining three charged
particles identified as kaons according to particle-identification (PID) criteria.  
 These particles must form a good-quality decay 
vertex, detached from the PV. The PV is chosen as that with the smallest value of \chisqip, where 
\chisqip\ is defined as the difference in the vertex-fit \chisq of the PV reconstructed with and without the 
particle under consideration, 
in this case the \Dp candidate. The selection of candidates is based on the distance between the PV and 
the \Dp decay 
vertex (the flight distance); the IP of the \Dp candidate; the angle between the reconstructed \Dp 
momentum vector and
the vector connecting the PV to the decay vertex; the $\chi^2$ of 
the \Dp decay vertex fit; the distance of closest approach between any two final-state tracks; and the 
momentum, the transverse 
momentum and the \chisqip of the \Dp candidate and of its decay products. The invariant mass of the \Dp 
candidate is required to be within the interval \mbox{1820--1920\mev}. In order to 
suppress the 
contamination from \mbox{$D^+_s \to K^-K^+\pi^+\pi^0$} decays, where the neutral pion is not 
reconstructed and the charged pion is misidentified as a kaon,
more stringent PID requirements are applied to the kaon candidates with the same charge.

A boosted decision tree (BDT) multivariate classifier~\cite{Breiman,AdaBoost} 
is used to further reduce the combinatorial background. In order to keep the selection efficiency uniform 
over the Dalitz plot,  the BDT uses only the quantities related to the $D^+$ candidate described above.
The BDT is trained using simulated  \DToKKK decays  for the signal, and data from the 
invariant-mass intervals  1820--1840\mev and 1900--1920\mev for the background. 
After the application of all selection requirements, approximately 
 0.5\% of the events include more than one signal candidate. All 
candidates are retained for further analysis.

\begin{figure}[!htb]
\centering
\includegraphics[width=0.6\linewidth]{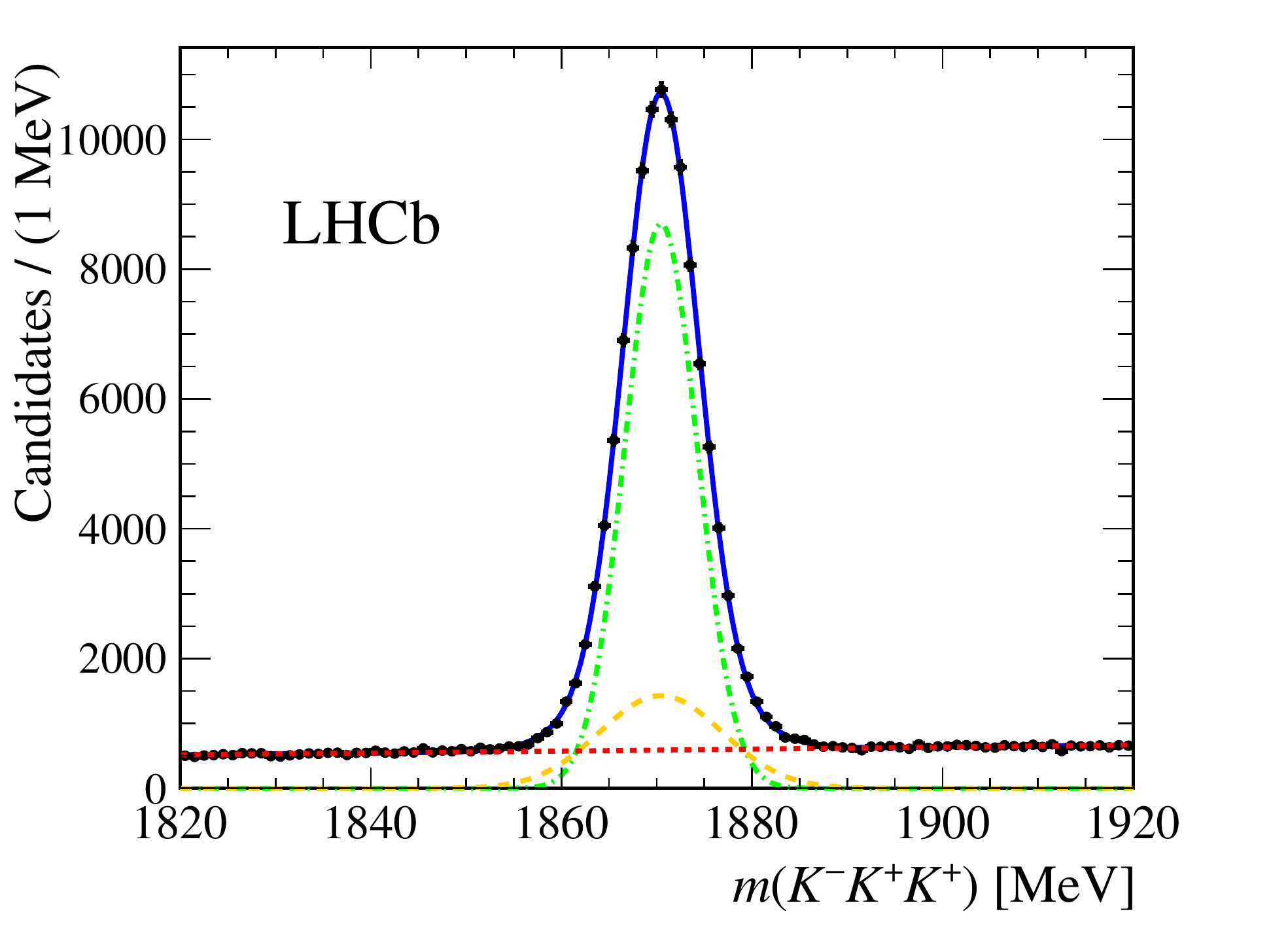}
\caption{Invariant-mass spectrum of the  $K^-K^+K^+$ candidates with the fit result overlaid (solid blue line). The 
orange and green dashed lines indicate the two Gaussian functions representing the signal and the red dashed line
is the background.}
\label{fig:fitmass}
\end{figure}

The invariant-mass spectrum of the selected \DToKKK sample is shown in Fig.~\ref{fig:fitmass}. 
To fit the invariant-mass distribution, the signal probability density function (PDF) is modeled by a 
sum of two Gaussian functions with a common mean and independent widths that are free parameters.
The signal model is validated with simulation.
The background PDF is parameterised by an exponential function. 
The fitted PDF is overlaid with the mass distribution in Fig.~\ref{fig:fitmass}.
For  the Dalitz plot analysis, only candidates within the range 1861.4--1879.5 \mev are considered.
This interval corresponds to four times the effective mass resolution, and contains 111 thousand
candidates,  of which $(90.45\pm0.07)\%$ correspond to signal.

\begin{figure}
\centering
\includegraphics[width=.49\linewidth]{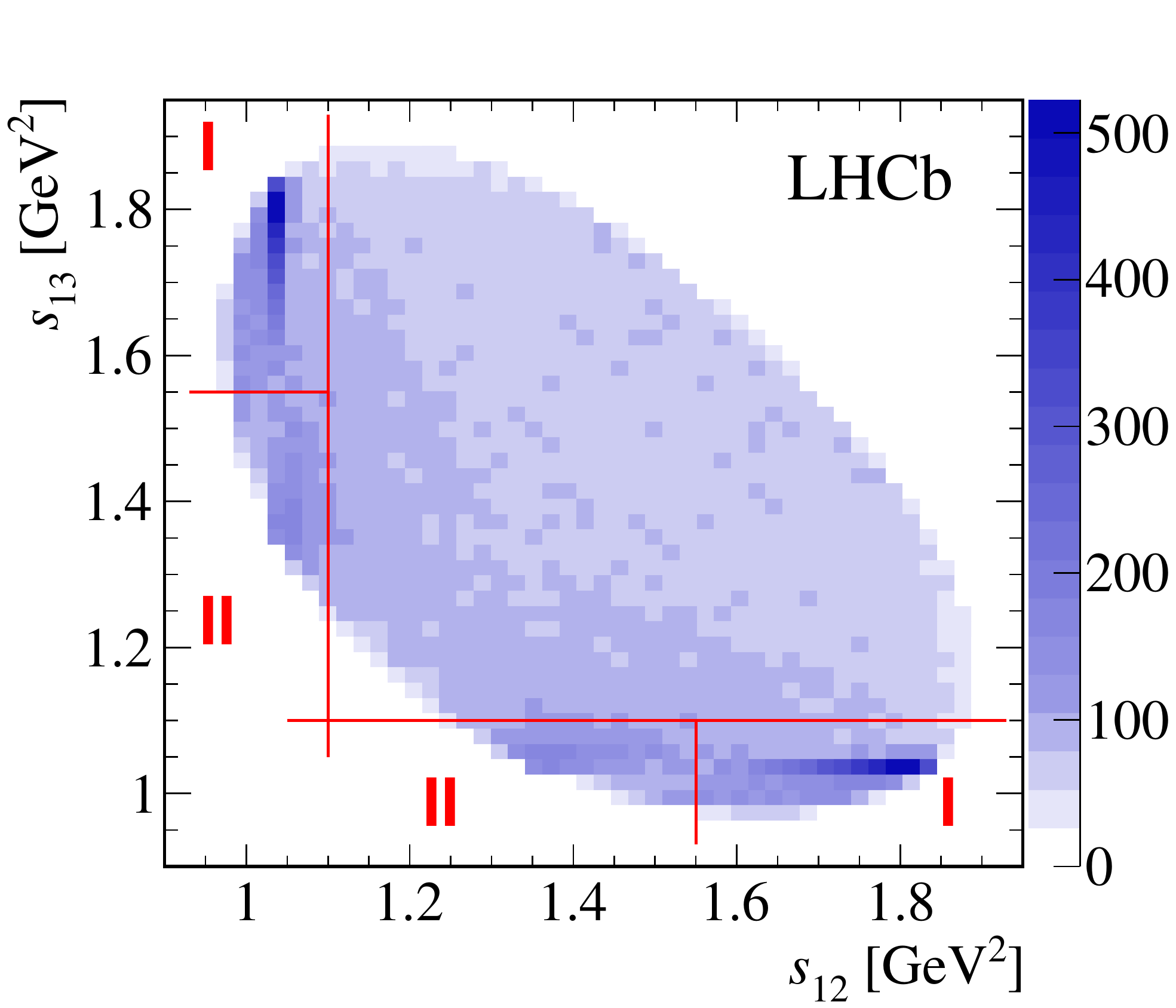}\hskip .3cm
\includegraphics[width=.49\linewidth]{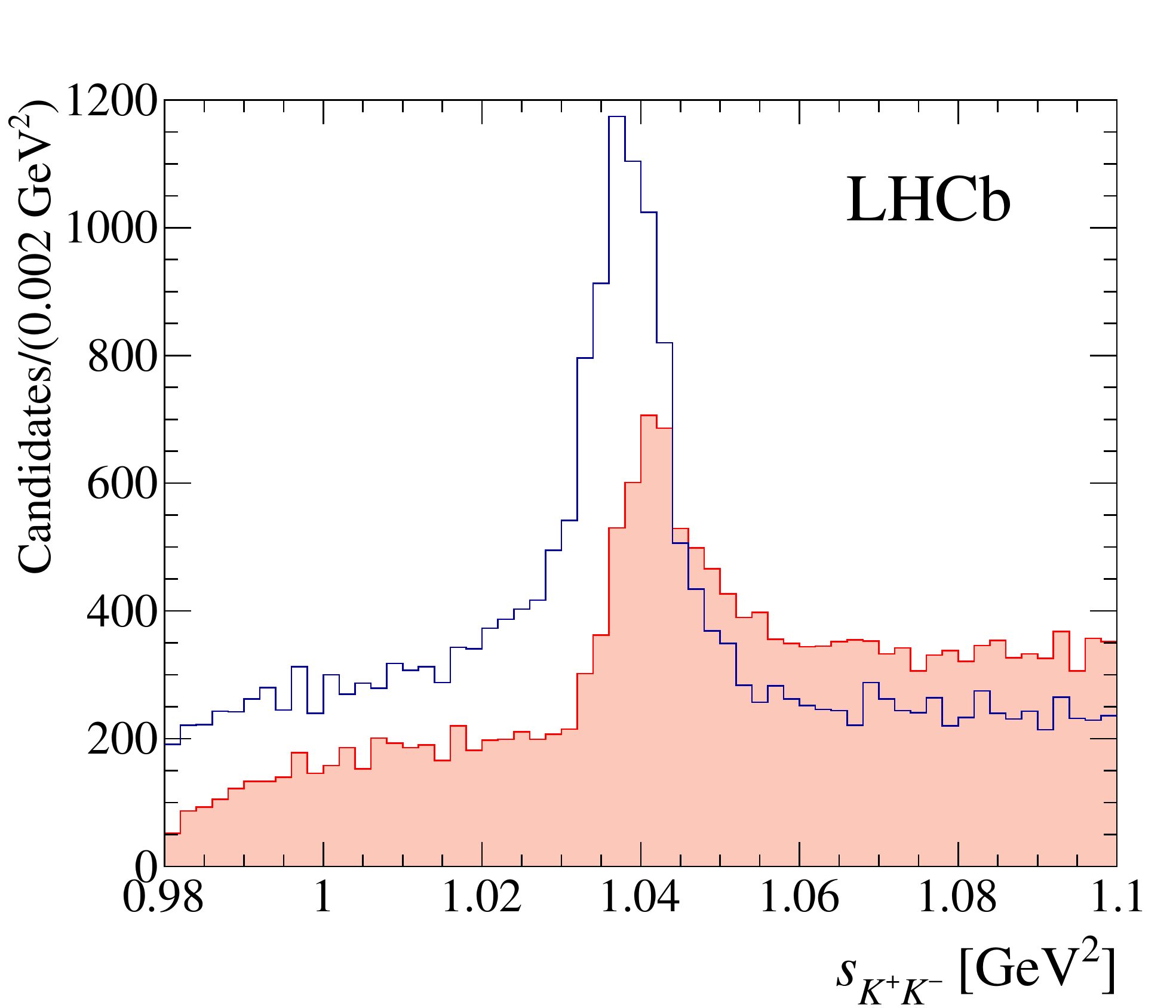}
\caption{(left) Dalitz plot of the selected sample, including background. 
(right) Dalitz plot projections for candidates from regions I (blue) 
and II (red), above and below 
   \mbox{$s_{K^-K^+}\!=\!1.5$\gevgev.}
The interference between the S- and P-wave amplitudes cause the asymmetry in the number of candidates in the
two regions, as well as the shift in the peak position. Both figures include all candidates in the selected mass range.}
\label{fig:DPfinal}
\end{figure}

The Dalitz plot of the candidates in the  signal region is shown in the left side of Fig~\ref{fig:DPfinal}.
The particle ordering is such that the  kaon with charge  opposite to that of  the $D^+$ meson
is always particle 1, 
and the same-sign kaons are randomly assigned particles 2 and 3, i.e.
$D^+\to K^-(p_1)K^+(p_2)K^+(p_3)$, where $p_i$ are the four-momenta. 
The Dalitz plot is represented in terms of the square of the invariant masses of the two 
$K^-K^+$ combinations, $s_{12}\equiv (p_1+p_2)^2$ and 
$s_{13}\equiv (p_1+p_3)^2$.  Throughout this paper, the symbol $s_{K^-K^+}$ is used to represent the 
invariant mass squared of both $K^-K^+$ combinations. These Lorentz-invariant quantities are 
computed constraining the invariant mass of the  candidate to the known $D^+$ mass~\cite{PDG2018}.  
An accumulation of candidates  is visible at \mbox{$s_{K^-K^+}$ $\sim$1.04\gevgev} which
corresponds to the  $\phi(1020)K^+$ component.
The difference in the number of candidates in the regions of the Dalitz plot above and below 1.55\gevgev 
(regions I and II in the left side of Fig.~\ref{fig:DPfinal}, respectively) is caused by
interference between the $\phi(1020)K^+$ and  S-wave amplitudes.  This interference also 
shifts the position of the peaks of the $s_{K^-K^+}$ distributions  in the two regions. 
These two effects are  
better illustrated in the projections of the Dalitz plot shown in the right side of Fig.~\ref{fig:DPfinal}.

\section{Efficiency and background model}
\label{sec:efficiencies}

\subsection{Efficiency variation over the Dalitz plot}
\label{sec:eff}
In the  fit to the Dalitz plot distribution, the variation of the total efficiency across the phase space must be 
taken into account. The total efficiency is determined from a combination of simulation and  methods
based on data,
and includes the geometrical acceptance of the detector and the reconstruction, selection, PID and trigger 
efficiencies. 

The geometrical acceptance,  reconstruction and selection efficiencies are obtained from simulation.
The PID efficiency of each $D^+$ candidate is determined by multiplying the  efficiencies for each of the final-state kaons.
The PID efficiencies for the kaons are evaluated from calibration samples of $D^{*+} \to D^0(\to K^-\pi^+)\pi^+$ 
decays~\cite{LHCb-PUB-2016-021} and depend on the particle momentum, pseudorapidity and event 
charged-particle multiplicity. 
The trigger efficiency is obtained from simulation, with a correction factor determined from data to
account for the small mismatch between the performance of the trigger in data and simulation.

\begin{figure}[!htb]
\centering
\includegraphics[width=0.49\linewidth]{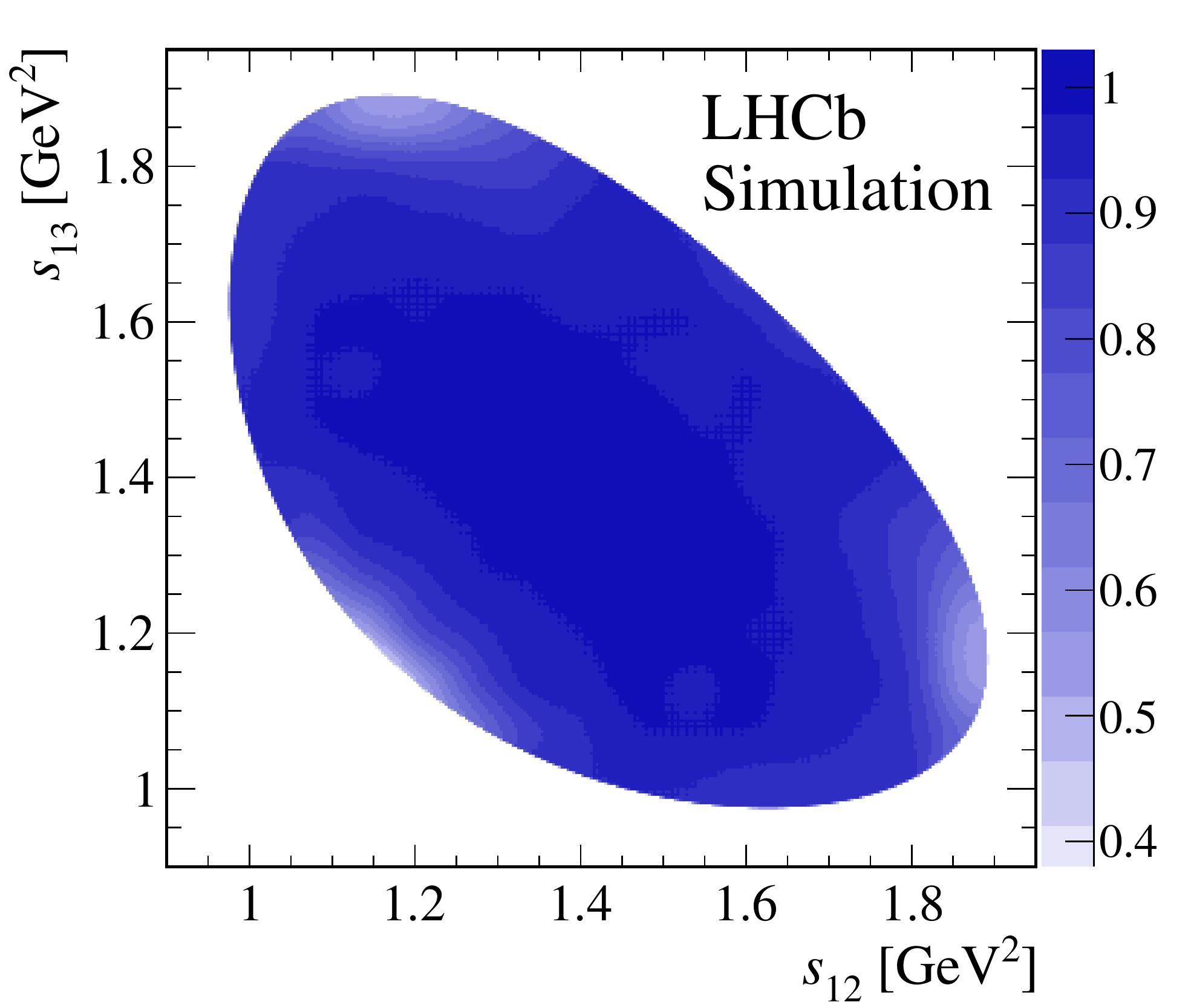}
\caption{Total efficiency, normalised to unity, for the \DToKKK signal over the Dalitz plot, including the geometrical acceptance and the reconstruction, selection, PID and trigger
efficiencies.}
\label{fig:finalacc}
\end{figure}

The total efficiency distribution is a two-dimensional histogram with $14\times 14$ uniform bins. 
 A two-dimensional cubic spline is used to smooth this distribution to avoid   binning discontinuities, 
yielding the high-resolution histogram ($300\times 300$ uniform bins), shown in Fig.~\ref{fig:finalacc}.
This histograms is used to weight the signal PDF in the Dalitz plot fit.
The binning scheme of the efficiency histogram is a source of systematic uncertainty.

\subsection{Background model}
\label{sec:bkg}
The background model is built from the inspection of the mass sidebands of the  \mbox{\DToKKK} signal.
The Dalitz plots of candidates from both sidebands, \mbox{1820--1840\mev} and 
1900--1920\mev,  are very similar,  with a clear peaking structure, corresponding to random 
\mbox{$\phi(1020)K^+$} combinations  over a smooth distribution.

The Dalitz plot variables are computed from the four-momenta determined 
by a $D^+$ mass constrained fit. This
constraint implies an unique boundary of the Dalitz plot, regardless of the value of the  invariant mass 
of the three-kaon system. It also improves the mass resolution of signal candidates, but has the
effect to distort and shift any structure present in the Dalitz plot of
the background candidates in the sidebands. This  effect depends strongly on the invariant mass 
of the three-kaon system and
prevents the determination of the background model from a two-dimensional parameterisation of the 
Dalitz plots from the sidebands.
An alternative method is used instead.  Each $m(K^-K^+K^+)$ sideband is divided into slices of 5\mev. For
each slice, the projections onto the $s_{K^+K^-}$ axis are fitted using a relativistic Breit--Wigner for the 
$\phi(1020)$ component (with floated mass and width) and a  phase-space distribution. The latter serves as a proxy
 for both the smooth component spread across the Dalitz plot and the projection of the 
$\phi$ candidates appearing in the other  $s_{K^+K^-}$ combination.
The fraction of the $\phi(1020)$ component is nearly constant in both sidebands, indicating that the
background composition is independent of $m(K^-K^+K^+)$. A  linear interpolation 
is used to obtain the fraction of peaking background in the signal region and  is found to be 
(20.67$\pm$0.28)\%. 

The fit to the $s_{K^+K^-}$ projection has the limitation of being less sensitive to the distribution near the $K^+ K^-$ threshold. The inspection of the Dalitz plot sidebands shows that the smooth background component has more candidates
at low values of $s_{K^+K^-}$ and fewer at low values of $s_{K^+K^+}\equiv(p_2+p_3)^2$, indicating that
this smooth distribution is not uniform over the phase space. 
A model for the smooth component of the background is built assuming a sum of two contributions, 
random $f_0(980)K^+$ candidates and a constant amplitude, with equal proportions.
The relative fractions of these two terms in the smooth component is treated as a source of systematic 
uncertainty.

\begin{figure}
\centering
\includegraphics[width=.49\linewidth]{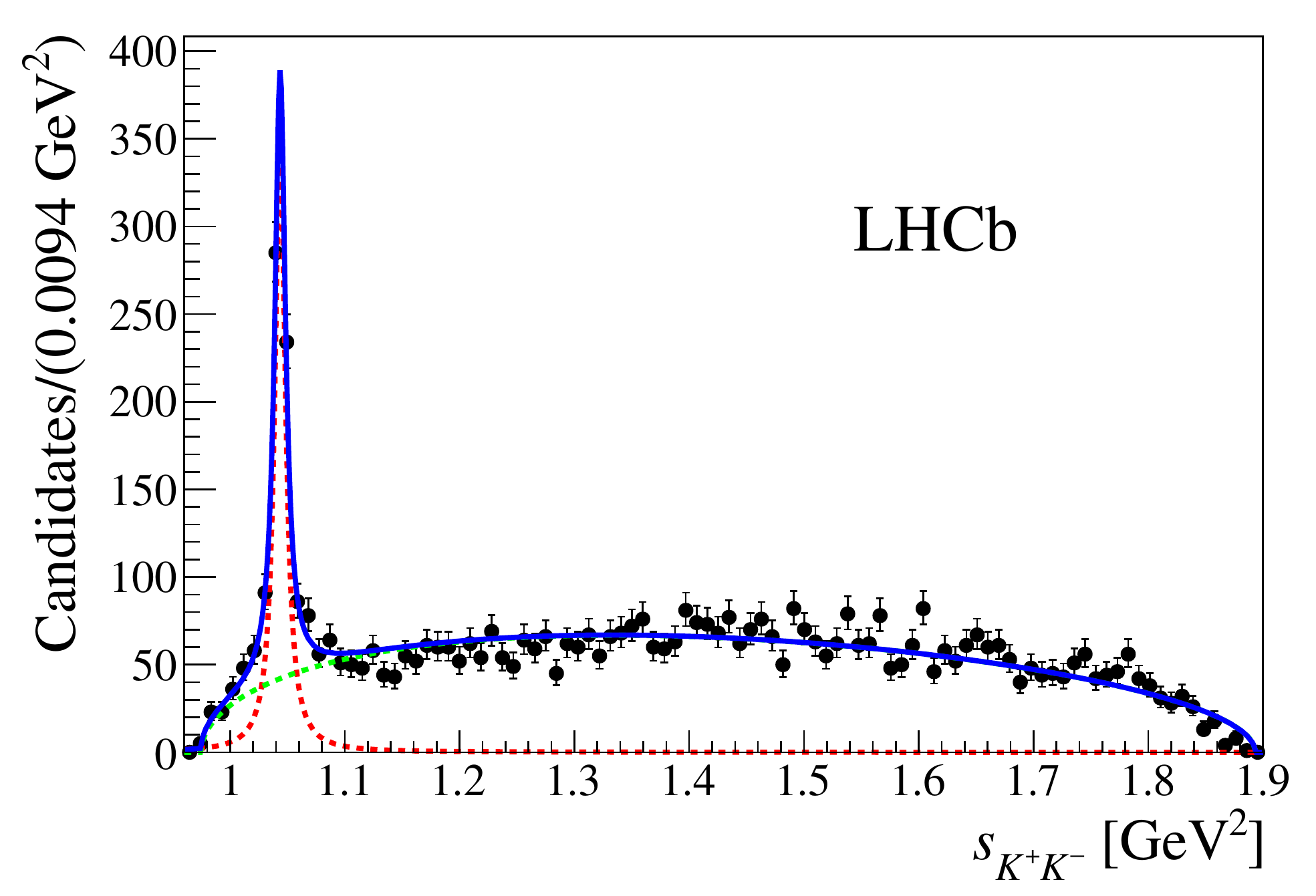}
\caption{ Projection onto $s_{K^+K^-}$  of $K^-K^+K^+$ candidates with
invariant mass in the range 1820--1830 MeV.}
\label{fig:bkg}
\end{figure}

\begin{figure}[!htb]
\centering
\includegraphics[width=0.49\linewidth]{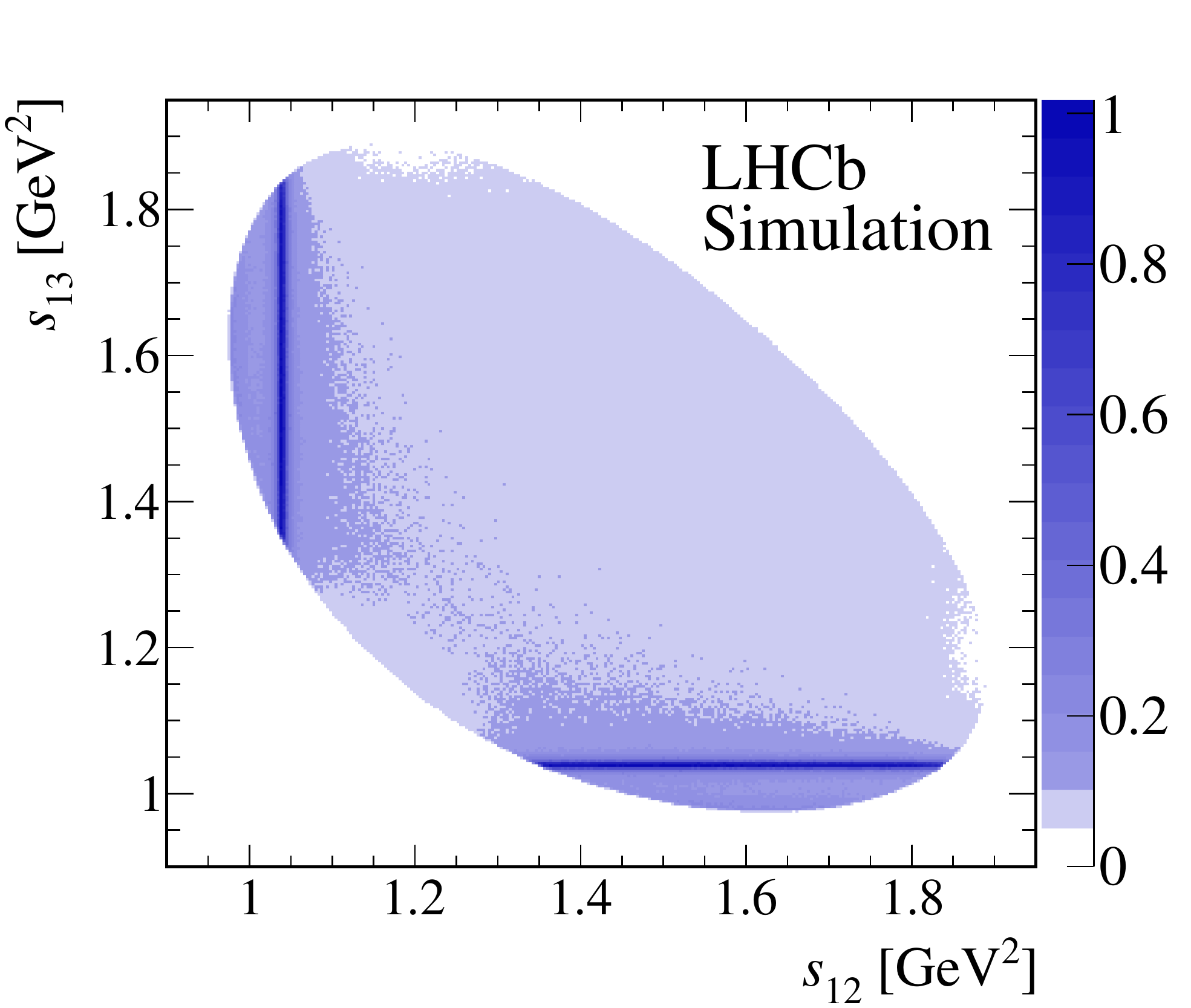}
\caption{High-resolution histogram representing the background model used in the Dalitz plot fits.}
\label{fig:finalbkg}
\end{figure}

A high-resolution normalised histogram ($300\times 300$ uniform bins) is used in the Dalitz plot fit  
to represent the background PDF,  and is shown in Fig.~\ref{fig:finalbkg}. 
This histogram is produced from a large simulated sample, using a PDF 
in which  the peaking and smooth components  are added incoherently
with the estimated relative fractions and weighted by the efficiency function.

\section{The Dalitz plot fit procedure}
\label{sec:dalitzprocedure}

The \DToKKK decays  are studied through an unbinned maximum-likelihood fit to the observed Dalitz plot distribution.  The total PDF is  constructed as a sum of signal and background components, and the likelihood function is given by
\begin{equation}
\mathcal{L}=\prod^{N_{\rm cand}} f_{\rm S} \times S_{\rm PDF}(s_{12},s_{13}) + 
(1-f_{\rm S}) \times B_{\rm PDF}(s_{12},s_{13}),
\end{equation} 
where $N_{\rm cand}$ is the total number of candidates and $f_S$ is the fraction of signal candidates 
in the sample, as obtained from the $m(K^-K^+K^+)$ fit described in Sec.~\ref{sec:selection}.
The background PDF, $B_{\rm PDF}(s_{12},s_{13})$, is described in Sec.~\ref{sec:bkg}.

The normalised signal PDF is written in terms of the total decay amplitude ${\mathcal{T}}(s_{12},s_{13})$, 
\begin{equation}
S_{\rm PDF}(s_{12},s_{13}) = \frac{1}{N_{\rm S}} {\left|\mathcal{T}(s_{12},s_{13})\right|}^2\varepsilon(s_{12},s_{13}) , \end{equation}
where $\varepsilon(s_{12},s_{13}) $ is the detection efficiency, described in Sec.~\ref{sec:eff}.
The normalisation factor, $N_{\rm S}$, is given by
\begin{equation}
N_{\rm S} = \int {\rm d}s_{12}\ {\rm d}s_{13}\left| \mathcal{T}(s_{12},s_{13})\right|^2\varepsilon(s_{13},s_{13}).
\end{equation}

\noindent For any given  model, the amplitude ${\mathcal{T}}(s_{12},s_{13})$ depends on a set of parameters that
are floated in the fit. The optimum values for these parameters are determined  by minimizing the quantity
$-2\ln \mathcal{L}$ using the \texttt{MINUIT} package~\cite{Minuit}.

In order to compare the fit results of a given model to the Dalitz plot distribution in data, a large 
simulated sample is generated
according to the model, including background and efficiency, normalised to the total number of 
data candidates. Since there are two identical kaons, the {\it folded} Dalitz plot is used, represented 
as $s_{K^+K^-}^{\rm high}$ 
versus $s_{K^+K^-}^{\rm low}$, which are respectively the higher and the lower values among 
$s_{12}$ and $s_{13}$. 
The Dalitz plot distribution  is divided into 1024 bins with approximately 110 candidates 
each and the normalised residuals are computed as
\begin{equation}
\Delta_i =  \frac{(N_{\rm pred}^i-N_{\rm obs}^i)}{\sigma_i}, 
\label{eq:chi2}
\end{equation}
where, for each bin $i$,  $N_{\rm pred}^i$ is the predicted number of candidates from the model, 
$N_{\rm obs}^i$ is the  
number of candidates in the data sample, 
and $\sigma_i$ is the statistical uncertainty from data and simulation added in quadrature.
The sum of the square values of $\Delta_i$ over all bins is the total $\chi^2$ and is used as a metric to compare fit results with different models.


\section{Dalitz plot analysis with the isobar model }
\label{sec:isobar}

In the isobar model, the  decay amplitude  is written as a coherent sum of a constant nonresonant  (NR) 
component and intermediate resonant amplitudes,
\begin{equation}
{\cal T}(s_{12},s_{13}) = c_{\rm NR} + \sum_k c_k T_k(s_{12},s_{13}).
\end{equation}

\noindent Each resonant amplitude,  $T_k$, is given by a product  of Blatt--Weisskopf penetration factors 
\cite{book:BlattWeisskopf}, 
$F_D^L$ and $F_R^L$, accounting for the finite size of the $D^+$ meson and the resonance, respectively, 
the  spin amplitude,  $\mathcal{S}$,  accounting for the conservation
of angular momentum,  and a  function, $M_R$, describing the resonance lineshape, which is either
a relativistic Breit--Wigner (Eq.~\ref{eq:BW})  or a Flatt\'e lineshape (Eq.~\ref{MRs12}).
The Zemach formalism~\cite{paper:Zemach} is used for  the spin amplitude $\mathcal{S}$.  
Details of each of these factors are given in Appendix~\ref{app:isobar}. 
Since there are two identical kaons in the final state, the resonant amplitudes are Bose-symmetrised,

\begin{equation}
T_k(s_{12},s_{13})  =  F_D^L(s_{12})F_R^L(s_{12})\,\times \, \mathcal{S}(s_{12},s_{13})\,\times \,M_R(s_{12}) + (2\lrar 3).
\label{eq:isobar}
\end{equation}

\noindent The fit parameters are the complex coefficients 
$c_{\rm NR} = a_{\rm NR} e^{i \delta_{\rm NR}}$ and $c_k = a_k e^{i \delta_k}$.
The results are expressed in terms of the magnitude and phase of the complex coefficient 
for each component, and the corresponding
fit fractions. The  fit fractions are computed by integrating the squared modulus of the 
corresponding amplitude over the phase space, and dividing by the integral of the total amplitude squared,
\begin{equation}
{\rm FF}_k = \frac{\int {\rm d}s_{12}\ {\rm d}s_{13} \  |c_k \ T_k (s_{12},s_{13})|^2}
{\int  {\rm d}s_{12}\ {\rm d}s_{13} \ \left|\sum_i c_i \ T _i (s_{12},s_{13})\right|^2 }.
\end{equation} 

The sum of fit fractions for all components is, in general, different from 100\% due to the presence of 
interference; it is less than 100\% in the case of net constructive interference or higher than 100\% otherwise.

\subsection{Signal models}

For the \DToKKK decay amplitude, contributions from following resonances are possible: 
the isoscalars $f_0(980)$, $f_0(1370)$ and $f_0(1500)$; 
the isovectors $a_0(980)$ and $a_0(1450)$;  the vector $\phi(1020)$;  the tensor $f_2(1270)$.
Contributions from resonances with  spin greater than one are suppressed due to the small phase 
space of the \DToKKK decay.
In the case of the $f_2(1270)$ state, a further suppression is expected due to  its small branching 
fraction to $K^-K^+$, 
$(4.6\pm 0.4)$\%~\cite{PDG2018}. The relatively narrow $f'_2(1525)$ state is neglected since it is well beyond the phase space.

Various combinations of the nonresonant and the possible resonant  amplitudes are considered. 
All models studied contain the $\phi(1020)K^+$, which is chosen as the reference amplitude, fixing 
the phase convention and setting the scale for the  magnitudes. The models tested differ by the 
composition of the S-wave. Near the $K^+K^-$ threshold, both the  $a_0(980)$ and $f_0(980)$ 
resonances can contribute. Similarly, at higher $K^+K^-$ invariant mass, contributions from  
several scalar resonances are possible.

The $\phi(1020)$ mass and width are fixed to the known values~\cite{PDG2018}; 
for the $f_0(980)$ state, a Flatt\'e 
lineshape is used, with parameters from the  BESII collaboration~\cite{besf0}.

\subsection{Results}

The simplest model that describes the data, referred to as model A,  consists of three 
intermediate components:   
$\phi(1020)K^+$,  $f_0(980)K^+$,  and $f_0(1370)K^+$. As the $f_0(1370)$ state has large 
uncertainties on its mass and width~\cite{PDG2018}, these parameters are allowed to float in the fit. 
Its contribution can also be interpreted, within the isobar formalism, as an effective representation 
for the overlap of  two or more broad structures at high $K^-K^+$ invariant mass. 

Further addition of scalar states does not improve the fit quality significantly, creates more complex 
interference effects,  and provides a very similar description of the lineshape and phase behaviour of 
the total S-wave. For example, in model B, a  constant nonresonant contribution is added to the 
resonant amplitudes of model A. The resulting  fit quality is essentially unchanged, with the total 
$\chi^2/{\rm ndof}$ being 1.15 and 1.14 for  models A and B, respectively. 
A similar situation occurs in model C, which has the same amplitudes as in model B plus the
$a_0(980)K^+$ component. In this model,
the contribution of the $f_0(1370)$ is found to be negligible and the value of $\chi^2/{\rm ndof}$ is 1.16.
Table~\ref{tab:isobar_results} summarizes the fit results for these three models.  The total S-wave fit
fraction includes the interference terms between the various S-wave components. 
In all cases, the total S-wave in the $K^+ K^-$ system is dominant, a notable feature also observed in 
other three-body $D$ decays with a pair of identical particles in the final state, 
such as the $D^+\to K^-\pi^+\pi^+$ and $D^+_{(s)}\to \pi^-\pi^+\pi^+$ decays\cite{PDG2018}.
The contribution from the $f_2(1270)K^+$ component is also tested and found to be consistent with zero in all models.

\begin{figure}[hbtp]
\begin{center}
\includegraphics [width=0.49\textwidth]{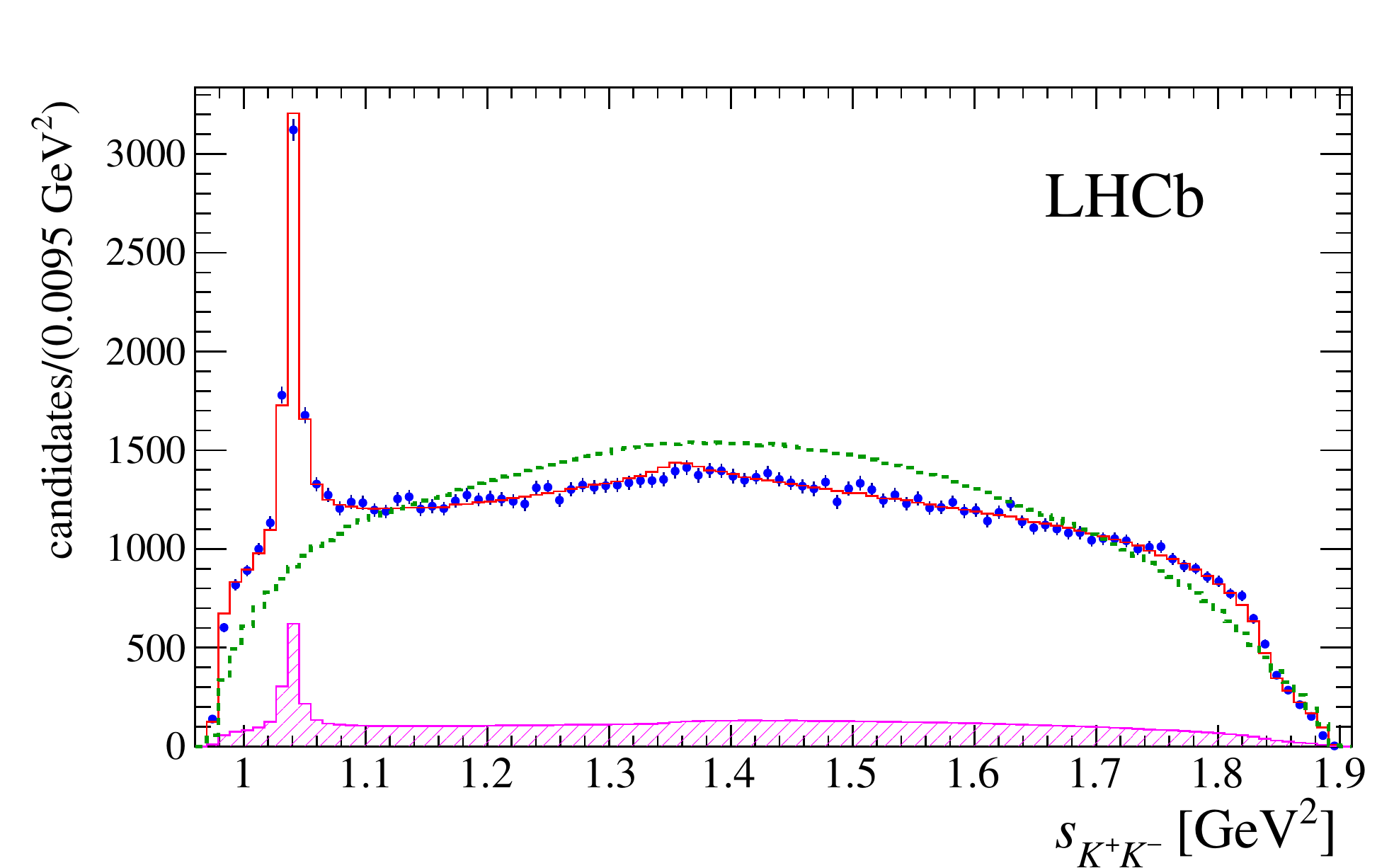}
\includegraphics [width=0.49\textwidth]{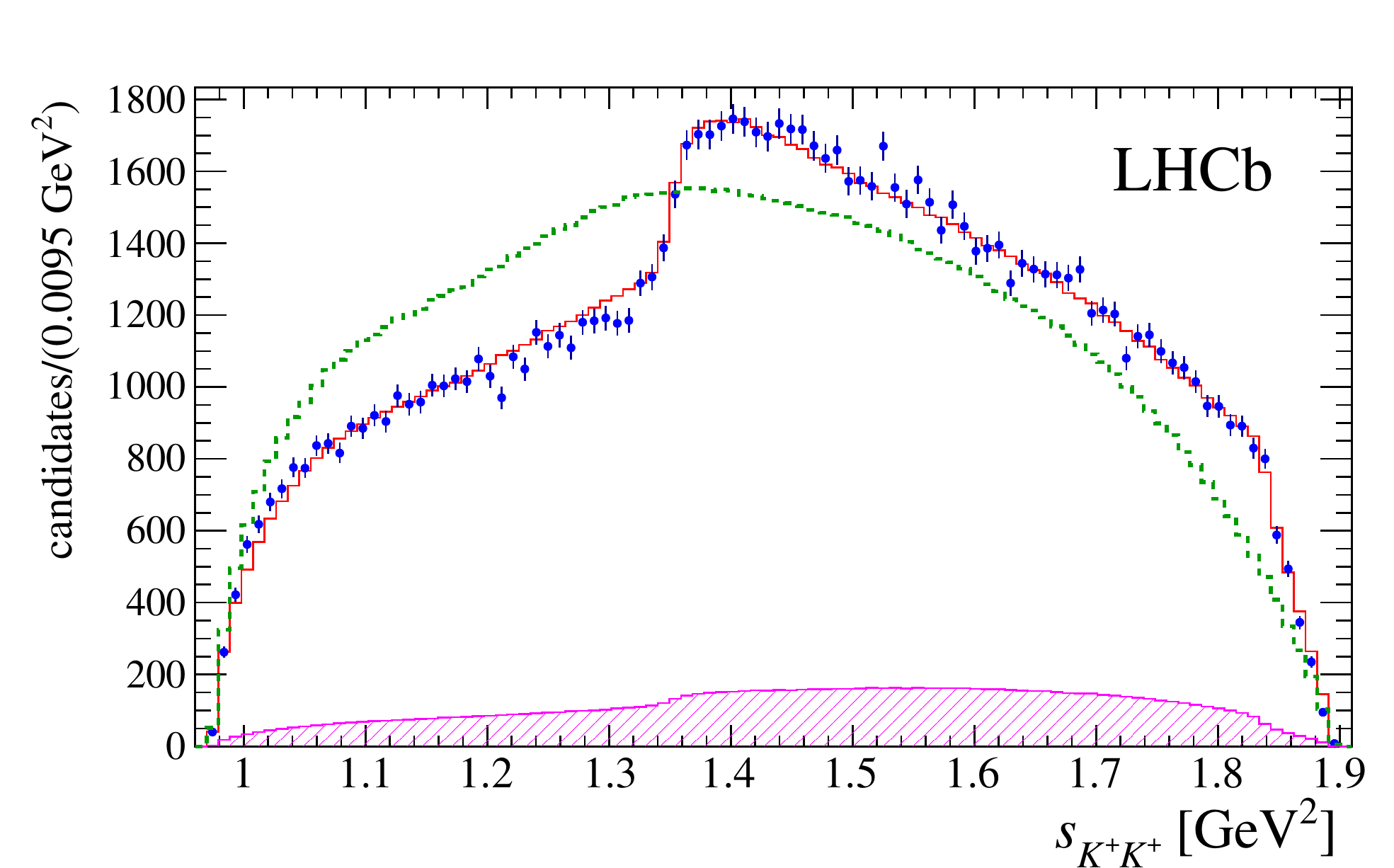}
                
\includegraphics [width=0.49\textwidth]{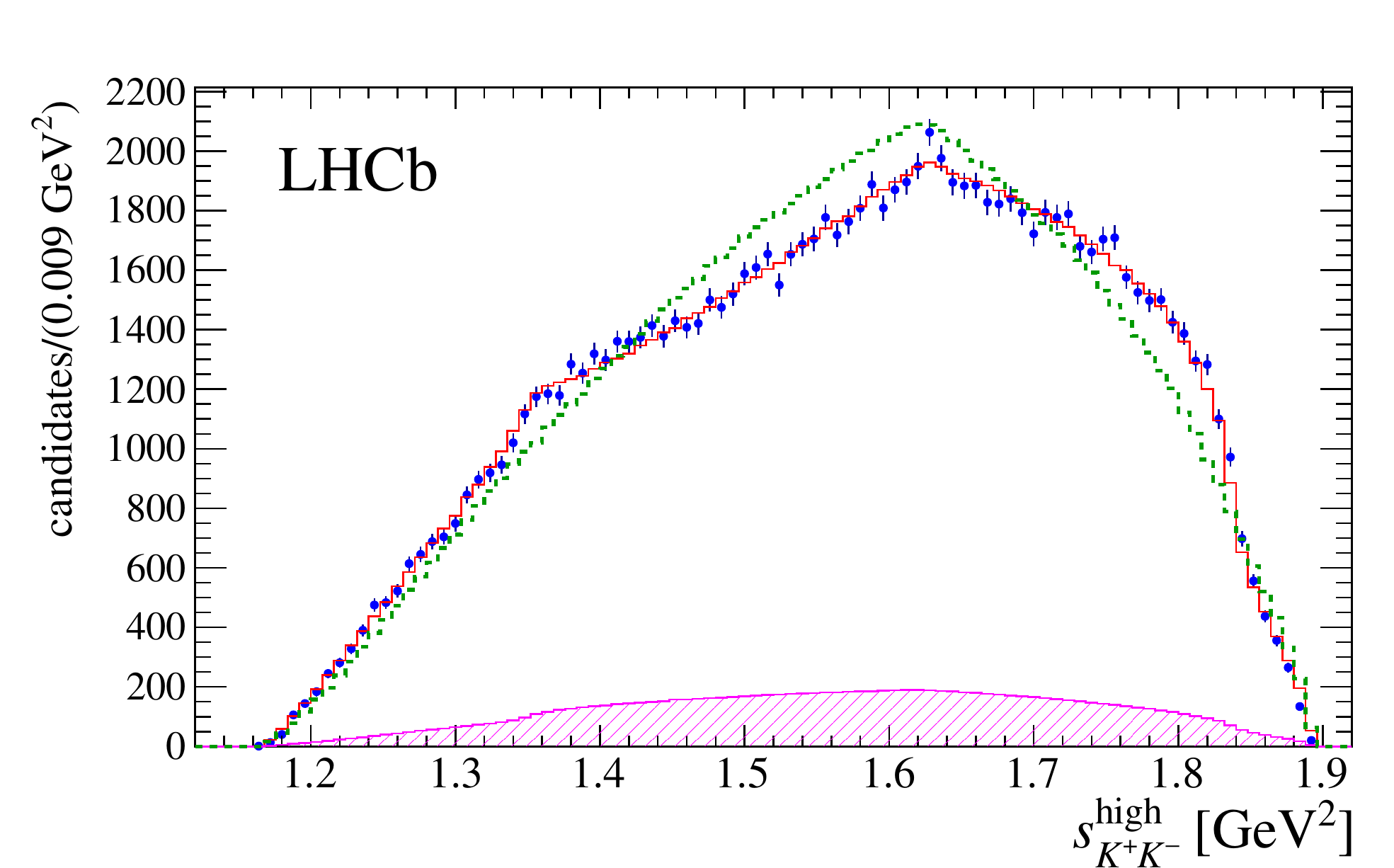}
\includegraphics [width=0.49\textwidth]{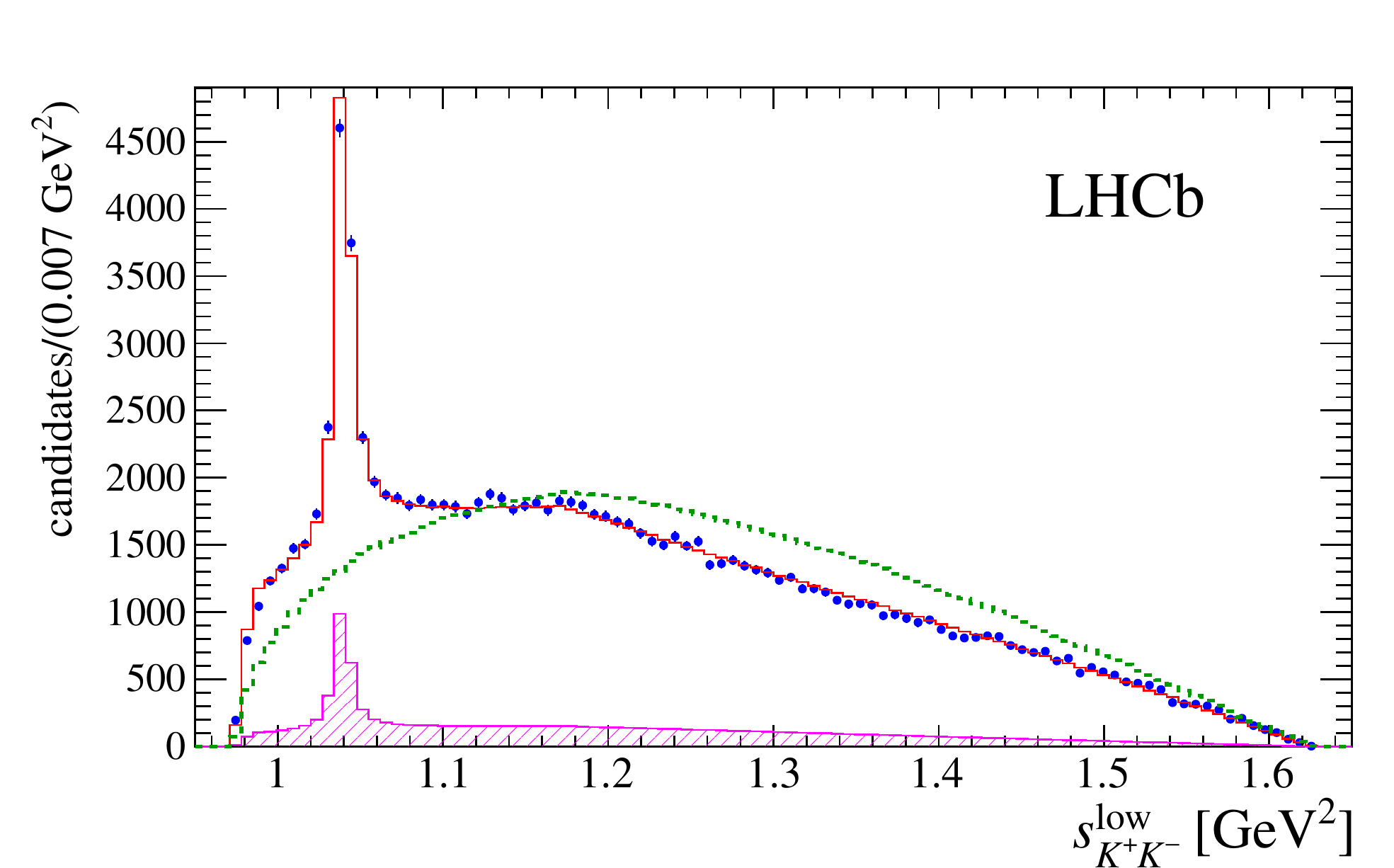}
\end{center}
    \vspace{-.7cm}
\caption{Projections of the Dalitz plot onto (top left) $s_{K^+K^-}$, (top right) $s_{K^+K^+}$, (bottom left)  
$s_{K^+K^-}^{\rm high}$ and (bottom right) $s_{K^+K^-}^{\rm low}$ axes, with the fit result with model A overlaid (red histogram). 
The histogram in magenta represents the contribution from the background, whereas the dashed green 
line is the phase-space distribution weighted by the efficiency.}
  \label{fig:model3c}
\end{figure}

Since model A is the simplest model describing all the general features of the observed Dalitz plot 
distribution, it is chosen  as the baseline result for the fit with the isobar model. 
The projections of the Dalitz plot, with the model A fit result overlaid, are shown in Fig.~\ref{fig:model3c}. 
The green dashed line represents the phase-space distribution, weighted by the efficiency, evidencing the presence of at 
least one broad, scalar  contribution not consistent with a  uniform distribution.

\begin{table}[hbtp]
\begin{center}
\caption{Results from the \DToKKK Dalitz plot fit with the isobar models  A, B and C. Magnitudes,  
$|c_k|$ , phases, $\arg(c_k)$ (in degrees),   and fit fractions (in \%) are given with statistical uncertainties only.}
\begin{tabular}{ll r@{\,$\pm$\,}l r@{\,$\pm$\,}l r@{\,$\pm$\,}l}\hline
 & &\multicolumn{2}{c}{Model A}&  \multicolumn{2}{c}{Model B} & \multicolumn{2}{c}{Model C}\\  

\hline
$\phi(1020)K^+$ &    Magnitude                   & \multicolumn{2}{c}{1 [fixed]} &\multicolumn{2}{c}{1 [fixed]} & \multicolumn{2}{c}{1 [fixed]} \\
                           &     Phase     & \multicolumn{2}{c}{0 [fixed]} &\multicolumn{2}{c}{0 [fixed]} & \multicolumn{2}{c}{0 [fixed]} \\
                           & Fraction        & 6.17 & 0.47  &  6.40 & 0.47  &  6.40 & 0.48 \\ \hline
$f_0(980)K^+$     &    Magnitude                 & 3.12 & 0.10& 2.64 & 0.08 & 2.84 &  0.13 \\
                           &     Phase     & $-58.9$ & 4.9& $-36.5$& 7.6  & $-25.9$&   8.4 \\
                           & Fraction       & 23.7 & 3.0 & 17.7 &  2.1 & 20.4 & 1.5  \\ \hline
$f_0(1370)K^+$     &    Magnitude                & 3.46 & 0.46& 2.33 & 0.35 & \multicolumn{2}{c}{--} \\
                         &     Phase       & 13.1 & 7.7 & 42   &  10  & \multicolumn{2}{c}{--} \\
                         & Fraction          & 25.4 & 5.0 & 18.7 & 1.5  & \multicolumn{2}{c}{--} \\ 
                         & $f_0(1370)$ mass [\gev] & 1.422 &  0.015 & 1.401 & 0.009 &  \multicolumn{2}{c}{--}\\
                         & $f_0(1370)$ width [\gev]& 0.324 &  0.038 & 0.178 & 0.031 &  \multicolumn{2}{c}{--}\\ \hline
NR                       &    Magnitude            & \multicolumn{2}{c}{--} & 8.8  &  1.3   & 11.7  &  1.8\\
                         &   Phase           & \multicolumn{2}{c}{--} & $-5.5$ &  6.5   & $-39.0$ &  4.4\\
                         & Fraction            & \multicolumn{2}{c}{--} & 18.4 &  5.9   & 32.7  &  8.2 \\ \hline
$a_0(980)K^+$  &    Magnitude                      & \multicolumn{2}{c}{--} & \multicolumn{2}{c}{--}& 5.9  &  0.4 \\
                         &     Phase         & \multicolumn{2}{c}{--} & \multicolumn{2}{c}{--}& 48.5 &  3.0\\
                         & Fraction            & \multicolumn{2}{c}{--} & \multicolumn{2}{c}{--}& 53.5 &  7.4\\ \hline
			& S-wave fraction     & 92  & 11    & 91 & 13 & 93 & 12\\ 
                         & Fractions sum           & 55.4 & 5.9    &61.2 & 6.4 & 113 & 11\\ 
\hline

\end{tabular}
\label{tab:isobar_results}
\end{center}
\end{table}

\begin{table}[h]
\begin{center}
\caption{Fit results with  model A, given in terms of the
magnitudes $|c_k|$, phases, $\arg(c_k)$ (in degrees),  and fit fractions (in \%). For each measurement, the first 
uncertainty is statistical, the second systematic and the third is a systematic uncertainty due to model. }
\begin{tabular}{l  r@{\,$\pm$\,}c@{\,$\pm$\,}c@{\,$\pm$\,}c   r@{\,$\pm$\,}c@{\,$\pm$\,}c@{\,$\pm$\,}c  r@{\,$\pm$\,}c@{\,$\pm$\,}c@{\,$\pm$\,}c    }\hline
 Component &  \multicolumn{4}{c}{Magnitude}  & \multicolumn{4}{c}{Phase [deg.]}  & \multicolumn{4}{c}{Fraction (\%)} \\ \hline
 $\phi(1020)K^+$         & \multicolumn{4}{c}{1.0 [fixed] }                       & \multicolumn{4}{c}{0.0 [fixed]}                      & 6.17 &  0.47  &  0.19  &  0.41 \\
 $f_0(980)K^+$           & 3.12 &  0.10  &  0.13  &  0.33   & $-58.9$  &  4.9  &  2.3  &  2.0  & 23.7 &  3.0 &  2.1  &  3.3 \\
 $f_0(1370)K^+$         & 3.46 & 0.46  &   0.32  &  0.73   &  13.1  &  7.7  & 1.6  & 3.2  & 25.4 &  5.0 &  3.4  &  3.8 \\
\hline
   sum         &      \multicolumn{4}{c}{}                          &      \multicolumn{4}{c}{}                           & 55.4  &  5.9  &  0.4  &  0.6 \\
\hline
\end{tabular}
\label{tab:model3}
\end{center}
\end{table}

\begin{figure}[!htb]
\centering
\includegraphics[width=0.49\linewidth]{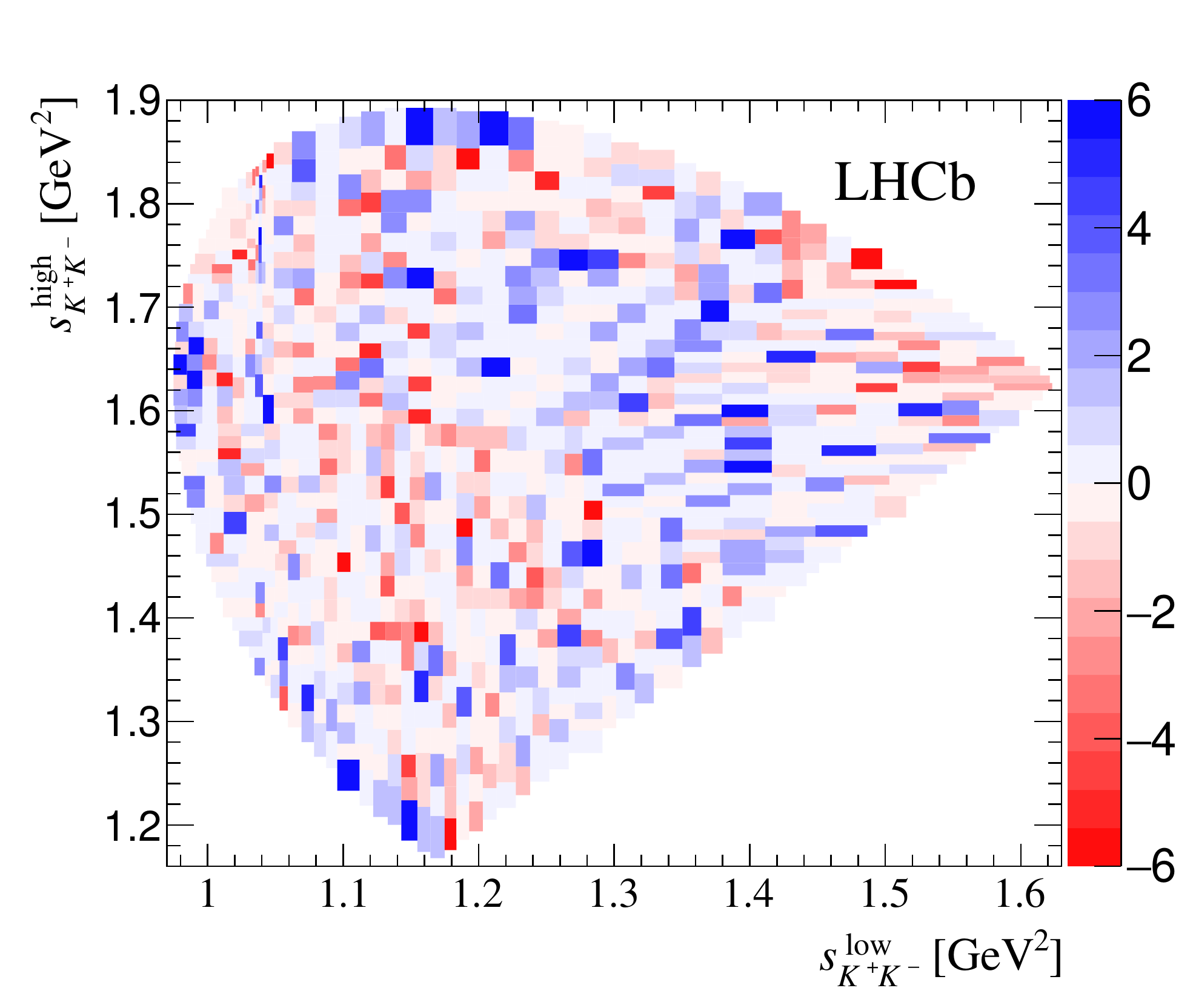}
\includegraphics[width=0.49\linewidth]{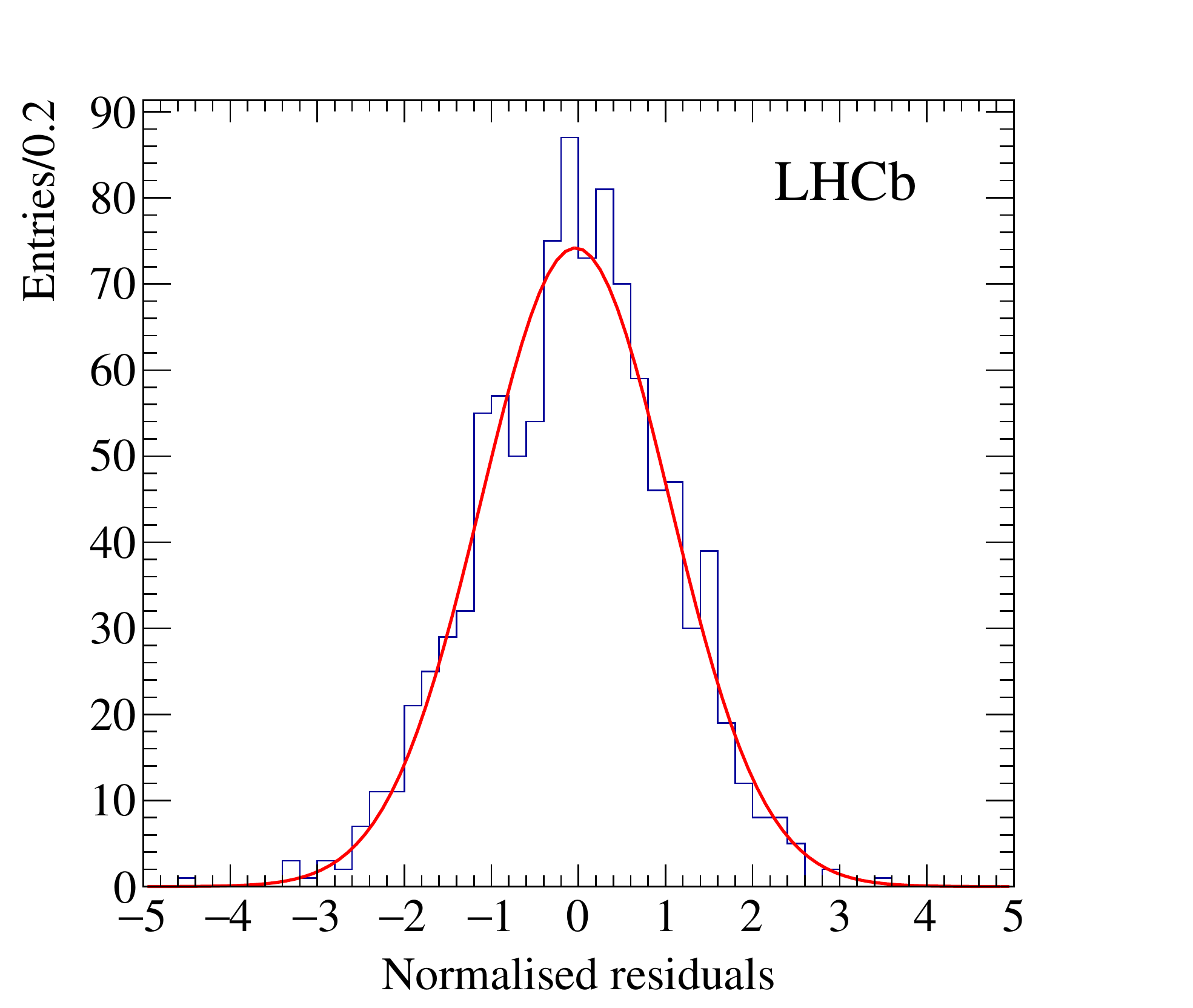}
\caption{(left) Normalised residuals $\Delta_i$ across the Dalitz plot, from the result of isobar fit. (right) Distribution of the normalised residuals with the fit result overlaid. The distribution is fitted with a Gaussian function and the fit result is consistent with the
standard normal distribution.}
\label{fig:model3a}
\end{figure}

The distribution of the normalised residuals $\Delta_i$ over the  Dalitz plot is shown in the left plot of 
Fig.~\ref{fig:model3a}, and their distribution is consistent with a normal Gaussian, as shown in the right 
plot of Fig.~\ref{fig:model3a}.
In Table~\ref{tab:model3} the results including the  systematic and model uncertainties, as discussed 
in Sec.~\ref{sec:systematics}, are presented. 

\begin{figure}[!htb]
\centering
 \includegraphics[width=0.49\linewidth]{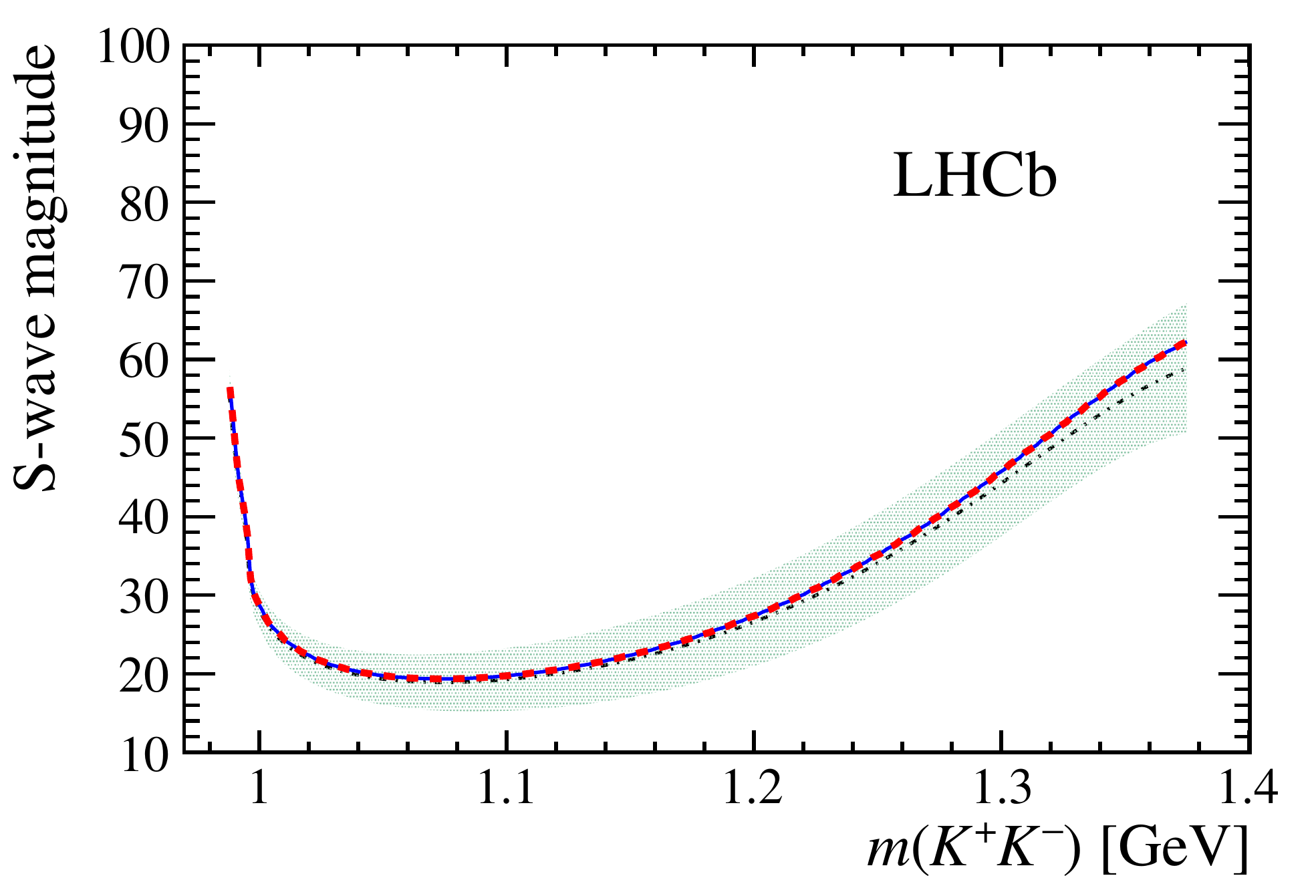}
 \includegraphics[width=0.49\linewidth]{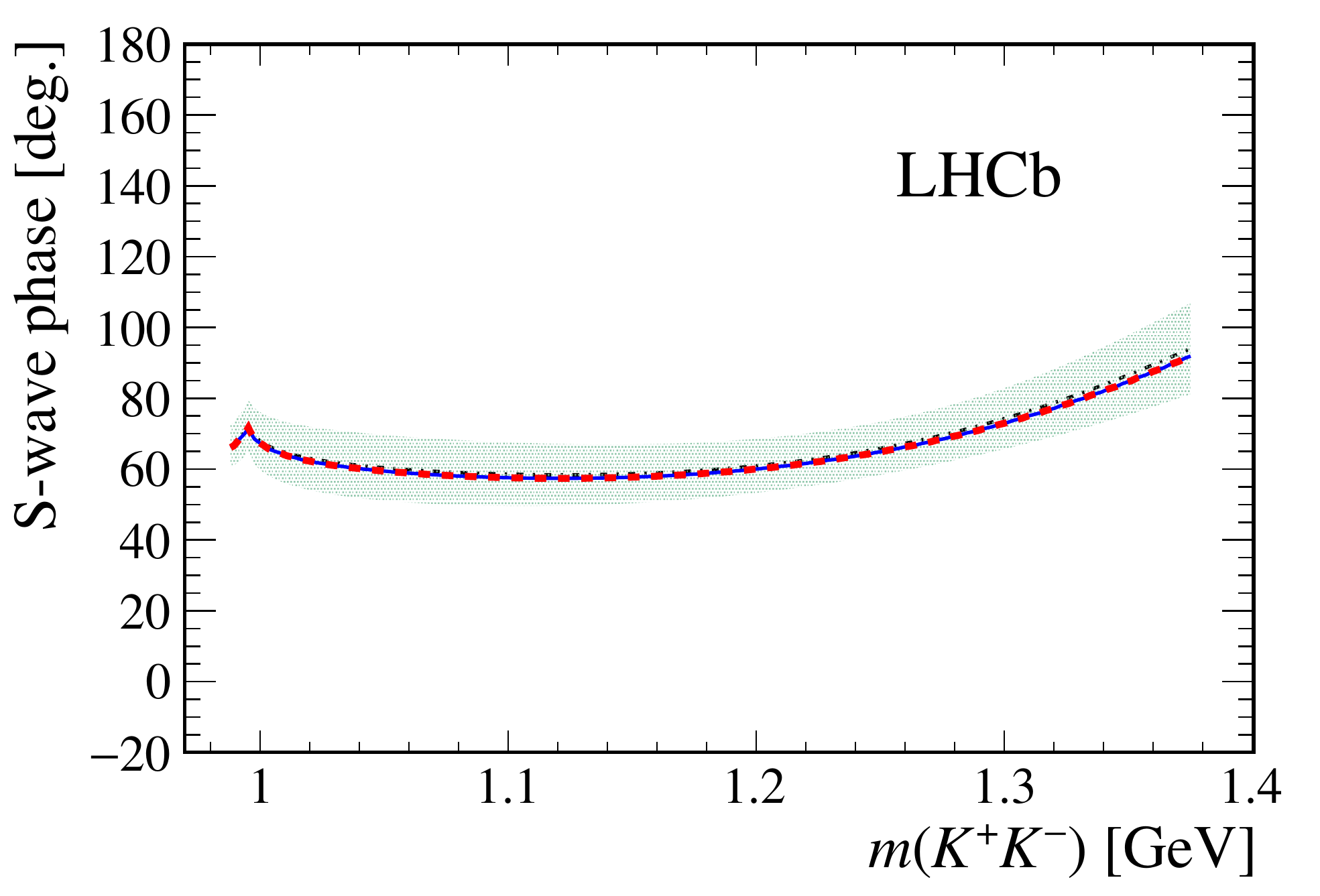}
\caption{(left) Magnitude and (right) phase of the total S-wave from the result of the Dalitz plot fit with the isobar model. 
The black line corresponds to model A and the green band represents the statistical and systematic uncertainties added 
in quadrature. For comparison, the results of
models B and C are shown as the blue solid and dashed thick red lines. Uncertainties on the 
S-wave magnitude and phase for models B and C are similar to those from model A and are not shown.}
\label{fig:modelA_MagPhase}
\end{figure}

The squared modulus and phase of the S-wave  amplitude from model A are shown in 
Fig.~\ref{fig:modelA_MagPhase} as a function of  the $K^+K^-$ mass,
with total uncertainties represented as bands. For comparison, the corresponding central results for 
models B and C are overlaid. Although the S-wave composition is different for these models,  
the total S-wave description is essentially the same, evidencing that the isobar model fails to
disentangle the individual contributions. 
The $f_0(1370)$ parameters are found to be \mbox{$m_0 = 1.422\pm 0.015 \pm 0.009\pm 0.028$\gev} and \mbox{$\Gamma_0 = 0.324\pm 0.038\pm 0.018\pm 0.038$\gev}, where the first uncertainties 
are statistical and the second systematic.

\clearpage

\section{Dalitz plot analysis with the Triple-M amplitude}
\label{sec:MMM}

\begin{figure}[!htb]
\centering
\includegraphics[width=0.6\linewidth]{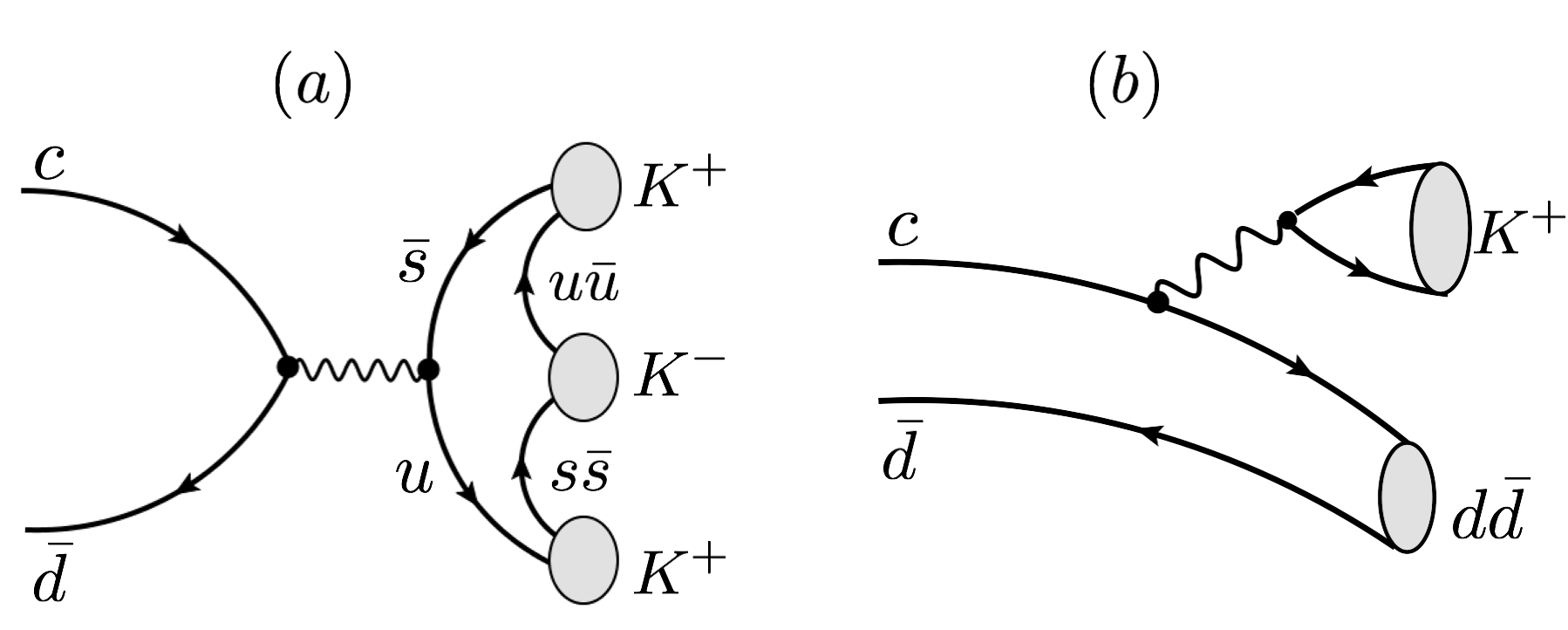}
\caption{Diagrams representing the two quark-level topologies for the \DToKKK decay. In the Triple-M~\cite{mmm},
diagram ($a$) is assumed to be the dominant mechanism of the decay, whereas  diagram ($b$) is 
suppressed since the production of a $K^+K^-$ pair from a $d\bar d$ pair requires rescattering.}
\label{fig:ann}
\end{figure}

The basic hypothesis of the Triple-M is the dominance of the annihilation diagram 
shown in Fig.~\ref{fig:ann}(a). The \DToKKK decay can also proceed via the diagram in 
Fig.~\ref{fig:ann}(b), but in this case a $K^+K^-$ pair could only be produced 
from the $d\bar d$ pair through rescattering, since charged kaons have no $d$-valence 
quark. The same holds for the production of the $\phi(1020)$ meson which is essentially an $s\bar s$ 
state~\cite{PDG2018}.

Assuming the annihilation diagram is the dominant mechanism for the \DToKKK decay, the Triple-M amplitude
is a product of two axial-vector currents,
\begin{equation}
\langle K^-K^+K^+|\mathcal{T}|D^+\rangle =   -  \lb \frac{G_F}{\rtw} \,\sin^2\th_C\rb \langle K^-K^+K^+ |A_{\mu}| 0 \rangle \langle 0|A_{\mu}| D^+
\rangle  \ ,
 \end{equation}
where $ G_F $ is the Fermi decay constant, $ \th_C $ is the Cabibbo angle and $ A^\m $ are the
axial currents. The weak vertex is $ \la \,0\,| A_\m |\, D^+(P) \ra  = -i\,\rtw\,f_D \,P_\m $,  where 
$P=p_1+p_2+p_3\,$ is the $D^+$ four-momentum and $f_D$  is the $D^+$ decay constant.

\begin{figure}[!htb] 
\begin{center}
\includegraphics[width=.60\linewidth,angle=0]{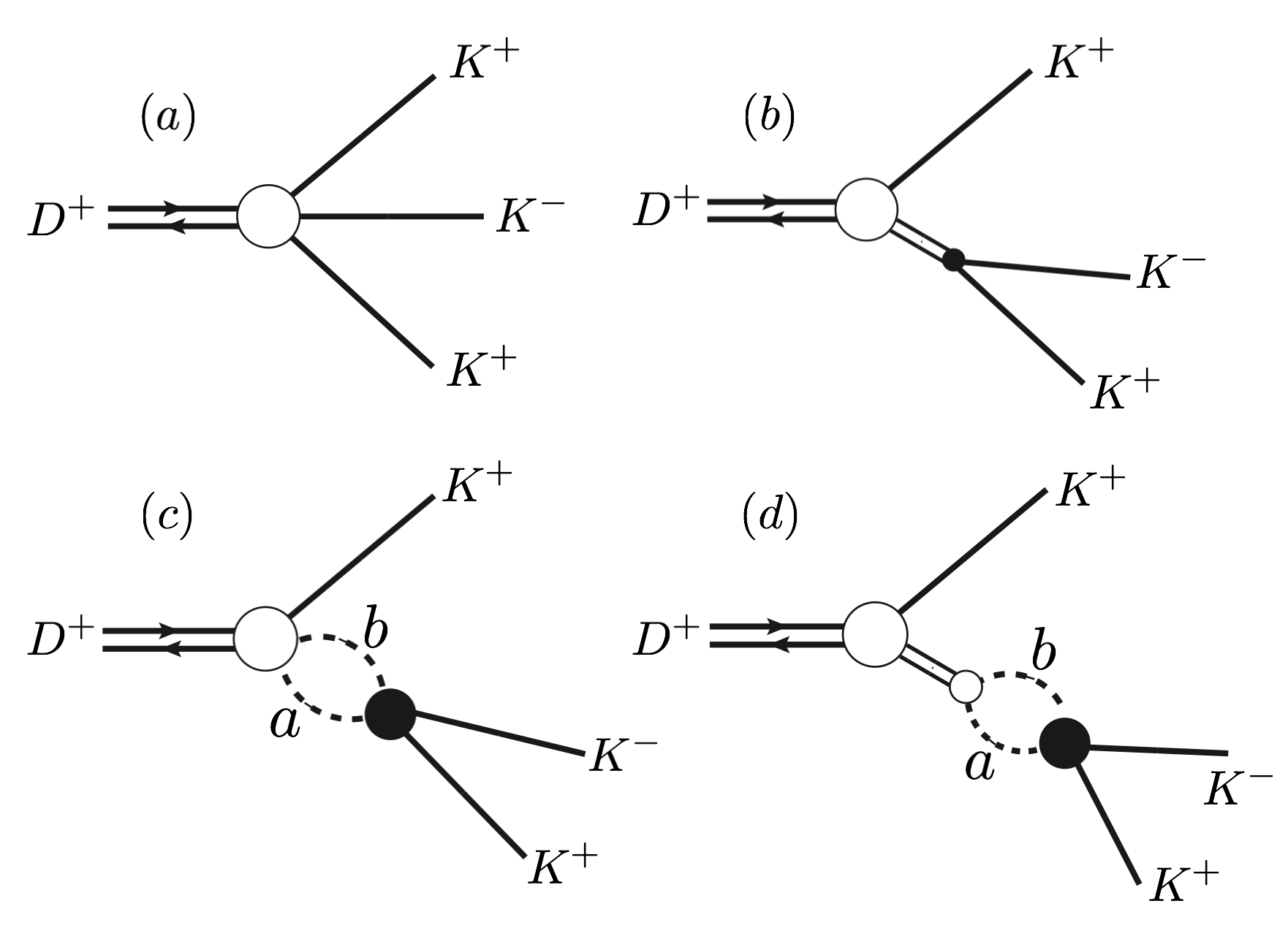}

\caption{Diagrams contributing to the amplitude $\mathcal{T}$ for the decay $D^+ \rar K^-\,K^+\, K^+$:
(a) the final state kaons are produced directly from the weak vertex; (b) a bare resonance 
is produced directly from the weak vertex; (c) particles produced at the weak vertex undergo  
final state interactions; (d) final state interactions endow finite widths to the resonances. 
The  full circle represents the unitary $ab\to K^+K^-$ scattering amplitude with angular 
momentum $J$ and isospin $I$, and  $ab=K\overline{K},\ \pi\pi,\ \eta\pi$ and $\eta\eta$.}
\label{decay}
\end{center}
\end{figure}

In the Triple-M,  the three-kaon system can  be produced in two ways, as illustrated in the 
diagrams in Fig.~\ref{decay}. Diagram (a) represents the production of the three kaons directly from 
the weak vertex, whereas in diagram (b) two of the three kaons result from the decay of a bare
intermediate resonance. Final state interactions are introduced in diagrams (c) and (d). The  full 
black circles indicate the unitarised scattering amplitudes, $A_{K^+ K^-}^{JI}$, representing the 
scattering $ab\to K^+K^-$ with the coupled channels $ab=K^+K^-,\ \pi\pi,\ \eta\pi$ and 
$\eta\eta$ in a well-defined spin ({\it J} ) and isospin ({\it I} ) state. The nonresonant component corresponds to  
diagram (a).  Due to the existence of two identical kaons, diagrams (b), (c) and (d) are 
symmetrised. As in the isobar analysis, contributions of D-wave are expected to be very small 
and are not included.

The Triple-M decay amplitude therefore has five components,
\begin{equation} 
\mathcal{T} = T_{\rm NR} + \sum_{J,I}  T^{JI},\hskip 0.6cm J,I=0,1.
\label{eqs:mmm}
\end{equation}
The free parameters in the Triple-M amplitude are the couplings and masses of the chiral 
Lagrangian. There are four couplings, 
$c_d,\ c_m,\ \tilde{c}_d,\   \tilde{c}_m$ in the scalar part, contributing to $T^{00}$ and $T^{01}$
terms; two masses, $m_{So}, \ m_{S1}$, for the scalar-isoscalar, $T^{00}$  contribution
and one, $m_{a_0}$, in the scalar-isovector  $T^{01}$ components;
one coupling, $G_V$, for the vector components, $T^{10}$ and $T^{11}$, and one mass, 
$m_{\phi}$, in the vector-isoscalar component. In the fit to the data, the combination 
$G_{\phi}\equiv G_V\sin\theta_{\omega-\phi}/F$ is used as free parameter, where 
$\theta_{\omega-\phi}$ is the $\omega-\phi$ mixing angle.
The parameter $F$ is the $SU(3)$ pseudoscalar  decay constant, common to all components. 
For convenience, the formulae of the various components of the Triple-M amplitude are reproduced 
from Ref.~\cite{mmm} in Appendix~\ref{app:triple-M}.

\clearpage

Equation~\ref{eqs:mmm} resembles that of the isobar model, but there are several significant
differences. The free parameters in the Triple-M amplitude are real quantities from the chiral
Lagrangian. Some of these parameters appear in different spin-isospin components of the model. 
In the isobar model the free parameters are the complex coefficients $c_k$, from which the
individual contributions of the resonances are determined. In the Triple-M amplitude, the relative
contributions of the various components  are fixed by theory. The nonresonant component is 
usually represented by an empirical constant in fits with the isobar model. In the Triple-M amplitude,
it is a function of the Dalitz plot coordinates and is fully determined by chiral symmetry.

\subsection{Fit results}

The optimum values of the Triple-M parameters are determined by an unbinned maximum-likelihood 
fit, as described in Sec.~\ref{sec:dalitzprocedure}. The fitted values of the Triple-M parameters  are 
listed in Table~\ref{tab:results}, with statistical and systematic uncertainties.  

\begin{figure}[hbtp]
\begin{center}
\includegraphics [width=0.49\textwidth]{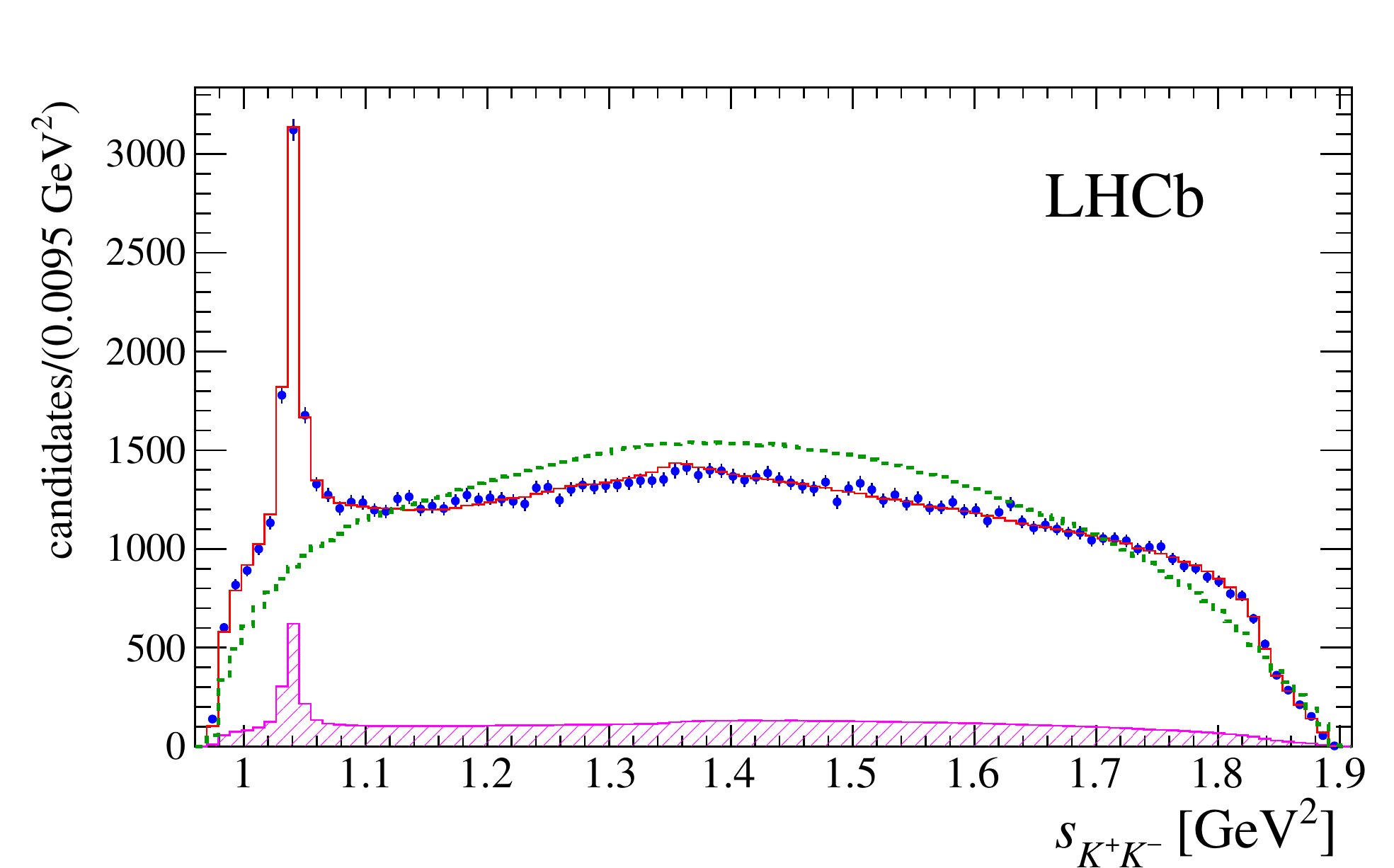}
\includegraphics [width=0.49\textwidth]{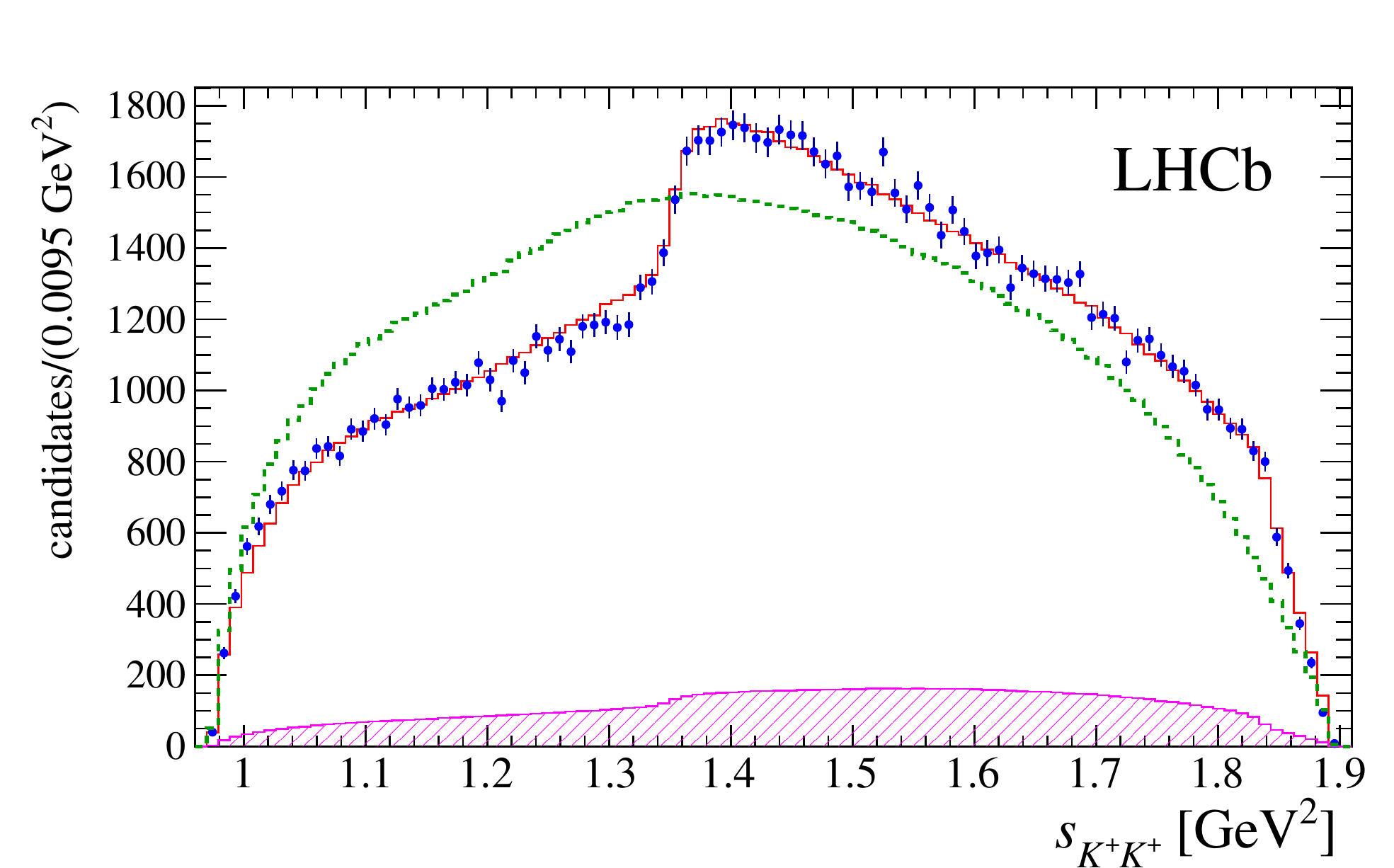}
                
\includegraphics [width=0.49\textwidth]{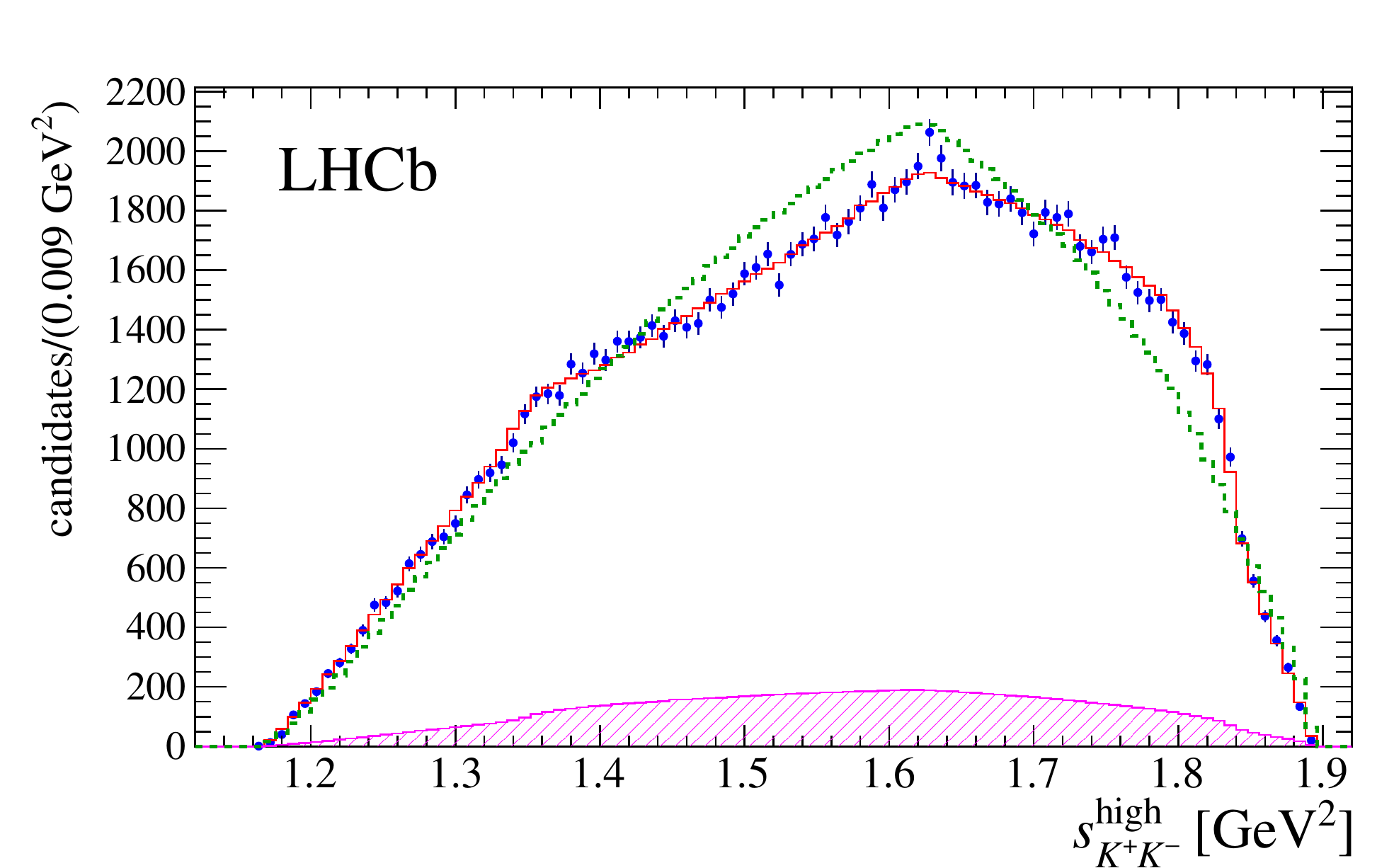}
\includegraphics [width=0.49\textwidth]{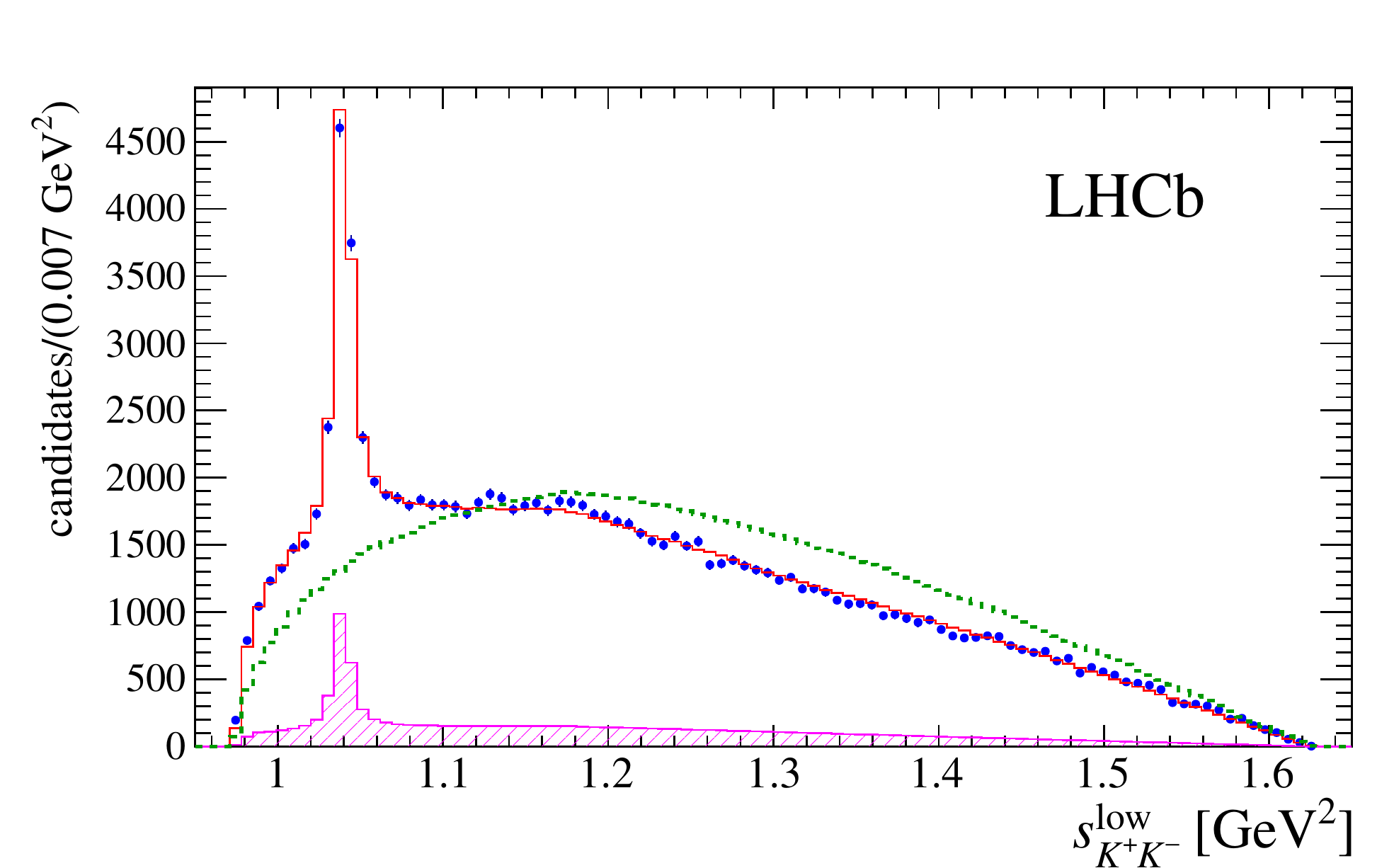}
\end{center}
    \vspace{-.7cm}
\caption{ Projections of  the Dalitz plot onto (top left) $s_{K^+K^-}$, (top right) $s_{K^+K^+}$, 
(bottom left)  $s_{K^+K^-}^{\rm high}$ and  (bottom right) $s_{K^+K^-}^{\rm low}$ axes, with the 
fit result with the Triple-M amplitude superimposed, whereas the dashed green line is the phase 
space distribution weighted by the efficiency. The magenta histogram represents the contribution
from the background.}
  \label{fig:proj}
\end{figure}

\begin{figure}[!htb]
\centering
\includegraphics[width=0.49\linewidth]{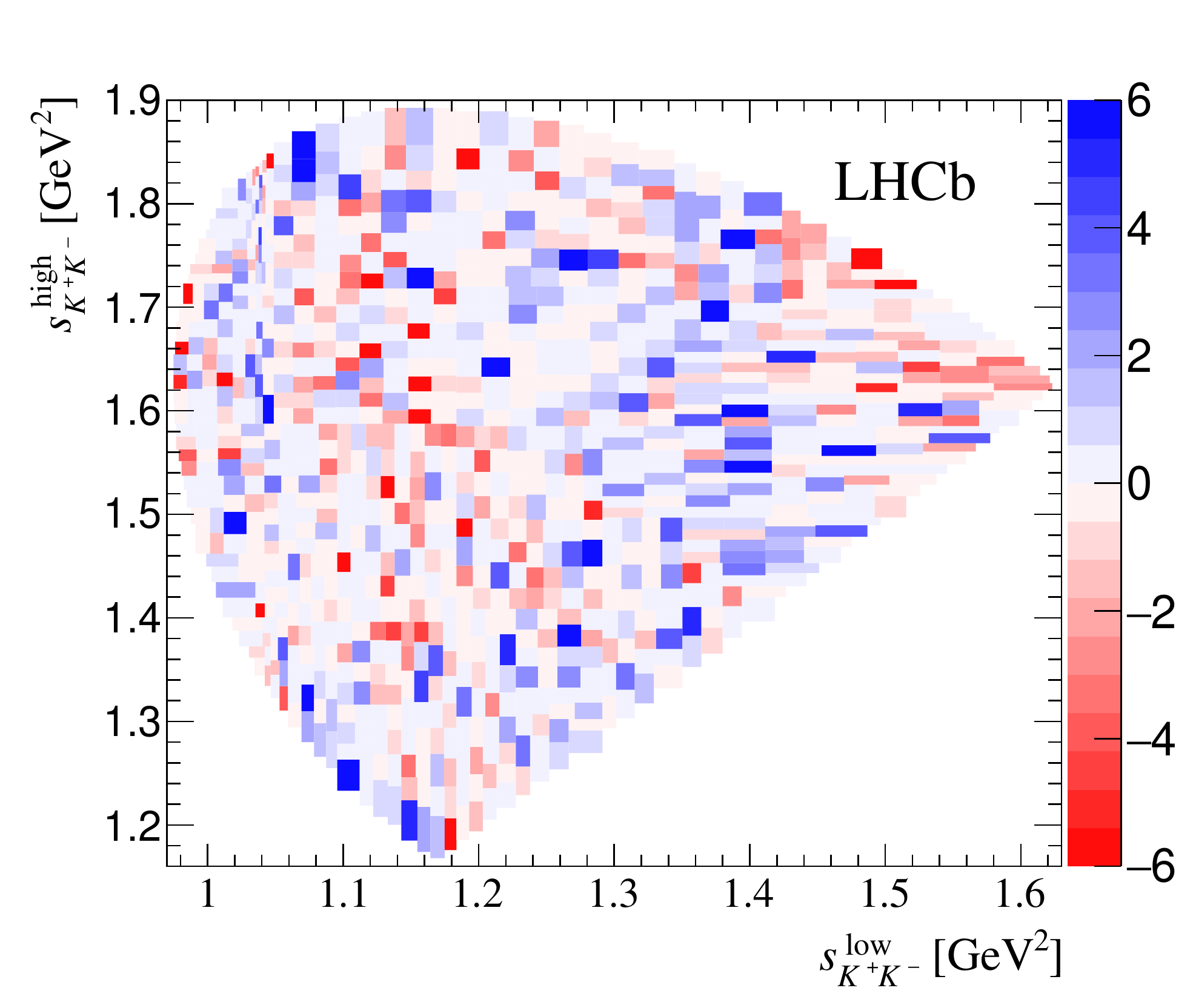}
\includegraphics[width=0.49\linewidth]{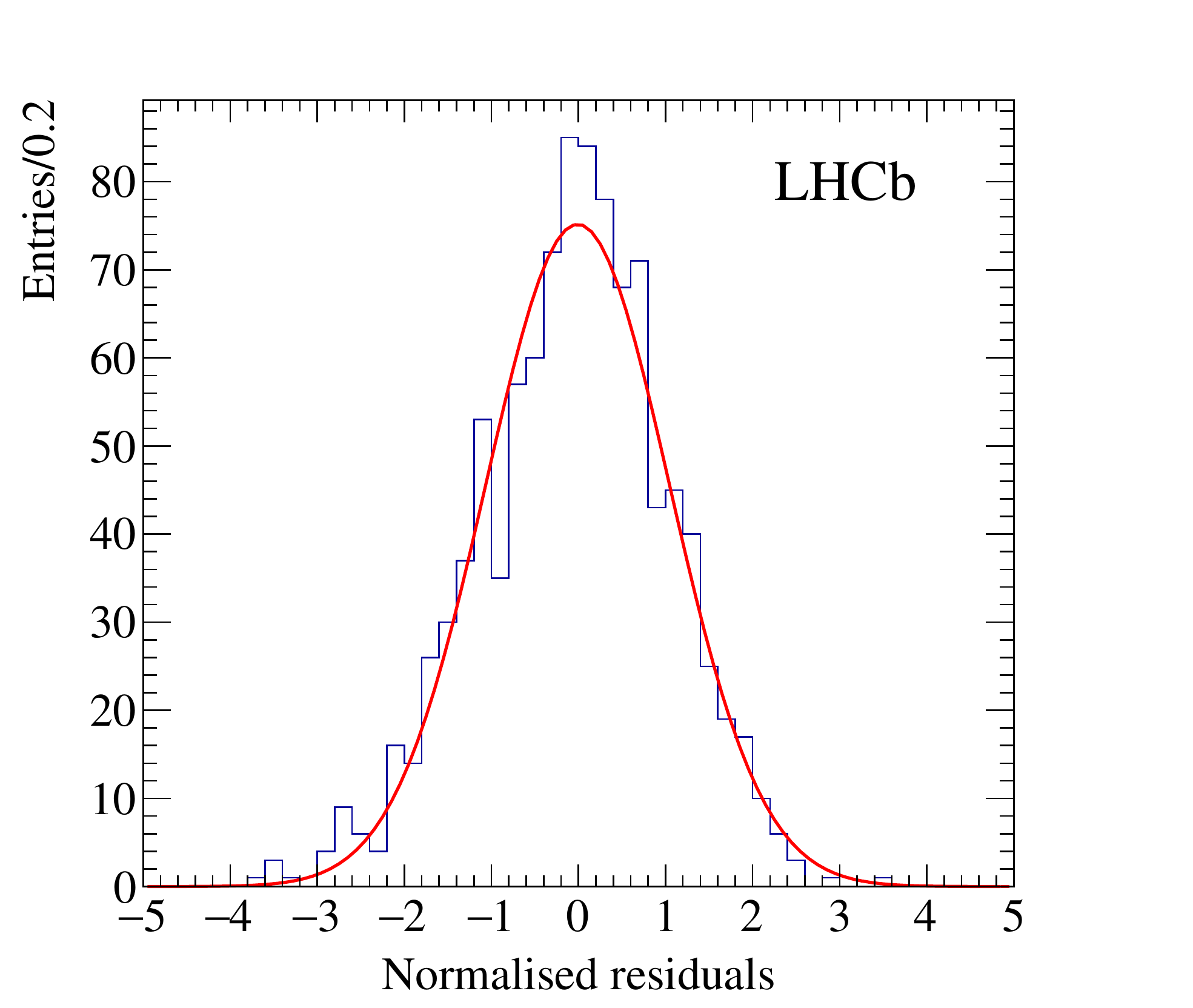}
\caption{(left) Two-dimensional distribution of the normalised residuals for the Triple-M fit. (right) 
Distribution of normalised residuals of each bin.}
\label{fig:chi2MMM}
\end{figure}

The quality of the fit with the Triple-M amplitude is tested with the metric defined in Eq.~\ref{eq:chi2}.
The  value of $\chi^2/{\rm ndof}$ is 1.12. The projections of the Dalitz plot onto the $s_{K^+K^-}$  and the 
$s_{K^+K^+}$ axes, as well as the projections onto the highest and lowest  invariant masses squared  
of the two $K^+K^-$ combinations, $s_{K^+K^-}^{\rm high}$ and  $s_{K^+K^-}^{\rm low}$, are shown in 
Fig.~\ref{fig:proj}, with the fit result superimposed. The projections indicate that the model is in 
good agreement with the data. The distribution of the normalised residuals  over the Dalitz plot
is shown in the right panel
of Fig.~\ref{fig:chi2MMM}.  The distribution of normalised residuals, shown in the left panel of 
Fig.~\ref{fig:chi2MMM}, is consistent with a normal Gaussian.

\begin{table}[!htb]
\caption{Results of the $D^+\to K^-K^+K^+$ Dalitz plot fit with the Triple-M amplitude.}
\begin{center}
\begin{tabular}{c|r@{\,$\pm$\,}l   }
\hline
parameter & \multicolumn{2}{c}{value} \\
\hline
$F$ &  $94.3^{+2.8}_{-1.7}$ & 1.5\mev  \\
$m_{a_0}$ & $947.7^{+5.5}_{-5.0}$ &  6.6\mev  \\
$m_{S_o} $&  $992.0^{+8.5}_{-7.5}$ &  8.6\mev\\
$m_{S_1}$ &  $1 330.2^{+5.9}_{-6.5}$ &  5.1\mev \\
$m_{\phi}$ & $1 019.54^{+0.10}_{-0.10}$ &  0.51\mev \\
$G_{\phi} $&  $0.464^{+0.013}_{-0.009}$ &  0.007   \phantom{0}\\
$c_d $&  $-78.9^{+4.2}_{-2.7}$  &  1.9\mev \\ 
$c_m$ & $106.0^{+7.7}_{-4.6}$ &  3.3\mev \\ 
$\tilde{c}_d$ &$ -6.15^{+0.55}_{-0.54}$ &  0.19\mev  \\ 
$\tilde{c}_m$ & $-10.8^{+2.0}_{-1.5}$ &   0.4\mev  \\
\hline
 \end{tabular}
\end{center}
\label{tab:results}
\end{table}

\subsection{Interpretation}

The resonance masses in the Triple-M are introduced in the denominators, D, of \mbox{Eqs.~\ref{un1}--\ref{un4}},
where the functions M are imaginary and proportional to interaction kernels which contain the bare 
masses of the effective chiral Lagrangian, $m_{a_0}$, $m_{S_o}$, $m_{S_1}$ and $m_{\phi}$.  
The Triple-M amplitude is derived assuming that only the imaginary part of the two-body 
propagators in Eqs.~\ref{un5}--\ref{un8} is relevant. In this approximation, the bare masses coincide 
with the masses of the physical states and the association $m_{S_o}=m_{f_0(980)}$ and 
$m_{S_1}=m_{f_0(1370)}$ can be made. As in the case of the isobar model, the masses in the 
Triple-M correspond to the values of $s_{K^+K^-}$ for which the real part of the denominator D of 
Eqs.~\ref{un1}--\ref{un4} vanishes. At these values of $s_{K^+K^-}$, only
the imaginary parts of the denominators remain, corresponding to the model prediction for the widths.
The denominators D would be very similar to those from the isobar model if no coupled channel was
considered. The inclusion of coupled channels is, therefore, 
the main difference between the Triple-M and Breit--Wigner denominators, resulting in widths with
different dynamical content.

\subsubsection{Resonant structure}

The  nonresonant contribution in the  Triple-M is a three-body  amplitude  predicted by chiral symmetry. 
It can be projected into  the S- and P-waves rewriting Eq.~\ref{dec2} as 
%
\bea
T_{\rm NR}  &=& \frac{C}{4} \lb  \, (m_D^2 - m_K^2+ s_{12}) + \,  (s_{13}-s_{23}) + 
(m_D^2 - m_K^2+ s_{13}) + \,  (s_{12}-s_{23}) \rb  \nonumber \\
&=&   T_{\rm NR}^S +  T_{\rm NR}^P \;,
\label{dXs.1}
\eea
where $C$ is a constant  common to all components of the Triple-M amplitude, and defined in 
Eq.~\ref{dec1a}. The decay amplitude can then be written as the sum of  scalar and vector components
\begin{equation}
\mathcal{T}  = \lb T^S + T^P + (2 \lrar 3) \rb ,
\label{dXs.2}
\end{equation}
with
\begin{equation}
T^S   = T_{NR}^S + T^{00}  + T^{01}  \; 
\label{dXs.3}
\end{equation}
and
\begin{equation}
T^P   = T_{NR}^P  + T^{11}  + T^{10}  \;.
\label{dXs.4}
\end{equation}

The relative contribution of each  individual component  of the Triple-M amplitude is determined by
integrating the modulus squared of
each term in the right-hand side of Eq.~\ref{eqs:mmm} over the phase space of the
 \DToKKK decay,
\begin{equation}
 {\rm FF}_{\rm NR} = \frac{\int {\rm d}s_{12}\ {\rm d}s_{13} \  |T_{\rm NR} (s_{12},s_{13})|^2}
{\int  {\rm d}s_{12}\ {\rm d}s_{13} \ \left|\mathcal{T} (s_{12},s_{13})\right|^2 }, \hskip .8cm
{\rm FF}^{JI} = \frac{\int {\rm d}s_{12}\ {\rm d}s_{13} \  |T^{JI} (s_{12},s_{13})|^2}
{\int  {\rm d}s_{12}\ {\rm d}s_{13} \ \left|\mathcal{T} (s_{12},s_{13})\right|^2 }.
\label{eqs:ff3m}
\end{equation}

Similarly, the S-wave contribution can be determined by the integral over the phase space of 
the modulus squared of the $T^S$ component, defined 
in Eq.~\ref{dXs.3}, and divided by the integral of the modulus squared of the
decay amplitude $\mathcal{T}$. The results are summarised
in Table~\ref{tab:fracs}. There is a large destructive interference between the two scalar 
below-threshold states, $a_0(980)$ and $f_0(980)$, yielding an S-wave contribution of  $(94\pm 1)$\%. 
The large $a_0(980)/f_0(980)$ interference may be, in part, due to the fact that in the $K^+K^-$ 
mass spectrum these two states have very similar lineshapes, since only the tails  are visible. 
This large interference is also observed in the fit with the isobar model C, yielding similar fit fractions for
the S-wave component.
A more accurate determination of the relative contribution of the
$a_0(980)$ and $f_0(980)$ resonances could be obtained from a simultaneous analysis of the 
$D^+\to \pi^+\pi^-\pi^+$ and $D^+\to \eta\pi^+\pi^0$. The contribution of the $\phi(1020)$ resonance,
$(7.1 \pm 0.5)\%$, is consistent to that observed in the fit with the isobar model.

\begin{table}[!htb]
\caption{Relative fractions (\%) of the various components of the Triple-M amplitude. The 
uncertainties correspond  to the combined statistical and systematic uncertainties.}
\begin{center}
\begin{tabular}{cccccc}
\hline
   ${\rm FF_{NR}}$ & ${\rm FF}^{00}$ &  ${\rm FF}^{01}$  &  ${\rm FF}^{10}$  & ${\rm FF}^{11}$ & $ {\rm FF_{S-wave}}$   \\
\hline
14 $\pm$ 1  & 29 $\pm$ 1 & 131 $\pm$ 2  & 7.1 $\pm$ 0.9  &  0.26 $\pm$ 0.01 & 94  $\pm$ 1 \\
\hline
\end{tabular}
\end{center}
\label{tab:fracs}
\end{table} 
 
\subsubsection{Decay and scattering amplitudes}
 
\begin{figure}[!htb]
\centering
\includegraphics[width=0.49\linewidth]{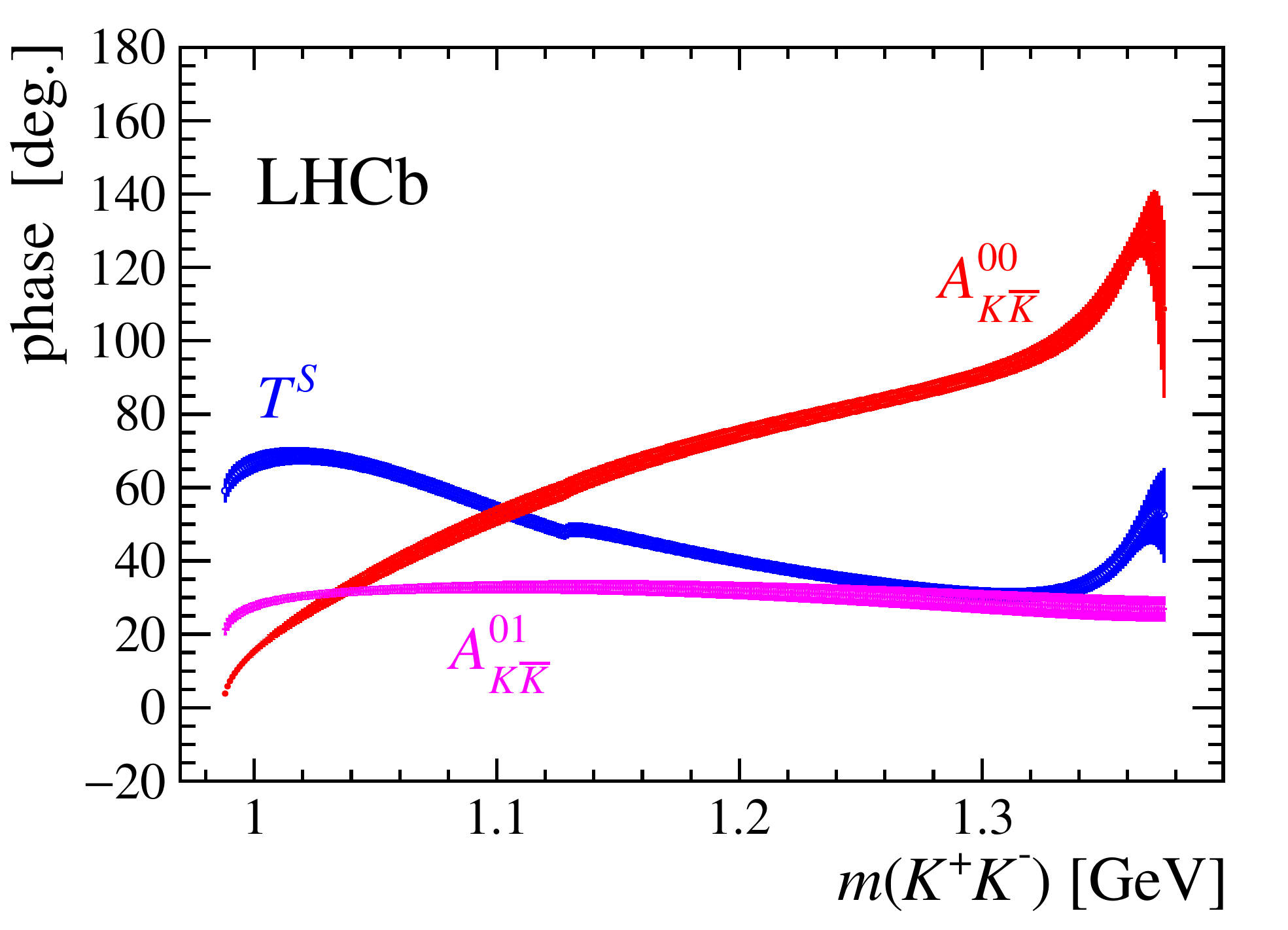}
\caption{Phase of the $J\!=\!0$ component of the decay amplitude (blue) 
\mbox{$T^S=T^{00}+T^{01}+T_{\rm NR}^S$},
compared to the phases of the scattering amplitudes, (red) $A_{K^+K^-}^{ 00 }$  and
 (magenta) $A_{K^+K^-}^{01}$ as a function of the $K^+K^-$ invariant mass. }
\label{fig:phsw}
\end{figure}

The phases of the S-wave amplitude, $T^S$, and  the $K^+K^-\to K^+K^-$ scattering amplitudes,
$A^{0I}_{K^+K^-}$, for the two allowed isospin states, are shown in  Fig.~\ref{fig:phsw}  as a function 
of the $K^+K^-$ invariant mass. The bands correspond to the statistical and systematic uncertainties 
added in quadrature. The kink in the phase of $T^S$ at $m(K^+K^-)\sim 1.25$ GeV is due to the opening of the 
$\eta\eta$ channel. The curves of Fig.~\ref{fig:phsw} illustrate the difference
between decay and scattering amplitudes. The latter, which depends on spin and isospin,
is a substructure of the former, which depends only on spin. The expressions of the various scattering 
amplitudes, derived in Ref.~\cite{mmm}, are reproduced in  Appendix~\ref{app:Scatteringamplitudes}.
 
\begin{figure}[hbtp]
\begin{center}
\includegraphics [width=0.4\textwidth]{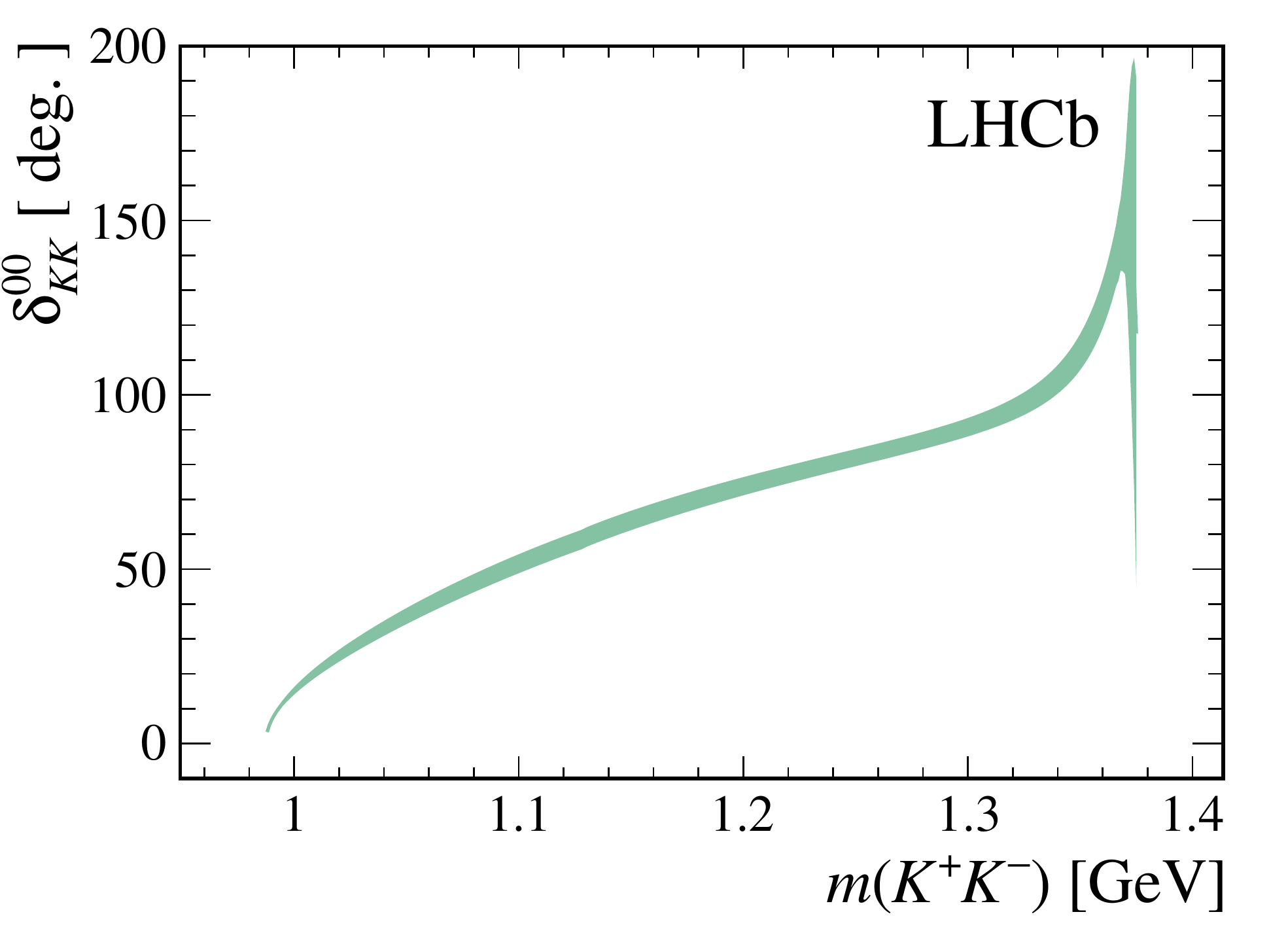}
\includegraphics [width=0.4\textwidth]{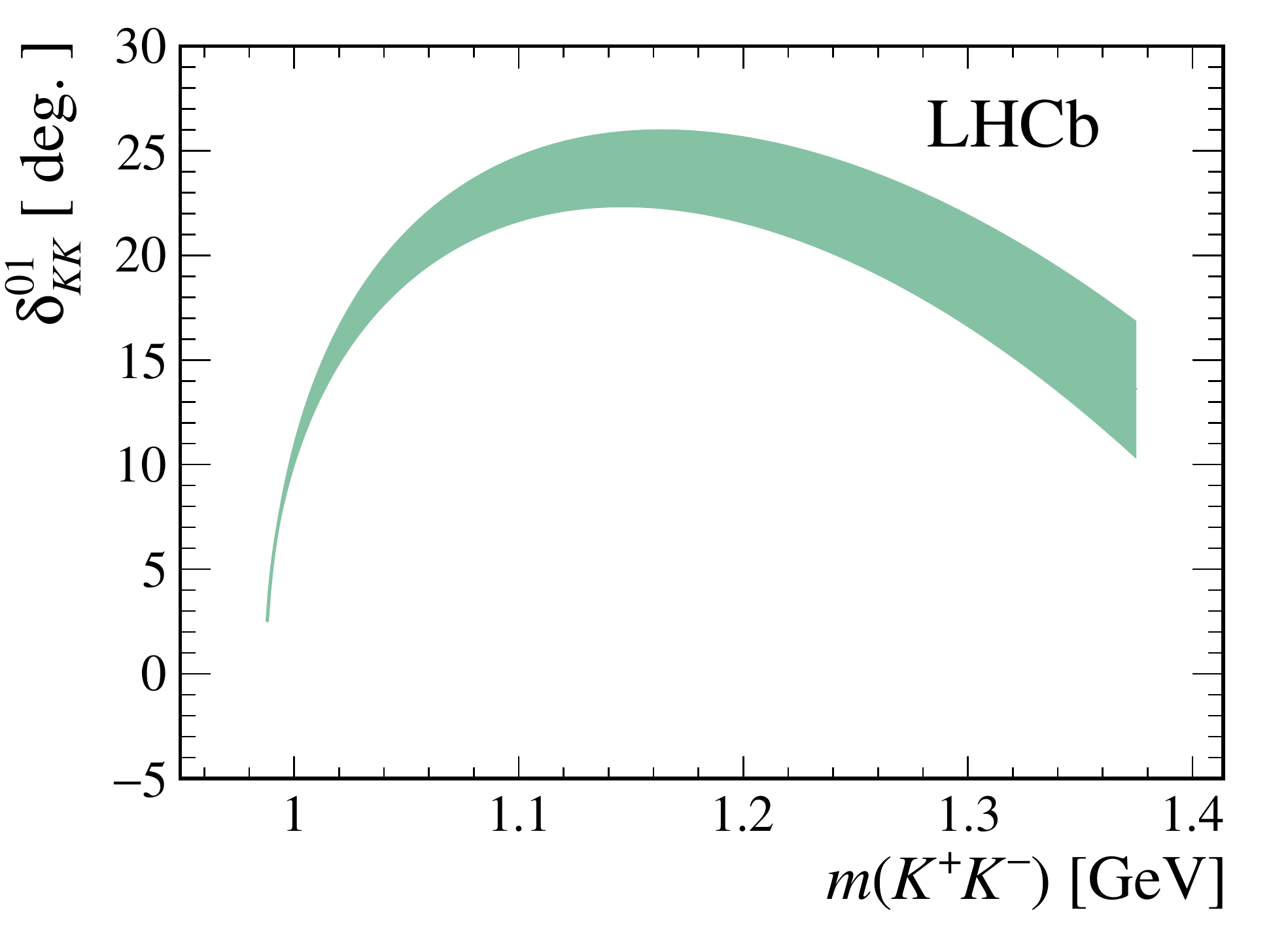}
                
\includegraphics [width=0.4\textwidth]{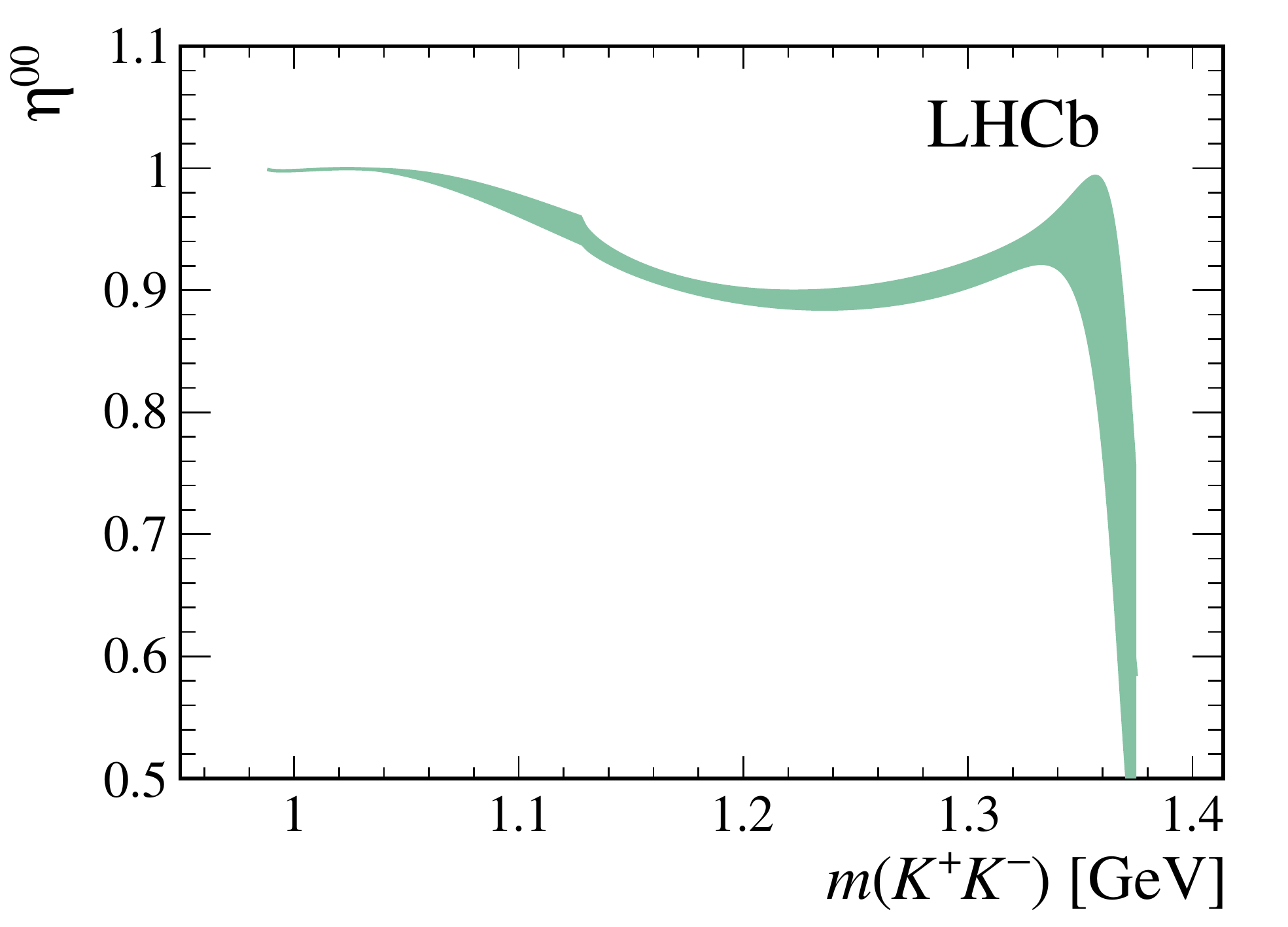}
\includegraphics [width=0.4\textwidth]{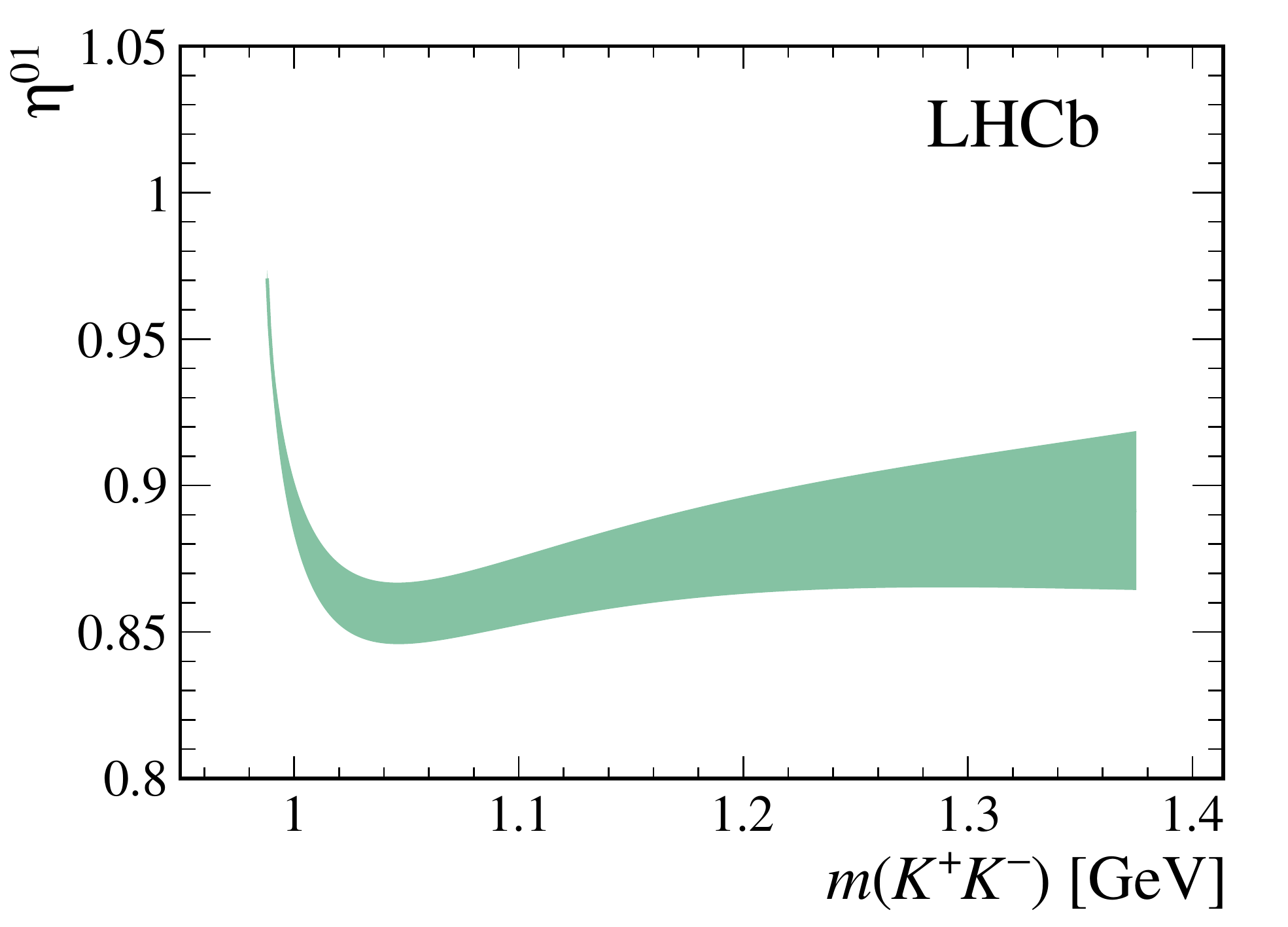}
\end{center}
    \vspace{-.7cm}
\caption{(top) Phase-shifts  $\delta^{0I}_{K^+K^-}$  and (bottom) inelasticities  $\eta^{0I}$ as a function 
of the $K^+K^-$ invariant mass, for both isospin states.}
  \label{fig:delta}
\end{figure}

The physics of two-body scattering is encompassed by the phase shifts and inelasticities. These 
quantities are obtained from  the scattering amplitudes, following the procedure described in 
Ref.~\cite{mmm}. The phase shifts, $\delta^{JI}_{K^+K^-}$,  and inelasticities $\eta^{JI}_{K^+K^- }$, 
are displayed in  Fig.~\ref{fig:delta}  for $J\!=\!0$ and $I\!=\!0,1$. 

The interpretation of the phase shifts for $K^+K^- $ scattering is not  as straightforward as in the case of 
elastic scattering, since for both isospin states, the $\pi\pi\to K^+K^-$ and $\pi\eta\to K^+K^-$  channels 
are already open  at the $K^+K^- $  threshold. An interesting feature of the results displayed is that 
the phase  variation of $\delta^{00}_{K^+K^-}$ is monotonic and spans over more than $180^\circ$, with
a fast variation starting at $m(K^+K^-)\sim 1.4$ GeV, close to the value of $m_{S_1}$ and typical of 
a resonance at high $K^+K^-$ mass. A fast variation of the phases is observed near threshold
for both $\delta^{00}_{K^+K^-}$ and $\delta^{01}_{K^+K^-}$, indicating the contribution from
the resonances  below  threshold.

The $\eta\eta$ channel contributes to $T^{00}$ but not to $T^{01}$ and 
its effect is visible in the bottom left plot of Fig.~\ref{fig:delta} as a kink at $m(K^+K^-)\sim 1.1$\gev.
As elastic scattering corresponds to $\eta^{JI}=1$, one sees that the 
isoscalar component becomes significantly more inelastic after the 
mass of the second scalar resonance.

\section{Systematic uncertainties}
\label{sec:systematics}

Sources of systematic uncertainties associated to the background model, to the efficiency correction and to
possible biases in the fitting procedure are common to the fits with the isobar model and the Triple-M.
They are summarized in 
Tables~\ref{tab:syst-iso} and \ref{tab:syst}, respectively. There is an additional source of systematic 
uncertainties  on the results of the fit with the isobar model due to the uncertainties on the parameters 
defining the $f_0(980)$  lineshape, which are fixed in the fit. This additional uncertainty, quoted separately 
from the experimental uncertainties, is estimated by repeating the fit varying the parameters $g_{\pi}$, 
$g_K$ and $m_0$ of Eq.~\ref{MRs12} by one standard deviation, one at a time, and taking the largest 
deviation as the systematic uncertainty. The radii of the Blatt--Weisskopf form factors
are also fixed in the fit. However, they impacts only the $\phi(1020)K^+$ amplitude. 
Fits with alternative values of these parameters are performed. 
The tested values of the radii are 4 and 6 GeV$^{-1}$, for $F_D^L$,
and 1 and 3 GeV$^{-1}$, for $F_R^L$. Since no significant
 deviation from the baseline fit is observed, no systematic uncertainty is assigned.

Two types of systematic uncertainties due to the background are investigated. First, the background
level is varied according to the uncertainty from the fit to the $K^-K^+K^+$ invariant mass. The data 
is fitted changing the fraction of the background  by $\pm 1\sigma$. No significant change in the 
fit parameters is found and no systematic uncertainty is assigned. Uncertainties due to the background
modelling are also investigated. The background model is built from inspection of the sidebands of the 
\DToKKK  signal. It is a combination of a peaking structure and a smooth component. The smooth 
component  corresponds to 80\% of the background and is modelled by a sum of a constant term and 
an $f_0(980)K^+$ contribution, in equal proportions. A systematic uncertainty due to the modelling of the
background is assigned by varying the relative fractions of these two components, fitting the data with 
these alternative background models and taking the largest variation as systematic uncertainty.


\begin{table}[!htb]
\caption{Systematic uncertainties (\%) on the results of the isobar model fit. For comparison, the statistical
uncertainties are listed in the last column.}
\begin{center}
\begin{tabular}{ccccc|c|c}
\hline
 parameter            & binning   & sim. stat.   &    bkg.   &    total  & model     & stat.    \\
\hline
$|c_{f_0(980)}|$               &  1.0     &  1.4       &   3.8     &  4.2   &   11    &   3.0    \\
$\delta_{f_0(980)}$          &  3.1     &  1.9       &   1.4     &  3.9   &   3.4   &   8.2    \\
$|c_{f_0(1370)}|$            &  3.5     &  3.5       &   7.8     &  9.2   &   21    &   13     \\
$\delta_{f_0(1370)}$       &  9.3     &  5.2       &   4.4     & 12     &   24    &   59     \\
$M_{f_0(1370)}$             &  0.1     &  0.3       &   0.6     &  0.7   &   2.0   &   1.0    \\
$\Gamma_{f_0(1370)}$  &  3.7     &  3.0       &   3.1     &  5.7   &   12    &   12     \\
\hline
\end{tabular}
\end{center}
\label{tab:syst-iso}
\end{table}

\begin{table}[!htb]
\caption{Systematic uncertainties (\%) on the results of  the Triple-M fit. For comparison, the statistcal
uncertainties are listed in the last column.}
\begin{center}
\begin{tabular}{ccccccc|c}
\hline
 parameter    & binning  & sim. stat.  &    PID     &  bkg.       &  fit bias & total      & stat.      \\
\hline
$F$                &  0.53     &  0.07       &   0.09     &   1.5        &  0.11    &   1.6      &   1.8      \\
$m_{a_0}$     &  0.54     &  0.14       &   -           &  0.40       &  0.16   & 0.70      &   0.54      \\
$m_{So}$      &  0.60     &  0.21        &   -          &   0.56       &  0.21   & 0.87      &    0.82      \\
$m_{S1}$      &  0.16     &  0.15        &   -          &   0.13       &  0.04   & 0.26      &    0.41    \\
$m_{\phi}$     &  0.002   & 0.001       &   -          &   -            & 0.002  & 0.003    &    0.005  \\
$G_{\phi}$     &  0.86     &  0.25        &   0.02     &  1.2         &  0.15   & 1.5        &    1.9      \\
$c_d$            &  0.18     &  0.08        &   0.09     &  2.4         &  0.13   & 2.4        &    3.3      \\
$c_m$           &  0.16     &  0.11         &   -           &  2.7         &  0.10  & 2.7        &    4.7      \\
$\tilde{c}_d$  &  0.13     &  0.15        &   -           &  2.6         &  1.1     & 3.1       &    8.8      \\
$\tilde{c}_m$ &  0.19     &  0.11        &  0.08       &  2.8         &   1.9    & 3.4       &    13     \\
\hline
\end{tabular}
\end{center}
\label{tab:syst}
\end{table} 

Systematic uncertainties are assigned to small biases in the fit  using  ensembles of 500 simulated 
samples. Two sets of samples are generated using the Triple-M amplitude and the isobar model,
both with the fitted values of the parameters. 
In the simulations the signal PDFs are weighted by the efficiency function and the background 
component is included. Each simulated sample is fitted independently, resulting in distributions of 
fitted values of the parameters and their respective uncertainties. 
For each parameter, the mean of the distribution of fitted values is compared to the input. The difference 
is assigned as the systematic uncertainty due  to the fit bias. A small bias is observed in the fit with the
Triple-M amplitude, whilst no bias is observed in the fit with the isobar model.

The systematic uncertainty associated to the efficiency variation across the Dalitz plot includes the effect 
of the uncertainties on the PID efficiency and  the hardware trigger correction factors, the effect of the finite
size of the  simulated sample, and the effect of the binning scheme of the efficiency histogram prior to 
the two-dimensional spline smoothing. The uncertainties on the PID efficiency are due to the finite size of 
the calibration samples and imply small systematic uncertainties  compared to the other sources of
systematics, in the fit with the Triple-M amplitude, and negligible uncertainties in the fit with the isobar
model. The uncertainty due to the hardware trigger correction factors is found to be negligible. 
The effect of the  finite size of the simulated sample is assessed by generating a set of alternative 
histograms from the selection efficiency histogram, prior to the hardware trigger correction and 
the PID efficiency weighting. The content of each bin of the selection efficiency histogram is varied 
according to a Poisson distribution.  For each of these alternative histograms, an efficiency map is 
produced and used to fit the data. For each parameter, the root mean square of the distribution of fitted 
values is assigned as a systematic uncertainty. The systematic uncertainty due to the binning scheme of 
the efficiency map is accessed by varying the number of bins of the final efficiency histogram. The 
histograms with alternative binnings are fitted by the two-dimensional cubic
spline. The data is fitted with these alternative efficiency maps and the largest variation of each parameter
is assigned as systematic uncertainty.

\section{Summary and conclusions}
\label{sec:conclusions}

In this paper, the first Dalitz plot analysis of the doubly Cabibbo-suppressed 
decay \DToKKK  is performed. The two goals of the analysis are the determinations of 
the resonant structure of the decay and the $K^+K^-$ scattering amplitudes. The resonant
structure is studied with two different approaches. In the fit with the isobar model, several
variations of the decay amplitude are tested. The Dalitz plot analysis is also performed with 
the Triple-M~\cite{mmm}, which is a model derived from a chiral effective Lagrangian. The Triple-M amplitude 
has a nonresonant component  plus the minimal $SU(3)$ content corresponding to four states, 
the $\phi(1020)$, the $a_0(980)$ and two isoscalar states, identified with the $f_0(980)$ and  
$f_0(1370)$ resonances. A good description of the data is achieved with both approaches.

The resonant structure of the \DToKKK is largely dominated by the S-wave,  with 
a  approximately 7\% contribution from the $\phi(1020)K^+$ component. The dominance of the S-wave  
contribution is also observed in other three-body $D^+_{(s)}$ decays with a pair of identical 
particles in the final state, such as the $D^+\to K^-\pi^+\pi^+$ and $D^+_{(s)}\to \pi^-\pi^+\pi^+$ 
decays\cite{PDG2018}. The possibility of determining the individual components of the S-wave,
however, is limited by the lack of structures in the Dalitz plot,
other than that from the $\phi(1020)$ resonance, and by the fact that the $f_0(980)$ and $a_0(980)$
mesons poles lie below the $K^+K^-$ threshold. In all the models tested,  large interference between
the various S-wave components is observed. In the fit with isobar model, different combinations of 
scalar resonances and nonresonant amplitudes yield fits of same quality and a very similar S-wave 
amplitude. In the fit with the Triple-M, a large $a_0(980)$ contribution
is observed, with a large destructive interference with the $f_0(980)$ component that yields an
S-wave fraction of about 94\%. The separation between the $f_0(980)$ and $a_0(980)$
contributions could better achieved with a simultaneous analyses of the \DToKKK,
$D^+\to \pi^-\pi^+\pi^+$ and $D^+\to \eta\pi^+\pi^0$ decays.

Predicitions for the $K^+K^-\to K^+K^-$ scattering amplitudes are obtained from the Dalitz plot
fit using the Triple-M amplitude. This is possible because the model incorporates explicitely
coupled channels and isospin degrees of freedom. In this respect, the chiral Lagrangian approach
represents an advance towards the description of the hadronic part of weak decays of $D$ mesons
in a more fundamental basis.

\section*{Acknowledgements}
%
%

\noindent We express our gratitude to our colleagues in the CERN
accelerator departments for the excellent performance of the LHC. We
thank the technical and administrative staff at the LHCb
institutes.
We acknowledge support from CERN and from the national agencies:
CAPES, CNPq, FAPERJ and FINEP (Brazil); 
MOST and NSFC (China); 
CNRS/IN2P3 (France); 
BMBF, DFG and MPG (Germany); 
INFN (Italy); 
NWO (Netherlands); 
MNiSW and NCN (Poland); 
MEN/IFA (Romania); 
MSHE (Russia); 
MinECo (Spain); 
SNSF and SER (Switzerland); 
NASU (Ukraine); 
STFC (United Kingdom); 
NSF (USA).
We acknowledge the computing resources that are provided by CERN, IN2P3
(France), KIT and DESY (Germany), INFN (Italy), SURF (Netherlands),
PIC (Spain), GridPP (United Kingdom), RRCKI and Yandex
LLC (Russia), CSCS (Switzerland), IFIN-HH (Romania), CBPF (Brazil),
PL-GRID (Poland) and OSC (USA).
We are indebted to the communities behind the multiple open-source
software packages on which we depend.
Individual groups or members have received support from
AvH Foundation (Germany);
EPLANET, Marie Sklodowska-Curie Actions and ERC (European Union);
ANR, Labex P2IO and OCEVU, and R\'{e}gion Auvergne-Rh\^{o}ne-Alpes (France);
Key Research Program of Frontier Sciences of CAS, CAS PIFI, and the Thousand Talents Program (China);
RFBR, RSF and Yandex LLC (Russia);
GVA, XuntaGal and GENCAT (Spain);
the Royal Society
and the Leverhulme Trust (United Kingdom);
Laboratory Directed Research and Development program of LANL (USA).



\clearpage

{\noindent\normalfont\bfseries\Large Appendix}

\appendix

\section{Decay amplitudes in the isobar model}
\label{app:isobar}
\setcounter{equation}{0}
\renewcommand{\theequation}{\thesection.\arabic{equation}}

Intermediate decay amplitudes within the Isobar model are given by Eq.~\ref{eq:isobar}.  Each factor appearing in that equation is presented here.

The form factors  $F_D^L$ and $F_R^L$, for the $D^+$ and the resonance decay, respectively, are 
parameterized by the Blatt--Weisskopf penetration factors~\cite{book:BlattWeisskopf}, and depend on $L$, the orbital angular momentum involved in   the transition. Since both the initial state (the $D^+$ meson) and the final state (three kaons) have spin 0, $L$ is equal to the spin of the resonance. In the rest frame of a resonance formed by particles 1  and 2 ,  {\em R}$_{12}$, $q$  is the modulus of the momentum of particle 1 or 2  (the decay momentum), $q_0$ is the decay momentum when $s_{12} = m_R^2$ ($m_R$ being the nominal resonance mass),  and $d$ is a measure of the effective radius of the decaying meson,  fixed in this work to $5.0 \mev^{-1}$ for the $D$ meson and $1.5 \mev^{-1}$ for the resonance.
Defining $z = (qd)^2$ and  $z_0 = (q_0d)^2$,  the Blatt--Weisskopf barrier factors are usually written with two different formulations,  $B_L$ and $B_L'$ \cite{PDG2018}, given in Table~\ref{tab:Blatt-Weisskopf}. The $B_L'$ formulation  is used in this analysis, consistent with the energy dependent width given below in Eq.~\ref{eq:width}, with the momenta in $F_D^L$ and $F_R^L$  computed in the rest frame of the respective decaying particle.

\begin{table}[!b]
\centering
\caption{Blatt--Weisskopf form factors for angular momentum $L=0,1,2$ with two distinct formulations. }
\begin{tabular}{ccc}\hline
L  &  $B_L$ & $B_L'$ \\ \hline
0  & 1 & 1     \\
1  & $\sqrt{\frac{2z}{1+z}}$ & $\sqrt{\frac{1+z_0}{1+z}}$   \\
2  & $\sqrt{\frac{13z^2}{1+z}}$ & $\sqrt{\frac{(z_0 - 3)^2+9z_0}{(z-3)^2+9z}}$    \\
\hline\hline
\end{tabular}
\label{tab:Blatt-Weisskopf}
\end{table}  

The function $ \mathcal{S}(\theta^{R12}_{13})$ 
describes the angular distribution of the decay particles, with $\theta^{R_{12}}_{13}=\theta^{R_{12}}_{13}(s_{12},s_{13})$
being the angle between particles 1 and 3 momenta measured in the rest frame {\em R}$_{12}$.
The Zemach formalism~\cite{paper:Zemach} is used for  the angular distribution
\begin{equation}
{\cal S} = (-2|p_1||p_3|)^L P_L(\cos{\theta}^{R_{12}}_{13}) ,
\label{eq:angular}
\end{equation}
where $P_L$ is the Legendre polynomial of order $L$. 
For vector and tensor resonances, this term introduces 
nodes in the Dalitz plot in regions where the helicity angle is either 90$^{\circ}$ or 270$^{\circ}$.

The relativistic Breit--Wigner function \cite{paper:BreitWigner} is used as the dynamical function,
\begin{equation}
M_R(s_{12}) = \frac{1}{s_{12} - m_R^2   + i m_R \Gamma(s_{12})} ,\label{eq:BW} 
\end{equation} 
where $m_R$ is the mass of the resonance and $\Gamma(s_{12})$ is the mass-dependent width, 
\begin{equation}
\Gamma(s_{12}) = \Gamma_R \left(\frac{q}{q_0}\right)^{2L+1} \frac{m_R}{\sqrt{s_{12}}} \left(\frac{F_R^L(z)}{F_R^L(z_0)}\right)^2 
\label{eq:width} 
\end{equation}
with $\Gamma_R$ being the nominal resonance width.

In the case of the $f_0(980)$ resonance, the relativistic Breit--Wigner is replaced by the Flatt\'e
formula\cite{paper:Flatte} 
\begin{equation}
M_R(s_{12}) = \frac{1}{s_{12} -m_R^2  + i m_R (\rho_{\pi\pi} \ g_{\pi}^2 + \rho_{KK}\  g_K^2)},
\label{MRs12}
\end{equation} 
where $g_{\pi}$ and $g_K$ are dimensionless coupling constants to the $K\Kbar$ and $\pi\pi$ channels, respectively,
 and $\rho_{\pi\pi}$ and $\rho_{KK}$ are the corresponding phase-space factors, 
\begin{equation}
\begin{split}
\rho_{\pi\pi} &= \sqrt{\left(\frac{s_{12}}{4} - m_\pi^2\right)} + \sqrt{\left(\frac{s_{12}}{4} - m_{\pi^0}^2\right)},\\
\rho_{KK} &= \sqrt{\left(\frac{s_{12}}{4} - m_K^2\right)} + \sqrt{\left(\frac{s_{12}}{4} - m_{K^0}^2\right)} .
\end{split}
\end{equation}

All the above formulation holds equally for the resonances in the system composed by particles 1 and 3, with $s_{12}\to s_{13}$ and
$\theta^{R_{12}}_{13}\to\theta^{R_{31}}_{12}$ (angular functions convention with cyclic permutation $(12)3 \to (31)2$).

\section{The Triple-M Decay amplitude}
\label{app:triple-M}
\setcounter{equation}{0}
\renewcommand{\theequation}{\thesection.\arabic{equation}}
All formulae presented in this appendix are reproduced from Ref.~\cite{mmm} for convenience.
The Triple-M decay amplitude for the $D^+ \rar K^- \, K^+ \, K^+ $ decay is given by
\bea
\mathcal{T}  &\!=\!& T_{\rm NR} + \lb T^{(1,1)}  + T^{(1,0)} + 
 T^{(0,1)} + T^{(0,0)} + (2\lrar 3) \rb \;,
\label{dec1}
\eea
where $T_{\rm NR} $ and the $T^{(J,I)} $ are the nonresonant and resonant contributions, respectively.
All components are proportional to the kaon mass squared, $m_K^2 $, included in the common factor
\bea
C  = \lc \lb  \frac{G_F}{\rtw} \, \sin^2\theta_C \rb \; 
\frac{2 F_D}{F} \, \frac{m_K^2}{(m_D^2-m_K^2)} \rc \,,
\label{dec1a}
\eea
where $F_D$ is the $D^+$ decay constant, $F$ is the $SU(3)$ pseudoscalar decay constant, $G_F$ is
the Fermi decay constant and $\theta_C$ is the Cabibbo angle. The nonresonant contribution is a three-body
amplitude, and therefore is not Bose-symmetrised. It is written as a real polynomial,
\bea
&& T_{\rm NR}  =  C\, \lb (s_{12}-m_K^2) + (s_{13}-m_K^2) \rb  \;.
\label{dec2}
\eea
The amplitudes $T^{(J,I)}$ are
\bea
&& T^{(1,1)}  = -\, \frac{1}{4}\, \lb \Gb_{KK}^{(1,1)}  - \G_{c|KK}^{(1,1)} \rb  \, (s_{13} \sm s_{23})\,,
\label{dec3} \\[4mm]
&& \Gb_{KK}^{(1,1)} = \frac{1}{D_\rho (s_{12})}
\lb M_{21}^{(1,1)}  \, \G_{(0)\, \p\p}^{(1,1)} + \lp 1 \sm M_{11}^{(1,1)} \rp \, \G_{(0)\, KK}^{(1,1)}  \rb \,,
\label{dec4}
\eea 
\bea
&& T^{(1,0)}  = -\,  \frac{1}{4}\,
\lb \Gb_{KK}^{(1,0)}  - \G_{c|KK}^{(1,0)}  \rb  \, (s_{13} \sm s_{23}) \,,
\label{dec5} \\[2mm]
&& \Gb_{KK}^{(1,0)} 
= \frac{1}{D_\phi (s_{12})}\; \G_{(0)\, KK}^{(1,0)} \,,
\label{dec6} \\[4mm]
&& T^{(0,1)}  =  -\, \frac{1}{2}\, \lb \Gb_{KK}^{(0,1)}  - \G_{c|KK}^{(0,1)} \rb  \;,
\label{dec7} \\[2mm]
&& \Gb_{KK}^{(0,1)} = \frac{1}{D_{a_0} (s_{12})}
\lb M_{21}^{(0,1)}  \, \G_{(0)\, \p 8}^{(0,1)} + \lp 1 \sm M_{11}^{(0,1)} \rp \, \G_{(0)\, KK}^{(0,1)}  \rb \,,
\label{dec8} \\[4mm] 
&& T^{(0,0)}  =  -\, \frac{1}{2}\, \lb \Gb_{KK}^{(0,0)}  - \G_{c|KK}^{(0,0)} \rb  \,,
\label{dec9}\\[2mm]
&& \Gb_{KK}^{(0,0)} = \frac{1}{D_S(s_{12}) }
\lc \lb M_{21}^{(0,0)} \lp 1\sm M_{33}^{(0,0)}\rp \sp M_{23}^{(0,0)} M_{31}^{(0,0)}\rb \, \G_{(0)\,\p\p}^{(0,0)} 
\right.
\nn \\[2mm]
&& \left. + \lb \lp 1\sm M_{11}^{(0,0)}\rp \lp 1\sm M_{33}^{(0,0)}\rp  
\sm M_{13}^{(0,0)} M_{31}^{(0,0)}\rb  \, \G_{(0)\,KK}^{(0,0)} 
\right.
\nn\\[2mm]
&&  \left. +\, 
\lb  M_{23}^{(0,0)} \lp 1\sm M_{11}^{(0,0)} \rp  \sp M_{13}^{(0,0)} M_{21}^{(0,0)} \rb  \, \G_{(0)\,88}^{(0,0)} \rc \;.
\label{dec10}
\eea
\vskip .4cm
The various functions $\G_{(0)ab}^{(J,I)} $ correspond to diagrams ($a$) and ($b$) of Fig.~\ref{decay}, and
represent the tree-level production of particles $abK^+$ from the weak vertex. The functions $\Gb_{KK}^{(J,I)}$ represent the full
decay vertex, from which the decay amplitude is obtained after subtracting the contribution of the contact terms $\G_{c|KK}^{(J,I)}$
to avoid double counting.  Their explicit form of the $\G_{(0)ab}^{(J,I)} $ functions are
\begin{equation}
 \G_{(0)\,\p\p}^{(1,1)}  =  C \lc 
\lb \frac{\rtw\, G_V^2}{ F^2} \rb \, \frac{s_{12}^2}{s_{12}^2 - m_\rho^2} 
+ \lb -\,  \frac{1}{\rtw}\rb_c \rc ,
\label{tsa2}
\end{equation}
\begin{equation}
 \G_{(0)\, KK}^{(1,1)}  = C \lc 
\lb \frac{G_V^2}{F^2} \rb \, \frac{s_{12}^2}{s_{12}^2 - m_\rho^2} 
+ \lb - \,  \frac{1}{2} \rb_c \rc .
\label{tsa3}
\end{equation}
\begin{equation}
 \G_{(0)\,KK}^{(1,0)} = C \lc 
\lb \frac{3\,G_V^2}{F^2}\, \sin^2\!\theta \rb \, \frac{s_{12}^2}{D_\f^{\p\rho}(s_{12}^2)} 
+ \lb -\,  \frac{3}{2}\rb_c \rc ,
\label{tsa5}
\end{equation}
\bea
&& \G_{(0)\,\p 8}^{(0,1)}  = C \lc 
\lb \frac{2\,\rtw}{\rth\,F^2} \rb  \;
\frac{ \lb - c_d \, P\cd p_3 + c_m \, m_D^2 \rb}{s_{12}^2 - m_{a_0}^2} 
\lb c_d \, \lp s_{12}^2 - m_\p^2 - m_8^2 \rp  
 + 2\, c_m \, m_\p^2 \rb 
 \right.
 \nn   \\[2mm]
 && \left.
+ \lb -\, \frac{\rth}{\rtw} \, \lb \, m_D^2/3 - P\cd p_3 \rb \rb_c \rc ,
 \label{tsa7}
 \eea
 \bea
&& \G_{(0)\,KK}^{(0,1)}  = C \lc 
\lb \frac{2}{F^2} \rb \;
\frac{ \lb - c_d \, P\cd p_3 + c_m \, m_D^2 \rb}{s_{12}^2 - m_{a_0}^2}
\lb c_d \, \lp s_{12}^2 - 2 m_K^2 \rp  
 + 2\, c_m \, m_K^2 \rb 
 \right.
 \nn\\[2mm]
 &&  \left.
+\lb -\, \frac{1}{2} \lb m_D^2 -  P \cd p_3 \rb \rb_c\rc ,
\label{tsa8} 
\eea
\bea
&& \G_{(0)\, \p\p}^{(0, 0)}  = C \lc 
\lb \frac{8 \rth}{F^2} \rb \; 
\frac{ \lb -\ct_d \, P\cd p_3 + \ct_m \, m_D^2 \rb}{s_{12}^2 - m_{S1}^2}\;
\lb \ct_d\, \lp s_{12}^2 - 2 m_\p^2 \rp  
 + 2\, \ct_m \, m_\p^2 \rb 
 \right.
\nn\\
&&  \left. 
- \;  \lb \frac{2}{\rth \, F^2} \rb  \;
\frac{ \lb - c_d \, P\cd p_3 + c_m \, m_D^2 \rb}{s_{12}^2 - m_{So}^2} \;
\lb c_d \, \lp s_{12}^2 - 2 m_\p^2 \rp  
 + 2\, c_m \, m_\p^2 \rb 
 \right.
 \nn\\
 && \left.
+\lb -\, \frac{\rth}{2} \, \lb m_D^2 - P\cd p_3 \rb \rb_c  \rc ,
\label{tsa10} 
\eea
\bea
&& \G_{(0)\,KK}^{(0,0)}  =  C \lc 
\lb \frac{16}{F^2} \rb \; 
\frac{ \lb - \ct_d \, P\cd p_3 + \ct_m \, m_D^2 \rb}{s_{12}^2 - m_{S1}^2}\;
\lb \ct_d \, \lp s_{12}^2 - 2 m_K^2 \rp  
 + 2\, \ct_m \, m_K^2 \rb 
 \right.
\nn\\[2mm]
&& \left.
+ \; \lb \frac{2}{3\, F^2} \rb  \;
\frac{ \lb -c_d \, P\cd p_3 + c_m \, m_D^2 \rb}{s_{12}^2 - m_{So}^2}
\lb c_d \, \lp s_{12}^2 - 2 m_K^2 \rp  
 + 2\, c_m \, m_K^2 \rb 
 \right.
 \nn\\[2mm]
 && \left.
+ \lb -\, \frac{3}{2} \, \lb m_D^2 - P\cd p_3 \rb \rb_c \rc ,
\label{tsa11} 
\eea
\bea
&& \G_{(0)\, 88}^{(0,0)}  = C \lc 
 \lb \frac{8}{F^2} \rb \;
\frac{ \lb -\ct_d \, P\cd p_3 + \ct_m \, m_D^2 \rb}{s_{12}^2 - m_{S1}^2}\;
\lb \ct_d \, \lp s_{12}^2 - 2 m_8^2 \rp  
 + 2\, \ct_m \, m_8^2  \rb 
 \right.
 \nn\\[4mm]
 && \left. 
 + \; \lb \frac{2}{3\, F^2} \rb  \;
\frac{ \lb -c_d \, P\cd p_3 + c_m \, m_D^2 \rb}{s_{12}^2 - m_{So}^2}
\lb c_d \, \lp s_{12}^2 - 2 m_8^2 \rp  
 + c_m \, \lp - 10 m_\p^2 + 16m_K^2 \rp/3 \rb 
 \right.
 \nn\\[2mm]
 && \left.
 + \lb - \, \frac{1}{2} \, \lb 5\,m_D^2/3  - 3\, P \cd p_3 \rb \rb_c \rc ,
\label{tsa12} 
\eea
with
\bea 
P\cd p_3 = \frac{1}{2} \lb m_D^2 + m_K^2 - s_{12}^2\rb \,.
\label{tsa13}
\eea
\vskip .4cm
In the above equations, $m_{\pi}$ and $m_D$ are the $\pi^+$ and the $D^+$ masses, respectively, and $\theta$ is the 
$\omega-\phi$ mixing angle.
The subscripts $8$ refer to the member of the $SU(3)$ octet with the quantum numbers of  the $\eta$.
The denominators in Eqs.~\ref{dec4}, \ref{dec6}, \ref{dec8} and \ref{dec10} are the model prediction for the
resonance line shapes: 
\bea
&& D_\rho = D^{(1,1)}
= \lb \lp 1\sm M_{11}^{(1,1)}\rp  \,  \lp 1\sm M_{22}^{(1,1)}\rp  - M_{12}^{(1,1)}\, M_{21}^{(1,1)} \rb \;,
\label{un1}\\[4mm]
&& D_\phi =  D^{(1,0)}  
= \lc 1\sm M^{(1,0)} \rc \,,
\label{un2}\\[4mm]
&& D_{a_0} = D^{(0,1)} 
= \lb \lp 1\sm M_{11}^{(0,1)} \rp \,  \lp 1\sm M_{22}^{(0,1)} \rp - M_{12}^{(0,1)}  M_{21}^{(0,1)}  \rb \,,
\label{un3} \\[4mm]
&& D_S = D^{(0,0)}
= [1\sm M_{11}^{(0,0)}][1\sm M_{22}^{(0,0)} ][1 \sm M_{33}^{(0,0)}]
-  [1 \sm M_{11}^{(0,0)}]  M_{23}^{(0,0)} M_{32}^{(0,0)} 
\nn\\[2mm]
&& - [1 \sm M_{22}^{(0,0)}] M_{13}^{(0,0)} M_{31}^{(0,0)} 
- [1 \sm M_{33}^{(0,0)}] M_{12}^{(0,0)} M_{21}^{(0,0)} 
\nn\\[2mm]
&& -  M_{12}^{(0,0)}  M_{23}^{(0,0)} M_{31}^{(0,0)}  
- M_{21}^{(0,0)}  M_{32}^{(0,0)}  M_{13}^{(0,0)} \;.
\label{un4}
\eea
\vskip .4cm
The functions $M_{ij}^{(J,I)}$ read
\bea
&& M_{11}^{(1,1)} =  - \cK_{\p\p|\p\p}^{(1, 1)} \, [ \Ob_{\p\p}^P /2 ]  \;,
\hspace{10mm}
M_{12}^{(1,1)} =  -  \cK_{\p\p|KK}^{(1, 1)}  \, [\Ob_{KK}^P/2] \;,
\nn \\[2mm]
&& M_{21}^{(1,1)}  = - \cK_{\p\p|KK}^{(1, 1)} \, [\Ob_{\p\p}^P/2] \;,
\hspace{10mm}
M_{22}^{(1,1)} = -  \cK_{KK|KK}^{(1, 1)} \, [\Ob_{KK}^P /2] \;,
\label{un5}
\eea
\bea
&& M^{(1,0)} = - \cK_{KK|KK}^{(1, 0)} \, [\Ob_{KK}^P/2]\;, 
\label{un6}
\eea
\bea
&& M_{11}^{(0,1)} =  - \cK_{\p 8|\p 8}^{(0, 1)} \, [\Ob_{\p 8}^S] \;,
\hspace{10mm} 
M_{12}^{(0,1)} =  - \cK_{\p 8 |KK}^{(0, 1)} \, [\Ob_{KK}^S/2] \;,
\nn \\[2mm]
&& M_{21}^{(0,1)}  = - \cK_{\p 8|KK}^{(0, 1)} \, [\Ob_{\p8}^S] \;,  
\hspace{10mm}
M_{22}^{(0,1)} = - \cK_{KK|KK}^{(0, 1)} \, [\Ob_{KK}^S/2 ]\;. 
\label{un7}
\eea
\bea
&& M_{11}^{(0,0)} =  - \cK_{\p\p|\p\p}^{(0,0)} \, [\Ob_{\p\p}^S/2] \;, 
\hspace{10mm}
M_{12}^{(0,0)} =  - \cK_{\p\p|KK}^{(0,0)} \, [\Ob_{KK}^S/2] \;,
\nn\\[2mm]
&& M_{13}^{(0,0)} =  - \cK_{\p\p|88}^{(0,0)} \, [\Ob_{88}^S/2] \;,
\hspace{10mm}
M_{21}^{(0,0)} =  - \cK_{\p\p|KK}^{(0,0)} \, [\Ob_{\p\p}^S/2] \;,
\nn\\[2mm]
&& M_{22}^{(0,0)} =  - \cK_{KK|KK}^{(0,0)} \, [\Ob_{KK}^S/2] \;,
\hspace{10mm}
M_{23} ^{(0,0)}=  - \cK_{KK|88}^{(0,0)} \, [\Ob_{88}^S/2] \;,
\nn \\[2mm]
&& M_{31}^{(0,0)} =  - \cK_{\p\p|88}^{(0,0)} \, [\Ob_{\p\p}^S/2] \;,
\hspace{10mm}
M_{32}^{(0,0)} =  - \cK_{KK|88}^{(0,0)} \, [\Ob_{KK}^S/2] \;,
\nn \\[2mm]
&& M_{33}^{(0,0)} =  - \cK_{88|88}^{(0,0)} \, [\Ob_{88}^S/2] \;.
\label{un8}
\eea
\vskip .4cm
The imaginary propagators $\Ob$ are given by
\bea 
&& \Ob_{ab}^S = -\,\frac{i}{8\p} \; \frac{Q_{ab}}{\sqrt{s}} \;
\theta(s \sm (M_a \sp M_b)^2) \;,
\label{un9}\\[2mm]
&& \Ob_{aa}^P = -\,\frac{i}{6\p} \; \frac{Q_{aa}^3}{\sqrt{s}} \;
\theta(s \sm4\, M_a^2) \;,
\label{un10}\\[2mm]
&& Q_{ab} = \frac{1}{2} \, \sqrt{s - 2\,(M_a^2 + M_b^2) + (M_a^2 - M_b^ 2)^2/s} \;.
\label{un11}
\eea 
\vskip .4cm
The functions $ \cK_{ab|cd}^{(J,I)} $ are the scattering kernels,
\begin{equation}
\cK_{(\p \p|\p\p)}^{(1,1)} =   - 2\, \lb\frac{G_V^2}{F^4} \rb \;\frac{s}{s-m_\rho^2} 
+ \lb \frac{1}{F^2} \rb_c ,
\label{k2}
\end{equation}

\begin{equation}
\cK_{(\p \p|KK)}^{(1,1)} = - \rtw \, \lb\frac{G_V^2}{F^4} \rb \;\frac{s}{s-m_\rho^2} 
+  \lb \frac{\rtw}{2\,F^2} \rb_c ,
\label{k3}
\end{equation}

\begin{equation}
 \cK_{(KK|KK)}^{(1,1)} = - \lb\frac{G_V^2}{F^4} \rb \;\frac{s}{s-m_\rho^2} 
+    \lb \frac{1}{2\,F^2} \rb_c ,
\label{k4} 
\end{equation}

\begin{equation} 
\cK_{(KK|KK)}^{(1,0)} = - 3\, \lb\frac{G_V^2 \; \sin^2\!\theta}{F^4} \rb \;\frac{s}{D_\phi^{\p\rho}} 
+  \lb \frac{3}{2\,F^2} \rb_c ,
\label{k6}
\end{equation}
\begin{equation} 
 \cK_{(\p 8|\p 8)}^{(0,1)} =  -\, \frac{1}{s-m_{a_0}^2}\,\lb\frac{4}{3\,F^4} \rb \;
\lb c_d \, (s\sm m_\p^2 \sm m_8^2)  + c_m \, 2m_\p^2 \rb^2 
+  \lb \frac{2 m_\p^2}{3 F^2} \rb_c ,
\label{k8}
\end{equation}
\bea
&& \cK_{(\p 8|KK)}^{(0,1)} =  -\, \frac{1}{s-m_{a_0}^2}\,\lb\frac{2\rtw}{\rth\,F^4} \rb 
\lb c_d \, (s\sm m_\p^2 \sm m_8^2)  + c_m \, 2m_\p^2 \rb \,
\lb c_d \, s - (c_d \sm c_m) \, 2m_K^2 \rb 
\nn\\[2mm]
&& + \lb \frac{(3s -4m_K^2)}{\rts\, F^2} \rb_c ,
\label{k9}
\eea
\begin{equation} 
 \cK_{(KK|KK)}^{(0,1)}  =  -\frac{1}{s-m_{a_0}^2}\,\lb\frac{2}{F^4} \rb \;
\lb c_d \, s - (c_d \sm c_m) \, 2m_K^2 \rb^2 
+ \lb \frac{s}{2F^2} \rb_c ,
\label{k10}
\end{equation}

\bea 
&& \cK_{(\p \p|\p \p)}^{(0,0)}  = -\, \frac{1}{s-m_{S1}^2}\,\lb\frac{12}{F^4} \rb \;
\lb \ct_d \, s - (\ct_d \sm \ct_m) \, 2m_\p^2\rb^2 
\nn\\[2mm]
&& -\, \frac{1}{s-m_{So}^2}\,\lb\frac{2}{F^4} \rb \;
\lb c_d \, s - (c_d \sm c_m) \, 2m_\p^2 \rb^2 
+ \lb \frac{2 s - m_\p^2}{ F^2} \rb_c ,
\label{k12}
\eea
\bea
&& \cK_{(\p \p|KK)}^{(0,0)}  = -\,\frac{1}{s-m_{S1}^2}\,\lb\frac{8\rth}{F^4} \rb \;
\lb \ct_d \, s - (\ct_d \sm \ct_m) \, 2m_\p^2\rb \,
\lb \ct_d \, s - (\ct_d \sm \ct_m) \, 2m_K^2 \rb 
\nn\\[2mm]
&& +   \frac{1}{s-m_{So}^2}\,\lb\frac{2}{\rth\,F^4} \rb \;
\lb c_d \, s - (c_d \sm c_m) \, 2m_\p^2 \rb 
\lb c_d \, s - (c_d \sm c_m) \, 2m_K^2 \rb  \nn\\[2mm]
&&+  \lb \frac{\rth\, s}{2 F^2} \rb_c ,
 \label{k13}\\[4mm]
&& \cK_{(\p \p|88)}^{(0,0)}  = -\, \frac{1}{s-m_{S1}^2}\,\lb\frac{4\rth}{F^4} \rb \;
\lb \ct_d \, s - (\ct_d \sm \ct_m) \, 2m_\p^2 \rb \,
\lb \ct_d \, s - (\ct_d \sm \ct_m) \, 2m_8^2\rb 
\nn\\[2mm]
&& +  \frac{1}{s-m_{So}^2}\,\lb\frac{2}{\rth\,F^4} \rb \;
\lb c_d \, s - (c_d \sm c_m) \, 2m_\p^2 \rb 
\lb c_d \, (s\sm 2 m_8^2)  + c_m \, (16m_K^2 \sm 10 m_\p^2)/3\rb \nn\\[2mm]
&& + \,  \lb \frac{\rth\, m_\p^2}{3 F^2} \rb_c ,
\label{k14}
\eea
%
%
%
\bea
&& \cK_{(KK|KK)}^{(0,0)}  = -\, \frac{1}{s-m_{S1}^2}\,\lb\frac{16}{F^4} \rb \;
\lb \ct_d \, s - (\ct_d \sm \ct_m) \, 2m_K^2\rb^2 
\nn\\[2mm]
&& -   \frac{1}{s-m_{So}^2}\,\lb\frac{2}{3\,F^4} \rb \;
\lb c_d \, s - (c_d \sm c_m) \, 2 m_K^2 \rb^2 
+  \lb  \frac{3 s}{2 F^2}  \rb_c ,
\label{k15}
\eea
%
%
%
%
\bea
&& \cK_{(KK|88)}^{(0,0)}  =  -\, \frac{1}{s-m_{S1}^2}\,\lb\frac{8}{F^4} \rb \;
\lb \ct_d \, s - (\ct_d \sm \ct_m) \, 2m_K^2 \rb \,
\lb \ct_d \, s - (\ct_d \sm \ct_m) \, 2m_8^2 \rb 
\nn\\[2mm]
&& -  \, \frac{1}{s-m_{So}^2}\,\lb\frac{2}{3\,F^4} \rb \;
\lb c_d \, s - (c_d \sm c_m) \, 2 m_K^2 \rb 
\lb c_d \, (s\sm 2 m_8^2)  + c_m \, (16m_K^2 \sm 10 m_\p^2)/3\rb 
\nn \\[2mm]
&& +  \,  \lb  \frac{9 s - 8m_K^2}{6 F^2} \rb_c ,
\label{k16}
\eea
%
%
%
%
\bea
&& \cK_{(88|88)}^{(0,0)}  =  -\, \frac{1}{s-m_{S1}^2}\,\lb\frac{4}{F^4} \rb \;
\lb \ct_d \, s - (\ct_d \sm \ct_m) \, 2m_8^2\rb^2 
\nn\\[2mm]
&& -  \, \frac{1}{s-m_{So}^2}\,\lb\frac{2}{3\, F^4} \rb \;
\lb c_d \, (s\sm 2 m_8^2)  + c_m \, (16m_K^2 \sm 10 m_\p^2)/3\rb^2 \nn\\[2mm]
&&+  \lb \frac{-7m_\p^2 +16 m_K^2}{9 F^2} \rb_c .
\label{k17}
\eea

\section{Scattering amplitudes}
\label{app:Scatteringamplitudes}
\setcounter{equation}{0}
\renewcommand{\theequation}{\thesection.\arabic{equation}}

The $ K^+K^- $ scattering amplitudes are written in terms of 
the denominators $ D^{(J,I)} $ as

\bea
&& A_{KK|KK}^{(1,1)} = \frac{1}{D_\rho (s_{12})}
\lb M_{21}^{(1,1)} \, \cK_{\p\p|KK}^{(1,1)} + \lp 1 \sm M_{11}^{(1,1)}\rp \, \cK_{KK|KK}^{(1,1)}  \rb \;,
\label{sca1}\\[4mm] 
&& A_{KK|KK}^{(1,0)} =  \frac{1}{D_\phi (m_{12}^2)}\;  \cK_{KK|KK}^{(1,0)} \;,
\label{sca2}\\[4mm]
&& A_{KK|KK}^{(0,1)} = \frac{ 1}{D_{a_0} (s_{12})}
\lb M_{21}^{(0,1)} \,  \cK_{\p 8|KK}^{(0,1)} + \lp 1 \sm M_{11}^{(0,1)} \rp \, \cK_{ KK|KK}^{(0,1)}  \rb ,
\label{sca3}\\[4mm] 
&& A_{KK|KK}^{(0,0)} = \frac{1}{D_S(s_{12} )}
\lc \lb M_{21}^{(0,0)} \lp 1\sm M_{33}^{(0,0)} \rp \sp M_{23}^{(0,0)} M_{31}^{(0,0)}\rb  \, \cK_{\p\p|KK}^{(0,0)}
\right.
\nn\\[2mm]
& & \left.  
\, + \lb \lp 1\sm M_{11}^{(0,0)} \rp \lp1\sm M_{33}^{(0,0)} \rp \sm M_{13}^{(0,0)} M_{31}^{(0,0)}\rb \, \cK_{KK|KK}^{(0,0)} 
\right.
\nn\\[2mm]
&& \left.  
\, + \lb M_{23}^{(0,0)} \lp 1\sm M_{11}^{(0,0)} \rp \sp M_{13}^{(0,0)} M_{21}^{(0,0)}\rb  \, \cK_{88|KK}^{(0,0)} \rc \;.
\label{sca4}
\eea


\addcontentsline{toc}{section}{References}
\setboolean{inbibliography}{true}
\bibliographystyle{LHCb}
\bibliography{main,standard,LHCb-PAPER,LHCb-CONF,LHCb-DP,LHCb-TDR}

\newpage
\centerline{\large\bf LHCb collaboration}
\begin{flushleft}
\small
R.~Aaij$^{30}$,
C.~Abell{\'a}n~Beteta$^{48}$,
B.~Adeva$^{45}$,
M.~Adinolfi$^{52}$,
C.A.~Aidala$^{80}$,
Z.~Ajaltouni$^{7}$,
S.~Akar$^{63}$,
P.~Albicocco$^{21}$,
J.~Albrecht$^{12}$,
F.~Alessio$^{46}$
% ...

{\footnotesize \it
% ...
$ ^{7}$Universit{\'e} Clermont Auvergne, CNRS/IN2P3, LPC, Clermont-Ferrand, France\\
% ...
$ ^{12}$Fakult{\"a}t Physik, Technische Universit{\"a}t Dortmund, Dortmund, Germany\\
% ...
$ ^{21}$INFN Laboratori Nazionali di Frascati, Frascati, Italy\\
% ...
$ ^{30}$Nikhef National Institute for Subatomic Physics, Amsterdam, Netherlands\\
% ...
$ ^{45}$Instituto Galego de F{\'\i}sica de Altas Enerx{\'\i}as (IGFAE), Universidade de Santiago de Compostela, Santiago de Compostela, Spain\\
$ ^{46}$European Organization for Nuclear Research (CERN), Geneva, Switzerland\\
% ...
$ ^{48}$Physik-Institut, Universit{\"a}t Z{\"u}rich, Z{\"u}rich, Switzerland\\
% ...
$ ^{52}$H.H. Wills Physics Laboratory, University of Bristol, Bristol, United Kingdom\\
% ...
$ ^{63}$University of Cincinnati, Cincinnati, OH, United States\\
% ...
$ ^{65}$Syracuse University, Syracuse, NY, United States\\
% ...
$ ^{80}$University of Michigan, Ann Arbor, United States, associated to $^{65}$\\
% ...
}
\end{flushleft}

\end{document}